\documentclass[11pt]{article}
\usepackage{amssymb}
\makeatletter\@addtoreset {equation}{section}\makeatother

\newtheorem{theorem}{Theorem}[section]
\newtheorem{lemma}[theorem]{Lemma}
\newtheorem{corollary}[theorem]{Corollary}
\newtheorem{proposition}[theorem]{Proposition}

\newtheorem{definition}[theorem]{Definition}
\newtheorem{remark}[theorem]{Remark}
\newtheorem{example}[theorem]{Example}
\newenvironment{proof}{
    \noindent {\it Proof.}}{\hfill$\Box$
}

\usepackage[dvips]{epsfig}
\usepackage{graphicx}

\begin{document}

\title{\bf Dark solitons in external potentials}
\author{Dmitry E. Pelinovsky$^{\dagger}$ and Panayotis G. Kevrekidis$^{\dagger  \dagger}$ \\
{\small $^{\dagger}$ Department of Mathematics, McMaster
University, Hamilton, Ontario, Canada, L8S 4K1} \\
{\small $^{\dagger \dagger}$ Department of Mathematics and
Statistics, University of Massachusetts, Amherst, MA 01003 } }
\date{\today}
\maketitle

\begin{abstract}
We consider the persistence and stability of dark solitons in the
Gross--Pitaevskii (GP) equation with a small decaying potential. We
show that families of black solitons with zero speed originate from
extremal points of an appropriately defined {\it effective
potential} and persist for sufficiently small strength of the
potential. We prove that families at the maximum points are
generally unstable with exactly one real positive eigenvalue, while
families at the minimum points are generally unstable with exactly
two complex-conjugated eigenvalues with positive real part. This
mechanism of destabilization of the black soliton is confirmed in
numerical approximations of eigenvalues of the linearized GP
equation and full numerical simulations of the nonlinear GP equation
with cubic nonlinearity. We illustrate the monotonic instability
associated with the real eigenvalues and the oscillatory instability
associated with the complex eigenvalues and compare the numerical
results of evolution of a dark soliton with the predictions of
Newton's particle law for its position.
\end{abstract}

\section{Introduction}

{\em Dark} solitons are solutions of nonlinear PDEs in the space of
one dimension with non-zero boundary conditions and non-zero phase
shift. They are represented by a family of traveling wave solutions
extending from the limit of zero speed (so-called {\em black}
solitons) to the limit of sound speed (so-called {\em grey}
solitons). From a physical point of view, dark solitons are waves in
defocusing nonlinear systems which move along the modulationally
stable continuous-wave background.

The original interest in studies of dark solitons emerged, roughly,
two decades ago
in the context of nonlinear optics, where dark solitons provide
modulations of the light intensity of an optical beam traveling in a
planar waveguide \cite{KL98}. The main model for dark solitons in
nonlinear optics is the generalized nonlinear Schr\"{o}dinger (NLS)
equation
\begin{equation}
\label{NLS} i u_t = - \frac{1}{2} u_{xx} + f(|u|^2) u,
\end{equation}
where $u(x,t) : \mathbb{R} \times \mathbb{R}_+ \mapsto \mathbb{C}$
and $f(q) : \mathbb{R}_+ \to \mathbb{R}$. We assume that $f(q)$ is a
smooth monotonically increasing function on $q \in {\cal I} \subset
\mathbb{R}_+$. Particular examples of $f(q)$ include the cubic NLS
with $f(q) = q$, the cubic-quintic NLS with $f(q) = \alpha q + q^2$,
$\alpha \in \mathbb{R}$ and the saturable NLS with $f = -1/(1 +
\beta q)$, $\beta \in \mathbb{R}_+$.

Among others, several analytical results were important in the
development of dark solitons in recent years: perturbation theory
based on renormalized power \cite{KV94} and momentum \cite{KY94},
orbital stability of dark solitons \cite{B96,BP93}, completeness of
eigenfunctions in the cubic NLS \cite{CH98,CH99}, inverse scattering
for the vector cubic NLS equation \cite{PAB}, construction of the
Evans function for dark solitons in the perturbed cubic NLS
\cite{KR00}, asymptotic analysis of the radiation and dynamics of
dark solitons \cite{L04,LF97,PKA96}, and spectral analysis of
transverse instabilities of one-dimensional dark solitons
\cite{KT88,PSK95}.

Subsequently, rigorous analysis of the existence and stability of
dark solitons was developed in the last decade based on the earlier
physical literature. In particular, Zhidkov \cite{Zh92} proved local
existence of the Cauchy problem and stability of kink solutions, de
Bouard \cite{B95} proved spectral and nonlinear instability of
stationary bubbles (black solitons with zero phase shift), Lin
\cite{Lin02} proved the criterion for orbital stability and
instability of dark solitons (for non-zero velocities), Maris
\cite{M03} studied bifurcation of dark solitons (for non-zero
velocities) in the presence of delta-function potential and its
generalizations, and Di Menza and Gallo \cite{MG06} proved recently
the stability criterion for kinks (black solitons with a non-zero
phase shift). Extensions of the existence and stability theory to
the generalized NLS equation in two and higher dimensions were also
developed by J.C. Saut and his co-workers.

While many mathematical results are now available for solutions of
the generalized NLS equation (\ref{NLS}) with non-zero boundary
conditions and non-zero phase shift, dark solitons have suffered
a decreasing popularity in the context of nonlinear
optics. This is not only because they possess infinite energy
due to non-zero boundary conditions but also because it is
difficult from the experimental point of view to separate the
effects of the dark soliton dynamics and the background dynamics.
Also no engineering or commercial applications of dark solitons
have been proposed, to the best of our knowledge.

Nevertheless, the interest in these nonlinear waveforms has recently
been rejuvenated by the rapid development of a new area of physics,
namely the field of Bose-Einstein condensates (BECs)
\cite{PS02,PS03}. In the latter setting, dark solitons typically
move along the localized ground state trapped by the external
potentials \cite{PPFA04}. The main model for BECs is a modification
of the NLS equation (\ref{NLS}) with an external potential, which is
called the Gross-Pitaevskii (GP) equation,
\begin{equation}
\label{GP} i u_t = - \frac{1}{2} u_{xx} + f(|u|^2) u + \epsilon V(x)
u,
\end{equation}
where $\epsilon \in \mathbb{R}$ is the strength of the potential
$V(x)$ and $V(x) : \mathbb{R} \to \mathbb{R}$ is assumed to be a
smooth function satisfying one of the three properties:
\begin{itemize}
\item[(i)] $V(x)$ is bounded and decaying, e.g.
\begin{equation}
\label{potential} \exists C > 0, \; \kappa > 0 : \quad |V(x)| \leq C
e^{-\kappa |x|}, \quad \forall x \in \mathbb{R}
\end{equation}

\item[(ii)] $V(x)$ is bounded but non-decaying,
e.g. $V(x+d) = V(x)$, $x \in \mathbb{R}$ with period $d > 0$

\item[(iii)] $V(x)$ is unbounded, e.g. $V(x) = x^2 + \tilde{V}(x)$,
where $\tilde{V}(x)$ is bounded on $x \in \mathbb{R}$.
\end{itemize}
The last two cases are of particular interest in the context of
Bose-Einstein condensates, where dynamics of dark solitons
(localized dips in the nonlinear ground state trapped by the
potential $V(x)$) was studied in many recent papers, see
\cite{Kon05,PFK05} for surveys of results. Although cases (ii) and
(iii) have been initially our primary motivations, this paper
covers only the case (i) when $V(x)$ is bounded and exponentially
decaying as in (\ref{potential}). In particular, we consider the
class of symmetric potentials $V(-x) = V(x)$ with two examples
\begin{equation}
\label{potential-explicit} V_1(x) = -{\rm sech}^2\left( \frac{\kappa
x}{2} \right), \qquad V_2(x) = x^2 e^{-\kappa |x|}, \qquad x \in
\mathbb{R}.
\end{equation}
The other cases (ii) and (iii) of the potential $V(x)$ will be the
subject of forthcoming studies.

We note in passing that while the potentials
(\ref{potential-explicit}) are, perhaps, less customary than the
standard magnetic (parabolic) and optical lattice (periodic)
potentials \cite{KF04}, they are nonetheless still quite physically
relevant. In particular, the potential $V_1(x)$ corresponds to a
red-detuned laser beam potential, analogous to the one used in
\cite{ORVACK00,RKDKHK99}. The potential $V_2(x)$ represents an
all-optically trapped BEC, as modeled in \cite{CHM05} and
experimentally implemented in \cite{BSC01}.

The strategy of our work is to exploit solutions of the GP
equation (\ref{GP}) in the limit of small strength $\epsilon$.
Starting with the limit $\epsilon = 0$, where both existence and
stability of dark solitons with non-zero and zero speeds are known
from the analysis of the NLS equation (\ref{NLS}), we shall use the
method of Lyapunov--Schmidt reductions and obtain existence and
stability results for small $\epsilon$. From the technical point of
view, we use local bifurcation analysis of solutions of the ODEs
with non-zero boundary conditions similarly to \cite{Maris},
persistence analysis of isolated eigenvalues in the problems with
small potentials similarly to \cite{Kapitula}, the count of
eigenvalues in the generalized eigenvalue problem for self-adjoint
operators with no spectral gaps similarly to \cite{Chugunova}, and
the construction of $L^2$ eigenfunctions of the stability problem
with fast and slow exponential decay similarly to \cite{PG97}.
Since our starting point is the case of $\epsilon = 0$, we will
also give alternative rigorous proofs to the existence and
stability of black solitons in the NLS equation (\ref{NLS}), which
complement the recent work of \cite{MG06}.

Our main results are listed as follows.

\begin{itemize}
\item[(i)] Let $u = \phi_0(x-s) e^{- i \omega t + i \theta}$ be a
solution of the NLS equation (\ref{NLS}) with $s \in \mathbb{R}$,
$\theta \in \mathbb{R}$, $\omega = f(q_0)$, $q_0 \in {\cal I}
\subset \mathbb{R}_+$ and $\phi_0(x) : \mathbb{R} \mapsto
\mathbb{R}$ such that $\phi_0(x) \to \pm \sqrt{q_0}$ as $x \to \pm
\infty$ exponentially fast. Let $s_0$ be the root of the function
$M'(s)$, such that
\begin{equation}
\label{effective-potential-M} M'(s_0) = \int_{\mathbb{R}} V'(x)
\left[ q_0 - \phi_0^2(x-s_0) \right] dx = 0
\end{equation}
and
\begin{equation}
\label{effective-potential-M-2} M''(s_0) = \int_{\mathbb{R}} V''(x)
\left[ q_0 - \phi_0^2(x-s_0) \right] dx \neq 0.
\end{equation}
Then, there exists a unique continuation of the solution $u =
\phi_{\epsilon}(x-s_{\epsilon}) e^{-i \omega t + i \theta}$ of the
GP equation (\ref{GP}) with $V(x)$ in (\ref{potential}) and
$\epsilon$ sufficiently small, where $\phi_{\epsilon} : \mathbb{R}
\mapsto \mathbb{R}$ and $\phi_{\epsilon}(x) \to \pm \sqrt{q_0}$ as
$x \to \pm \infty$ exponentially fast, such that
$\phi_{\epsilon}(x)$ and $s_{\epsilon}$ are $\epsilon$-close to
$\phi_0(x)$ and $s_0$ in $L^{\infty}$-norm.

\item[(ii)] Let the solution $\phi_0(x)$ be spectrally stable in
the time evolution of the NLS equation (\ref{NLS}). Then, the
solution $\phi_{\epsilon}(x)$ is spectrally unstable in the time
evolution of the GP equation (\ref{GP}) for sufficiently small
$\epsilon$ with exactly one real positive eigenvalue if $M''(s_0) <
0$ and exactly two complex-conjugate eigenvalues with positive real
part if $M''(s_0) > 0$.

\item[(iii)] Let the function $u_0(x) : \mathbb{R} \mapsto
\mathbb{R}$ be close to the solution $\phi_{\epsilon}(x)$ in
$L^{\infty}$-norm and satisfy the same boundary conditions $u_0(x)
\to \pm \sqrt{q_0}$ as $x \to \pm \infty$. Then, the solution
$u(x,t) : \mathbb{R} \times \mathbb{R}_+ \mapsto \mathbb{C}$ of
the Cauchy problem to the GP equation (\ref{GP}) with $u(x,0) =
u_0(x)$ remains close to the dark soliton
$\phi_{\epsilon}(x-s(t))$ with varying coordinate $s(t)$ for $0
\leq t < T$ such that $s(t)$ solves the Newton's particle equation
\begin{equation}
\mu_0 \ddot{s} - \epsilon \lambda_0 M''(s) \dot{s} = - \epsilon
M'(s),  \qquad 0 \leq t < T, \label{Newton-law}
\end{equation}
where $M(s)$ is the effective potential in
(\ref{effective-potential-M}) and (\ref{effective-potential-M-2}),
$(\mu_0,\lambda_0)$ are constants representing the soliton's mass
and gain terms, and the time $T > 0$ is of the order of ${\rm
O}(1/\epsilon)$.
\end{itemize}

Statement (i) is formulated and proved in Section 2 (see Theorem
\ref{theorem-continuation}). Statement (ii) is formulated and proved
in Section 3 (see Theorems \ref{theorem-stability-kink},
\ref{theorem-stability-kink-mode}, and \ref{theorem-persistence})
under some non-degeneracy assumptions (see Corollary
\ref{corollary-persistence} and Remark \ref{remark-persistence}).
The two complex-conjugate eigenvalues with positive real part in the
statement (ii) for $M''(s_0) > 0$ follows from the linearized
version of the Newton's particle equation (\ref{Newton-law}) with
$\mu_0 > 0$ and $\lambda_0 > 0$, which is rigorously derived in
Section 4 (see Theorem \ref{theorem-splitting}). Constants
$(\mu_0,\lambda_0)$ are defined by Remark \ref{remark-constants}.
The linearization around a black soliton is investigated in Section
5 using numerical bifurcation theory and approximations of
eigenvalues. Statement (iii) is a conjecture which is tested against
appropriately crafted numerical experiments in Section 6. The
summary and open problems are discussed in the concluding Section 7.

\section{Existence analysis of dark and black solitons}

We first consider the family of traveling solutions of the NLS
equation (\ref{NLS}). After we formulate the conditions for
existence of dark and black solitons, we address persistence of
stationary solutions in the Gross--Pitaevskii equation (\ref{GP})
for small $\epsilon$ and the potential $V(x)$ in (\ref{potential}).
Applications of the persistence analysis to the two potentials
(\ref{potential-explicit}) indicate that the families of black
solitons bifurcate from the extremal points of the effective
potential $M(s)$ in
(\ref{effective-potential-M})--(\ref{effective-potential-M-2}).

\begin{definition}
A dark soliton is the traveling solution of the NLS equation
(\ref{NLS}) in the form:
\begin{equation}
\label{dark-soliton} u(x,t) = U(x-vt) e^{-i\omega t}, \quad U(z) =
\Phi(z) e^{i \Theta(z)}, \quad z = x-vt,
\end{equation}
where $U : \mathbb{R} \mapsto \mathbb{C}$, $\Phi : \mathbb{R}
\mapsto \mathbb{R}_+$, and $\Theta : \mathbb{R} \mapsto
[-\pi,\pi]$ are smooth functions of their arguments, which
converge exponentially fast to the boundary conditions
\begin{equation}
\label{bc-dark-soliton} \lim_{z \to \pm \infty} \Phi(z) =
\sqrt{q_0}, \qquad \lim_{z \to \pm \infty} \Theta(z) = \Theta_{\pm}.
\end{equation}
Here $(\omega,v) \in {\cal D} \subset \mathbb{R}^2$, $q_0 \in {\cal
I} \subset \mathbb{R}_+$ and $\Theta_{\pm} \in [-\pi,\pi]$ are
parameters of the solution. Moreover, the functions $\Phi(z)$ and
$\Theta(z)$ can be chosen to satisfy the normalization conditions
$\Phi'(0) = 0$ and $\Theta_+ = 0$. Additionally, we require that
$\Phi(z) < \sqrt{q_0}$ on $z \in \mathbb{R}$.
\label{definition-dark-soliton}
\end{definition}

\begin{remark}
{\rm The normalization conditions for $\Phi(z)$ and $\Theta(z)$
use the gauge [$u(x,t) \to u(x,t) e^{i \theta}$, $\forall \theta
\in \mathbb{R}$] and translational [$u(x,t) \to u(x-s,t)$,
$\forall s \in \mathbb{R}$] invariance of the NLS equation, while
the linear growth of $\Theta(z)$ in $z$ is excluded by the Galileo
invariance [$u(x,t) \to u(x-kt,t) e^{i kx - ik^2 t/2}$, $\forall k
\in \mathbb{R}$].}
\end{remark}

\begin{theorem}
Let $f(q)$ be $C^1(\mathbb{R}_+)$ and fix $q_0 \in \mathbb{R}_+$
such that $f'(q_0) > 0$ and $c = \sqrt{q_0 f'(q_0)} > 0$. The dark
soliton $U(z)$ of Definition \ref{definition-dark-soliton} exists if
$\omega = f(q_0)$ and $v \in (-c,0) \cup (0,c)$. Moreover, for these
solutions, $\Phi(z)$ has a global minimum at $z = 0$ with $0 <
\Phi(0) < \sqrt{q_0}$ and $\Theta(z)$ is monotonically decreasing
for $v > 0$ and increasing for $v < 0$.
\label{lemma-existence-dark-soliton}
\end{theorem}

\begin{proof}
It follows immediately that the function $U(z)$ satisfies the
second-order ODE:
\begin{equation}
\label{second-order-ode-u} - i v U' + \frac{1}{2} U'' + (\omega -
f(|U|^2)) U = 0,
\end{equation}
while the functions $\Phi(z)$ and $\Theta(z)$ satisfy the ODE system
in the hydrodynamic form:
\begin{eqnarray}
\label{Phi-equation-1} \Phi'' - (\Theta')^2 \Phi + 2 v \Theta' \Phi
+ 2 (\omega - f(\Phi^2)) \Phi & = & 0 \\
\label{Theta-equation-2} (\Theta' \Phi^2 - v \Phi^2)' & = & 0
\end{eqnarray}
Integrating the ODE (\ref{Theta-equation-2}) under the boundary
conditions (\ref{bc-dark-soliton}), we obtain
\begin{equation}
\label{theta-ODE} \Theta' = v \frac{\Phi^2 - q_0}{\Phi^2}.
\end{equation}
As a result, the ODE (\ref{Phi-equation-1}) reads as follows
\begin{equation}
\label{second-order-ode-Phi} \Phi'' + 2 (\omega - f(\Phi^2)) \Phi +
v^2 \frac{\Phi^4 - q_0^2}{\Phi^3} = 0.
\end{equation}
The equilibrium point $\Phi = \sqrt{q_0}$ exists if and only if
$\omega = f(q_0)$ and it is a non-degenerate hyperbolic point if and
only if $v^2 < c^2 = q_0 f'(q_0)$. Integrating the second-order ODE
(\ref{second-order-ode-Phi}) subject to the boundary conditions
(\ref{bc-dark-soliton}), we obtain the quadrature
\begin{equation}
\label{quadrature} (\Phi')^2 - 2 W(\Phi^2) + v^2 \frac{(\Phi^2 -
q_0)^2}{\Phi^2} = 0,
\end{equation}
where
\begin{equation}
\label{W-function} W(q) = \int_{q_0}^{q} \left( f(q_0) - f(q)
\right) dq.
\end{equation}
The non-degenerate turning point $0 < \Phi_0 < \sqrt{q_0}$ of the
effective potential $-2W(q)+v^2(q-q_0)/q$ exists for any $0 < v^2 <
c^2$, such that the trajectory from the hyperbolic point turns at
$\Phi(0) = \Phi_0$ and returns back to the hyperbolic point forming
a homoclinic orbit, thus proving the statement.
\end{proof}

\begin{remark}
{\rm It is not difficult to prove from the ODE analysis that
$U(z)$ is in fact $C^{\infty}(\mathbb{R})$ for $v \in (-c,0) \cup
(0,c)$ and $U(z)$ converges to the limiting values $\sqrt{q_0}
e^{i \Theta_{\pm}}$ as $z \to \pm \infty$ at the exponential rate
${\rm O}(e^{-\sqrt{c^2 - v^2}{z}})$. Although Lemma
\ref{lemma-existence-dark-soliton} formulates only the sufficient
condition for existence of dark solitons, it is possible to
consider all other solutions of the ODE system for $\Phi(z)$ and
$\Theta(z)$. The case $v^2 > c^2$ corresponds to the elliptic
point $\Phi = \sqrt{q_0}$ and has no homoclinic orbits at all. The
marginal case $v^2 = c^2$ may admit an algebraically decaying
homoclinic orbit, which is excluded by Definition
\ref{definition-dark-soliton}. In the case $v^2 < c^2$, other
homoclinic orbits can exist with $\Phi(z) > \sqrt{q_0}$ but they
are also excluded by Definition \ref{definition-dark-soliton}. The
only case, which may be included into the necessary and sufficient
condition of existence of dark soliton, is the limit $v \to 0$.
This limit leads to two different kinds of solutions which are
classified in the following definition.}
\label{remark-convergence-dark-soliton}
\end{remark}

\begin{definition}
The limiting solution $\phi_0(x) = \lim\limits_{v \downarrow 0}
U(x)$ in the family of dark solitons of Definition
\ref{definition-dark-soliton} is said to be the black soliton if
$\phi_0(x)$ is a real-valued smooth function on $x \in
\mathbb{R}$. The black soliton is called the bubble if $\phi_0(-x)
= \phi_0(x)$ with $0 \leq \phi_0(0) < \sqrt{q_0}$ and
$\lim\limits_{x \to \pm \infty} \phi_0(x) = \sqrt{q_0}$, while it
is called the kink if $\phi_0(-x) = - \phi_0(x)$ with $\phi_0(0) =
0$ and $\lim\limits_{x \to \pm \infty} \phi_0(x) = \pm
\sqrt{q_0}$, where $\phi_0(x)$ converges exponentially fast to the
limits. \label{definition-black-soliton}
\end{definition}

\begin{theorem}
Let $f(q)$ satisfy the same conditions as in Lemma
\ref{lemma-existence-dark-soliton} and $W(q)$ be defined by
(\ref{W-function}). The kink solution of Definition
\ref{definition-black-soliton} exists if and only if $\omega =
f(q_0)$, $v = 0$, and $W(q) > 0$ for all $0 \leq q < q_0$. The
bubble solution of Definition \ref{definition-black-soliton} exists
if and only if $\omega = f(q_0)$, $v = 0$, and there exists the
largest value $q_1 \in [0,q_0)$ that satisfies $W(q_1) = 0$ and
$W'(q_1) > 0$. \label{lemma-existence-black-soliton}
\end{theorem}

\begin{proof}
By Lemma \ref{lemma-existence-dark-soliton}, we have $0 < \Phi(z)
< \sqrt{q_0}$ and $\Theta'(z) < 0$ on $z \in \mathbb{R}$ for $v
> 0$ and hence no kink or bubble solutions may exist for $v
> 0$. Let us consider the real-valued solution $\phi_0(x)$ of the second-order ODE
\begin{equation}
\label{second-order-ode-phi-0} \frac{1}{2} \phi_0'' + (\omega -
f(\phi_0^2)) \phi_0 = 0.
\end{equation}
The equilibrium points $\phi_0 = \pm \sqrt{q_0}$ exist if and only
if $\omega = f(q_0)$ and they are non-degenerate hyperbolic
points. Integrating the second-order ODE
(\ref{second-order-ode-phi-0}) under the boundary conditions in
Definition \ref{definition-black-soliton}, we obtain
$$
(\phi_0')^2 - 2 W(\phi_0^2) = 0,
$$
where $W(q)$ is defined by (\ref{W-function}). No turning point of
$W(q)$ exists if $W(q) > 0$ for any $q \in [0,q_0)$, and the
outgoing trajectory from the hyperbolic point $\phi_0 =
\sqrt{q_0}$ connects the incoming trajectory to the hyperbolic
point $\phi_0 = -\sqrt{q_0}$ and vice versa forming a pair of
heteroclinic orbits. If there exists a non-degenerate largest
turning point $q_1$ of $W(q)$ on $q_1 \in [0,q_0)$, such that
$W(q_1) = 0$ and $W'(q_1) > 0$, then the trajectory from the
hyperbolic point $\phi_0 = \sqrt{q_0}$ turns at $\phi_0 =
\sqrt{q_1}$ and returns back to the point $\phi_0 = \sqrt{q_0}$
forming a homoclinic orbit. If the point $q_1$ is degenerate, such
that $W(q_1) = W'(q_1) = 0$, there exists a front solution
$\phi_0(x)$ from $\lim\limits_{x \to -\infty} \phi_0(x) =
\sqrt{q_1}$ to $\lim\limits_{x \to \infty} \phi_0(x) = \sqrt{q_0}$
and vice versa forming a pair of fronts, which are excluded by
Definition \ref{definition-black-soliton}.

It remains to prove that the family of dark solitons $U(x)$ for $0
< v < c$ converges to the black soliton $\phi_0(x)$ as $v
\downarrow 0$. The proof follows from the quadrature
(\ref{quadrature}). There exist unique classical solutions
$\phi_{\pm}(z)$ on $z \in \mathbb{R}_{\pm}$ for any $v
> 0$ with $\Phi_{\pm}(0) = \sqrt{q_*}$ found from the largest root of
$-2 W(q_*) + v^2(q_*-q_0)/q_* = 0$ on $q \in [0,q_0)$. In the
limit $v \downarrow 0$, the root $q_*$ converges to $0$ if $W(0) >
0$ and no other roots of $W(q)$ exists on $q \in [0,q_0)$.
Otherwise, the root $q_*$ converges to $q_1$, where $q_1$ is the
largest root of $W(q)$ on $q \in [0,q_0)$. It follows from the ODE
(\ref{theta-ODE}) in the limit $v \downarrow 0$ that $\Theta(z)$
is piecewise constant function on $z \in \mathbb{R}$ with a
possible jump discontinuity at $z = 0$.

Let us write $x = z$ for $v = 0$. In the case $q_* \downarrow 0$
as $v \downarrow 0$, the two smooth solutions $\Phi_{\pm}(x)$ are
glued into one smooth real-valued solution $\phi_0(x)$ if and only
if $\Theta(x) = 0$ on $x \in \mathbb{R}_+$ and $\Theta(x) = \pi$
on $x \in \mathbb{R}_-$ (under the normalization $\Theta_+ = 0$).
This limiting solution becomes the kink of Definition
\ref{definition-black-soliton}. In the case $q_* \downarrow q_1$
as $v \downarrow 0$ and $q_1$ is a non-degenerate root of $W(q)$,
the two smooth solutions $\Phi_{\pm}(x)$ are glued into a smooth
real-valued solution $\phi_0(x)$ if and only if $\Theta(x) = 0$ on
$x \in \mathbb{R}$ (under the same normalization). This limiting
solution becomes the bubble of Definition
\ref{definition-black-soliton}.
\end{proof}

\begin{remark}
{\rm Due to their potential stability in the time evolution of the
NLS equation (\ref{NLS}), only kinks are considered in the GP
equation (\ref{GP}) for sufficiently small $\epsilon$. The bubbles
are always unstable in the time evolution of the NLS equation
(\ref{NLS}) \cite{B95}.}
\end{remark}

\begin{definition}
The kink mode is the stationary solution of the GP equation
(\ref{GP}) in the form:
$$
u(x,t) = \phi_{\epsilon}(x) e^{-i f(q_0) t + i \theta}, \qquad
\theta \in \mathbb{R},
$$
where $\phi_{\epsilon}(x)$ is a real-valued smooth function on $x
\in \mathbb{R}$, which converges exponentially fast to the
boundary conditions $\lim\limits_{x \to \pm \infty}
\phi_{\epsilon}(x) = \pm \sqrt{q_0}$, for any $\epsilon \in
\mathbb{R}$. \label{definition-kink}
\end{definition}

\begin{lemma}
Let $\phi_{\epsilon}(x)$ be the kink mode of Definition
\ref{definition-kink} and $V(x)$ be the potential satisfying
(\ref{potential}). Then, for any $\epsilon \neq 0$,
\begin{equation}
\int_{\mathbb{R}} V'(x) \left[ q_0 - \phi_{\epsilon}^2(x) \right] dx
= 0. \label{nec-cond-kink}
\end{equation}
\label{lemma-necessary-condition}
\end{lemma}

\begin{proof}
The stationary solutions of Definition \ref{definition-kink}
satisfy the second-order ODE:
\begin{equation}
\label{second-order-ode-phi-epsilon} \frac{1}{2} \phi_{\epsilon}'' +
\left( f(q_0) - f(\phi_{\epsilon}^2)  \right) \phi_{\epsilon} =
\epsilon V(x) \phi_{\epsilon},
\end{equation}
which is generated by the Hamiltonian function
$$
E(\phi_{\epsilon},\phi_{\epsilon}',x) = \frac{1}{2}
(\phi_{\epsilon}')^2 - W(\phi_{\epsilon}^2) + \frac{\epsilon}{2}
V(x) \left[ q_0 - \phi_{\epsilon}^2 \right],
$$
where $W(q)$ is given by (\ref{W-function}). Therefore, the change
of $E(\phi_{\epsilon}(x),\phi_{\epsilon}'(x),x)$ at the classical
solution $\phi_{\epsilon}(x)$ of the second-order ODE
(\ref{second-order-ode-phi-epsilon}) is given by
$$
\frac{d E}{d x} = \epsilon V'(x) \left[ q_0 - \phi_{\epsilon}^2(x)
\right].
$$
Integrating this equation on $x \in \mathbb{R}$ and using the
boundary conditions for $\phi_{\epsilon}(x)$ of Definition
\ref{definition-kink}, such that $\lim\limits_{x \to \pm \infty}
E(\phi_{\epsilon}(x),\phi_{\epsilon}'(x),x) = 0$, we derive the
condition (\ref{nec-cond-kink}).
\end{proof}

\begin{remark}
If $V(-x) = V(x)$ and $\phi_{\epsilon}(-x) = -\phi_{\epsilon}(x)$ on
$x \in \mathbb{R}$, the necessary condition (\ref{nec-cond-kink}) is
always satisfied. The center of the kink is located at $x = 0$,
which is the minimal point of $V(x)$ if $V''(0) > 0$ and maximal
point if $V''(0) < 0$. \label{remark-symmetry-kink-mode}
\end{remark}

\begin{remark}
{\rm The necessary condition (\ref{nec-cond-kink}) specifies
restrictions on the shape of the kink mode $\phi_{\epsilon}(x)$
but does not give us any information about its existence. By using
the smallness of the external parameter $\epsilon$ in front of
$V(x)$, we will show that this condition is equivalent to the
bifurcation equation in the Lyapunov--Schmidt reduction technique.
A similar result for bright solitons was obtained in
\cite{Kapitula}. A different role of the condition
(\ref{nec-cond-kink}) was exploited in \cite{PSK04} in the context
of local bifurcations of small gap solitons in finite periodic
potentials $V(x)$.  }
\end{remark}

\begin{theorem}
Let $\phi_0(x)$ be the kink of Definition
\ref{definition-black-soliton}. Let $s_0$ be a simple root of the
function
\begin{equation}
\label{M-function} M'(s) = \int_{\mathbb{R}} V'(x) \left[ q_0 -
\phi_0^2(x-s) \right] dx, \qquad s \in \mathbb{R},
\end{equation}
such that $M'(s_0) = 0$ and $M''(s_0) \neq 0$. Let $f(q)$ be
$C^1(\mathbb{R}_+)$ and $V(x)$ be $C^2(\mathbb{R})$ satisfying
(\ref{potential}). Then, there exists a unique continuation of
$\phi_0(x-s_0)$ to the kink mode $\phi_{\epsilon}(x-s_{\epsilon})$
of Definition \ref{definition-kink} for sufficiently small
$\epsilon$, such that $\phi_{\epsilon}(x)$ and $s_{\epsilon}$ are
$\epsilon$-close to $\phi_0(x)$ and $s_0$ in the
$L^{\infty}$-norm. \label{theorem-continuation}
\end{theorem}

\begin{proof}
We use the decomposition $\phi_{\epsilon}(x) = \phi_0(x-s) +
\varphi(x,\epsilon,s)$ and rewrite the second-order ODE
(\ref{second-order-ode-phi-epsilon}) for $\phi_{\epsilon}(x)$ as
the root of the nonlinear operator-valued function
\begin{equation}
\label{lyapunov-schmidt} F(\varphi,\epsilon,s) = L_+ \varphi +
N(\varphi,s,\epsilon) + \epsilon V(x) \left[ \phi_0(x-s) + \varphi
\right],
\end{equation}
where $L_+ : H^2(\mathbb{R}) \mapsto L^2(\mathbb{R})$ is the
self-adjoint operator parameterized by $s$
$$
L_+ = -\frac{1}{2} \partial_x^2 + f(\phi_0^2) - f(q_0) + 2 \phi_0^2
f'(\phi_0^2),
$$
and $N(\varphi,\epsilon,s) : H^2(\mathbb{R}) \mapsto
H^2(\mathbb{R})$ is the nonlinear vector field
$$
N = \phi_0 \left[ f((\phi_0 +\varphi)^2) - f(\phi_0^2) - 2 \phi_0
\varphi f'(\phi_0^2) \right] + \varphi \left[ f((\phi_0 +
\varphi)^2) - f(\phi_0^2) \right],
$$
such that $N(\varphi,\epsilon,s) = {\rm o}(\|\varphi\|_{H^2})$ as
$\|\varphi\|_{H^2} \to 0$ (since $f \in C^1(\mathbb{R}_+)$).
Because $\phi_0(x)$ converges to $\pm \sqrt{q_0}$ as $x \to \pm
\infty$ exponentially fast, the essential spectrum of $L_+$ is
bounded from below by $2 c^2 > 0$. The operator may have isolated
positive eigenvalues and no negative eigenvalues since the kernel
$L_+ \phi'(x-s) = 0$ is a positive definite ground state.
Therefore, $H^2(\mathbb{R}) = {\rm Ker}(L_+) \oplus {\rm
Ker}(L_+)^{\perp}$ and the method of Lyapunov--Schmidt reductions
can be applied. Projection of $F(\varphi,\epsilon,s)$ onto ${\rm
Ker}(L_+)^{\perp}$ defines a unique smooth map $(x,\epsilon,s)
\mapsto \varphi \in {\rm Ker}(L_+)^{\perp} \subset
H^2(\mathbb{R})$ such that $\| \varphi \|_{H^2} = {\rm
O}(\epsilon)$ as $\epsilon \to 0$. By the Sobolev Embedded
Theorem, $\| \varphi \|_{L^{\infty}} = {\rm O}(\epsilon)$, such
that $\varphi(x)$ can be decomposed as follows:
$$
\varphi = \epsilon \varphi_1(x) + \tilde{\varphi}(x,\epsilon,s),
$$
where $\varphi_1(x)$ is specified below and $\|
\tilde{\varphi}\|_{L^{\infty}} = {\rm o}(\epsilon)$. Projection of
$F(\varphi,\epsilon,s)$ onto ${\rm Ker}(L_+)$ defines the
bifurcation equation:
\begin{equation}
\label{implicit-function} G(\epsilon,s) = \epsilon \left( \phi_0',
V(x) (\phi_0 + \varphi) \right) + \left( \phi_0',
N(\varphi,\epsilon,s)\right) = \frac{\epsilon}{2} M'(s) +
\tilde{G}(\epsilon,s),
\end{equation}
where $(\cdot,\cdot)$ is the standard inner product in
$L^2(\mathbb{R})$ and $\tilde{G}(\epsilon,s) = {\rm o}(\epsilon)$
as $\epsilon \to 0$.  By the Implicit Function Theorem for the
root of (\ref{implicit-function}), we obtain that $s = s_0 +
\tilde{s}(\epsilon)$, where $M'(s_0) = 0$, $M''(s_0) \neq 0$, and
$\tilde{s} = {\rm o}(1)$ as $\epsilon \to 0$. By the
Lyapunov--Schmidt Reduction Theorem for the root of
(\ref{lyapunov-schmidt}), a unique continuation of $\phi_0(x-s_0)$
into $\phi_{\epsilon}(x-s_{\epsilon})$ exists. In particular, the
correction term $\varphi_1(x)$ satisfies the inhomogeneous problem
\begin{equation}
\label{varphi-1-equation} L_+ \varphi_1 = -V(x) \phi_0(x-s_0),
\end{equation}
which has a unique solution $\varphi_1 \in {\rm Ker}(L_+)^{\perp}
\subset H^2(\mathbb{R})$ by the Fredholm Alternative (since
$M'(s_0) = 0$).
\end{proof}

\begin{corollary}
(i) Let $f(q)$ be $C^2(\mathbb{R}_+)$. Then,
$N(\varphi,\epsilon,s) = {\rm O}(\|\varphi\|_{H^2})$ and
$\tilde{G}(\epsilon,s) = {\rm O}(\epsilon)$ as $\epsilon \to 0$,
such that $\tilde{s} = {\rm O}(\epsilon)$ and $\|
\tilde{\varphi}\|_{L^{\infty}} = {\rm O}(\epsilon^2)$ as $\epsilon
\to 0$.

\noindent (ii) Let $f(q)$ be $C^{\infty}(\mathbb{R}_+)$. Then,
$\tilde{s}(\epsilon)$ and $\tilde{\varphi}(x,\epsilon,s)$ are
$C^{\infty}$-functions of $\epsilon$ near $\epsilon = 0$.
\end{corollary}

\begin{remark}
{\rm The renormalization of $\phi^2 \mapsto (\phi^2 - q_0)$ is not
needed if the potential $V(x)$ satisfies the condition
(\ref{potential}). We expect that the same quantity $M(s)$ with
the renormalization above can be useful to treat the other cases
(ii) and (iii) of the potential term $V(x)$. However, the method
of Lyapunov--Schmidt reductions does not work for these cases since
the essential spectrum of $L_+$ is structurally deformed if
$\epsilon \neq 0$ and $V(x)$ is not decaying. In this case, the
root finding problem (\ref{lyapunov-schmidt}) is not defined in
$H^2(\mathbb{R})$ because the term $V(x) \phi_{\epsilon}(x)$ does
not decay to zero exponentially fast as $|x| \to \infty$.}
\end{remark}

\begin{figure}[htbp]
\begin{center}
\includegraphics[height=5cm]{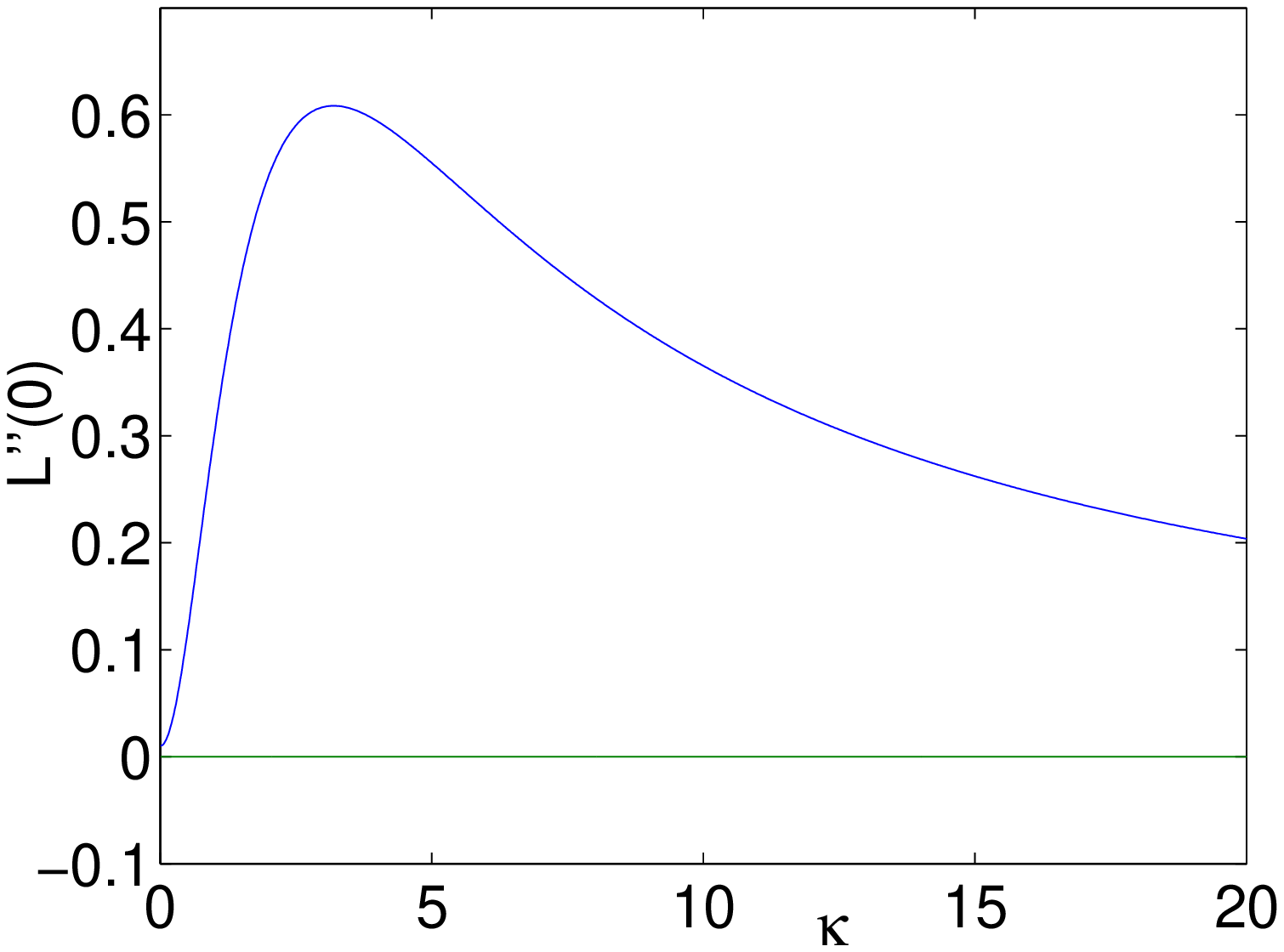}
\includegraphics[height=5cm]{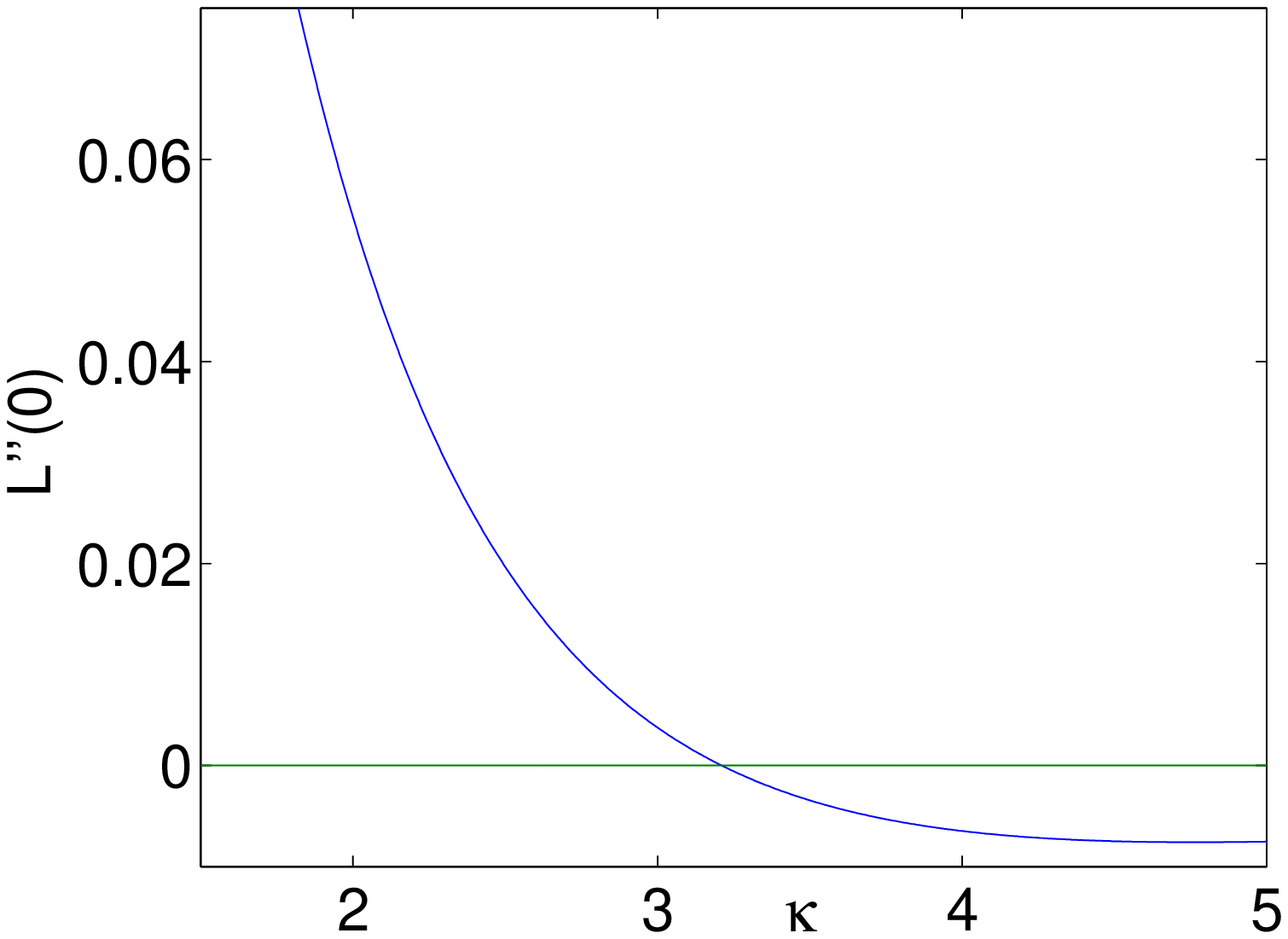}
\includegraphics[height=5cm]{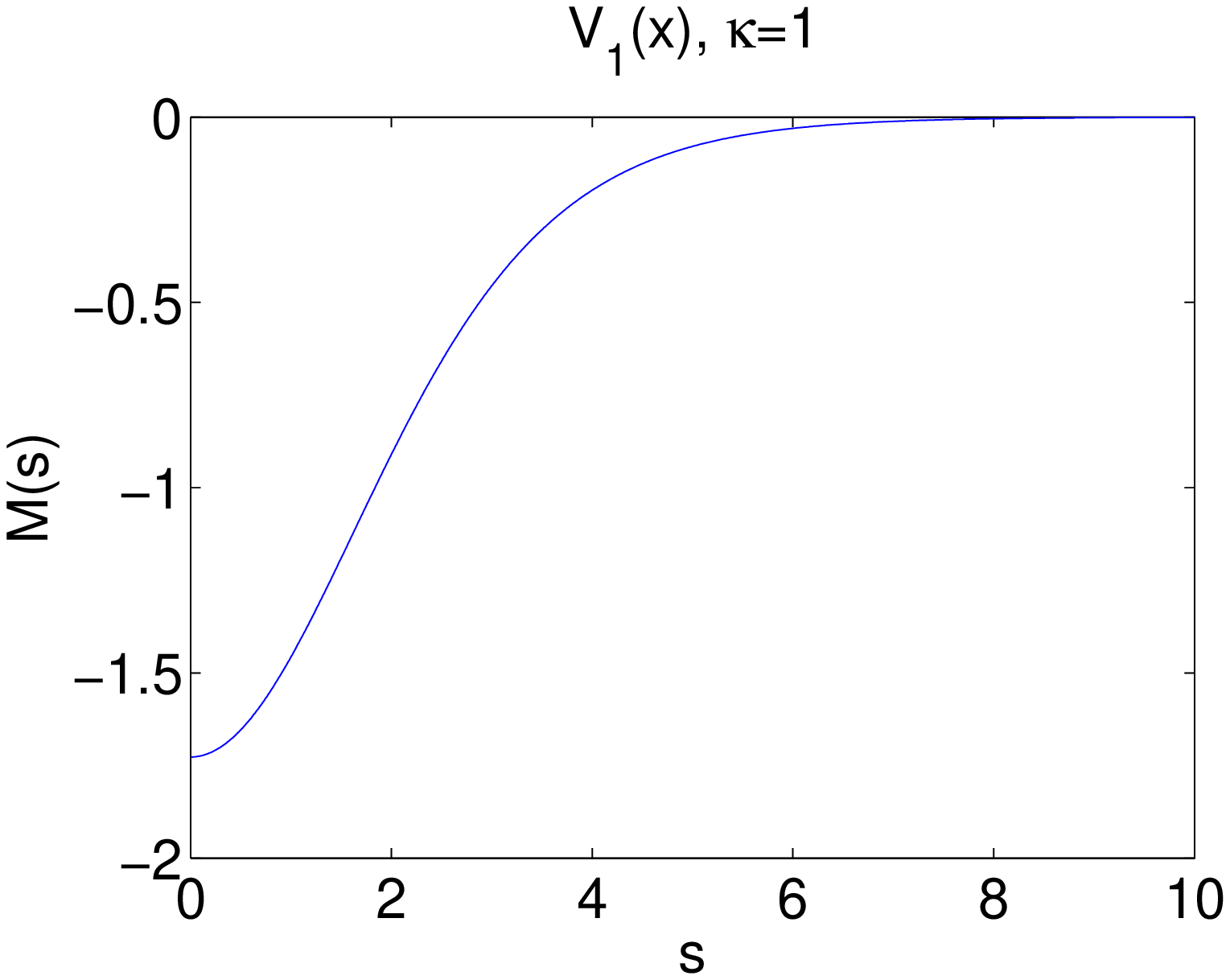}
\includegraphics[height=5cm]{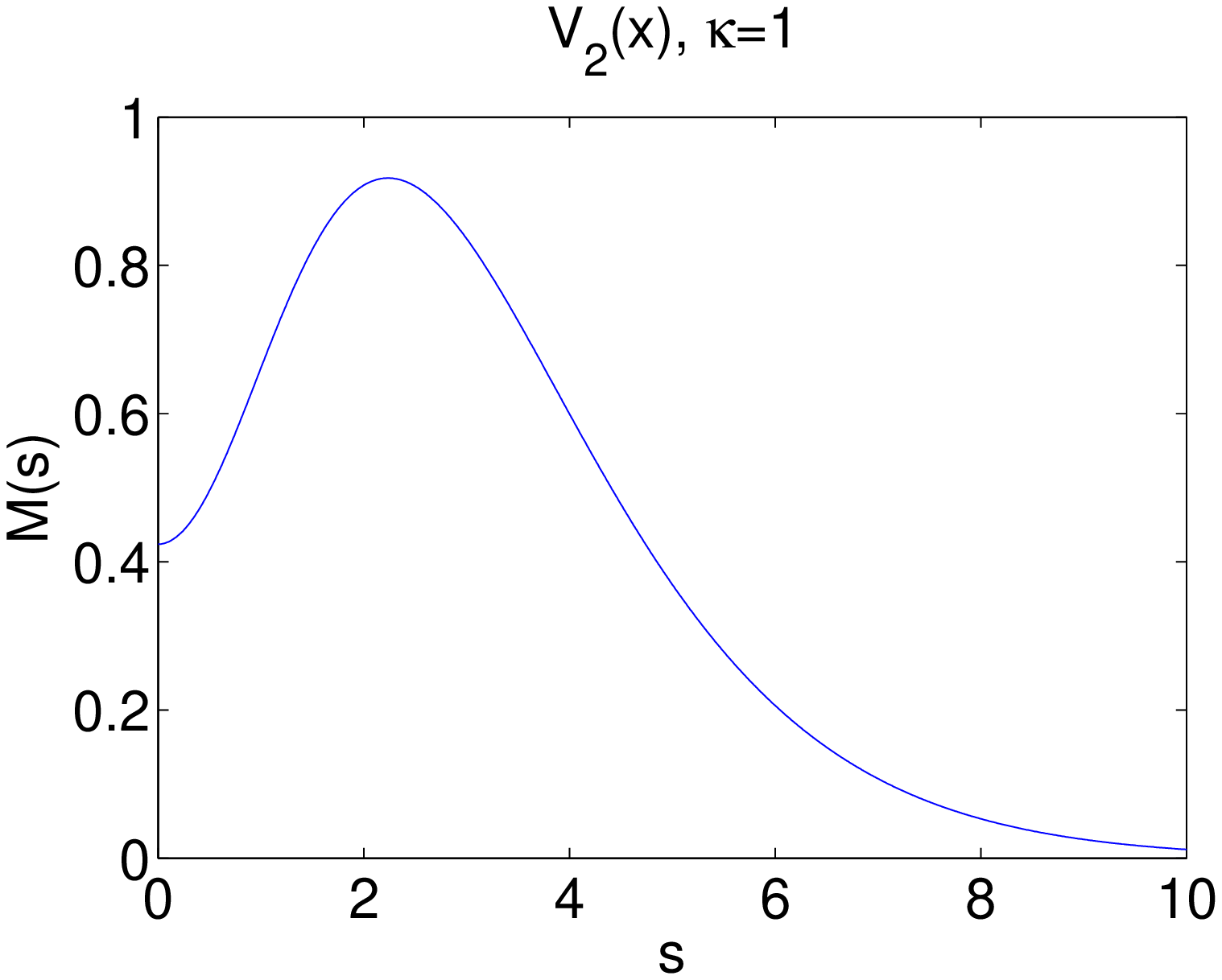}
\includegraphics[height=5cm]{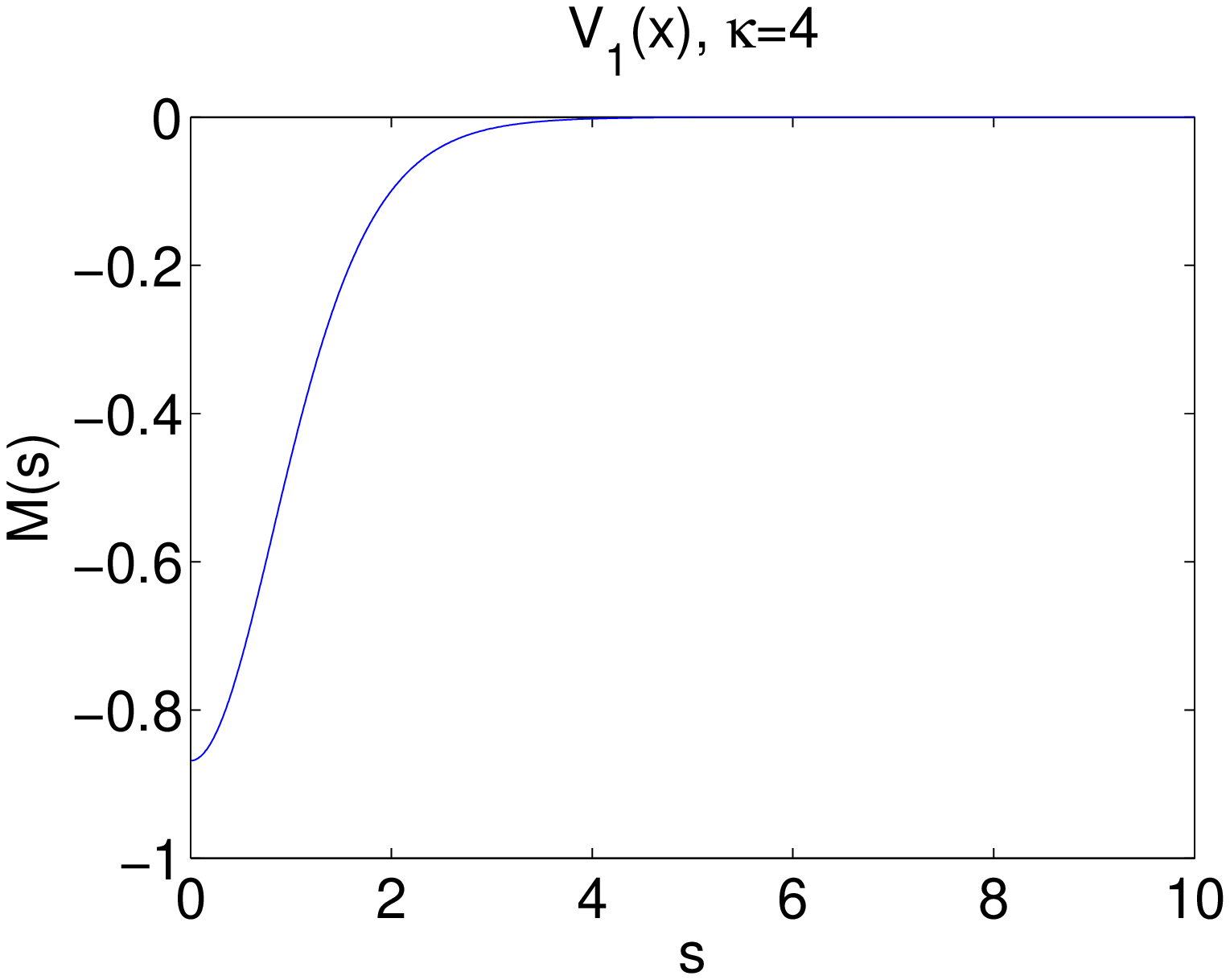}
\includegraphics[height=5cm]{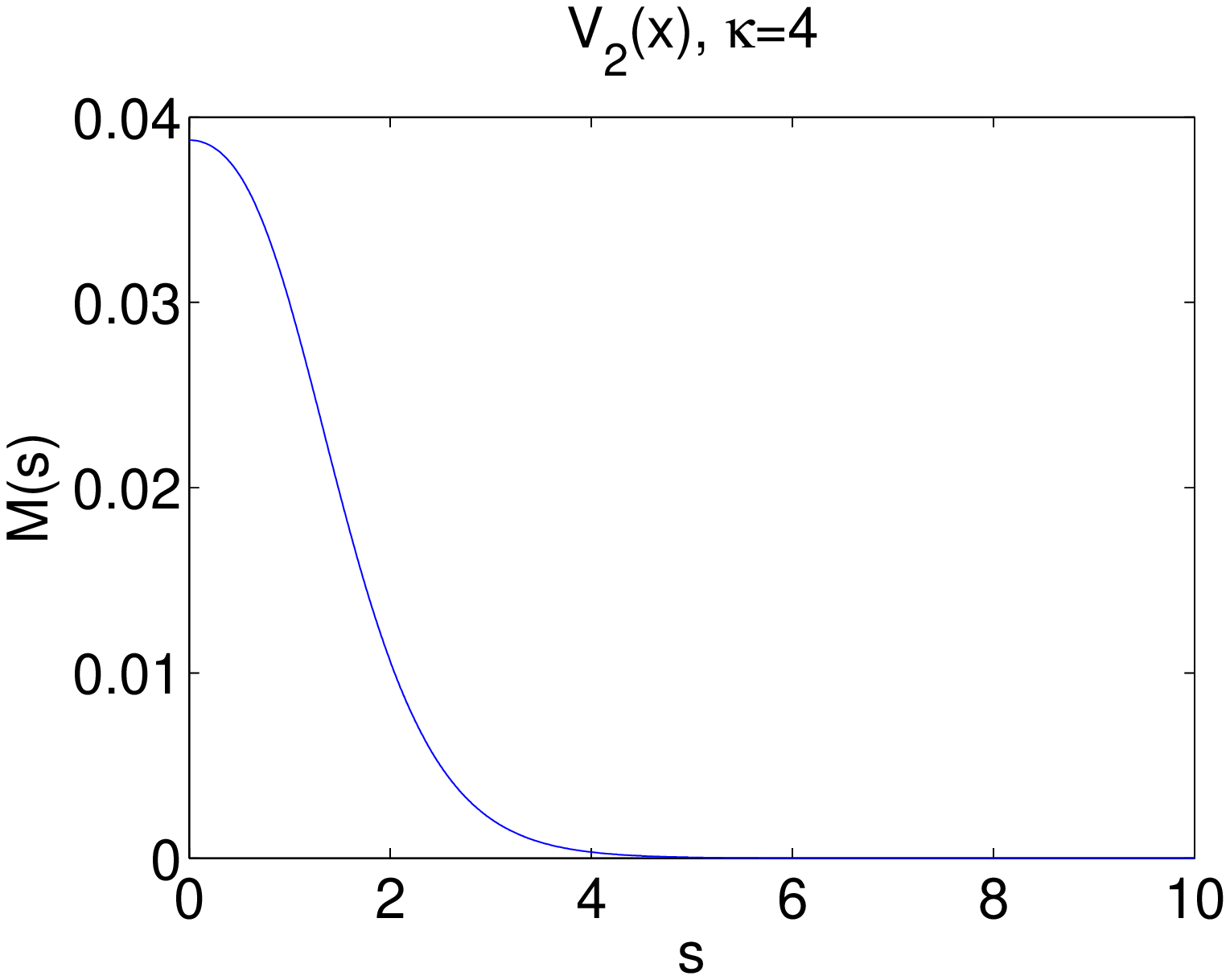}
\end{center}
\caption{The effective potential evaluated numerically for
$V_1(x)$ (left) and $V_2(x)$ (right): the quantity $L''(0)$ versus
$\kappa$ (top panels) and the function $M(s)$ for $\kappa=1$
(middle panels) and $\kappa=4$ (bottom panels).} \label{fig0}
\end{figure}

\begin{example}
{\rm When the cubic NLS is considered with $f(s) = s$, the value
$q_0 \in \mathbb{R}_+$ can be normalized by $q_0 = 1$. In this case,
the second-order ODE (\ref{second-order-ode-u}) admits an exact
solution
\begin{equation}
\label{dark-soliton-exact} U(z) = k \tanh(kz) + i v, \qquad k =
\sqrt{1 - v^2},
\end{equation}
where $v \in (-1,1)$. The other components $\Phi(z)$ and
$\Theta(z)$ of the polar form (\ref{dark-soliton}) can also be
evaluated explicitly. The black soliton corresponds to the kink
$\phi_0(x) = \tanh x$. When the potential $V(x)$ is even $V(-x) =
V(x)$, the function $M'(s)$ in (\ref{M-function}) can be split
into two parts:
$$
M'(s) = \int_{\mathbb{R}_+} V'(x) \left[ \phi_0^2(x+s) -
\phi_0^2(x-s) \right] dx = L'(s) - L'(-s),
$$
where
\begin{equation}
\label{L-function} L(s) = \int_{\mathbb{R}_+} V(x) \left[ q_0 -
\phi_0^2(x-s) \right] dx.
\end{equation}
If $V(x)$ is $C^2(\mathbb{R})$ and satisfies the decay condition
(\ref{potential}), the function $L(s)$ is $C^2(\mathbb{R})$ and
$L(s) \to 0$ exponentially fast as $|s| \to \infty$. Therefore,
$M'(0) = L'(0) - L'(0) = 0$ and one family of kink modes
bifurcates from $s_0 = 0$. Additional families of kink modes of
the GP equation (\ref{GP}) may bifurcate if $L(0)$ and $L''(0)$
are of the same sign. In this case, two global extrema of $L(s) +
L(-s)$ exist at $s_0 = \pm s_*$ with $s_* > 0$, such that two
other families of kink modes bifurcate from $s_0 = \pm s_*$.

When $V = V_1(x)$, $\phi_0 = \tanh x$, and $q_0 = 1$, the function
$L(s)$ is computed in the implicit form
$$
L(s) = - \int_{\mathbb{R}_+} {\rm sech}^2\left( \frac{\kappa x}{2}
\right) {\rm sech}^2(x-s) dx.
$$
Clearly $L(0) < 0$ and, as can be seen from Fig. \ref{fig0} (top
left panel), $L''(0) > 0$ for any $\kappa \neq 0$. Additionally,
Fig. \ref{fig0} (middle and bottom left panels) suggests that
$M(s) < 0$ and $M(s) \to 0$ as $s \to \infty$ for any $\kappa$.
Therefore, there is only one kink mode that bifurcates from $s_0 =
0$, where $V_1(x)$ has a minimum.

When $V = V_2(x)$, the function $L(s)$ is computed in the implicit
form:
$$
L(s) = \int_{\mathbb{R}_+} x^2 e^{- \kappa x} {\rm sech}^2(x-s)
dx,
$$
where $\kappa \in \mathbb{R}_+$. The above $L(s)$ can be expressed
as a generalized hypergeometric function, however, we will not
reproduce the resulting expression here. Instead, we note that $L(0)
> 0$ and
\begin{eqnarray*}
L''(0) = \int_{\mathbb{R}_+} x^2 e^{- \kappa x} \left[ 4 {\rm
sech}^2 x - 6 {\rm sech}^4 x \right] dx.
\end{eqnarray*}
When $\kappa = 0$, $L''(0) = 2$. By using the Laplace method for
computations of the integrals, one can find that $L''(0) = -4
\kappa^{-3} + {\rm O}(\kappa^{-5})$ as $\kappa \to \infty$.
Therefore, there exists $\kappa_0^{\pm} \in \mathbb{R}_+$ with
$\kappa_0^- \leq \kappa_0^+$, such that $L''(0)
> 0$ for $0 < \kappa < \kappa_0^-$ and $L''(0) < 0$ for $\kappa >
\kappa_0^+$. As can be seen from Fig. \ref{fig0} (top right
panel), $\kappa_0^- = \kappa_0^+ \approx 3.21$. Additionally, Fig.
\ref{fig0} (middle and bottom right panels) suggest that no other
extremal points of $M(s)$ exist on $s \in \mathbb{R}$ for any
$\kappa \in \mathbb{R}_+$. When $0 < \kappa < \kappa_0$, there
exist three kink modes: two modes with $s_0 = \pm s_*$ are
associated with the global maxima of the effective potential
$M(s)$, while the mode with $s_0 = 0$ is associated with the local
minimum of $M(s)$. When $\kappa > \kappa_0$, no kink modes with
$s_0 = \pm s_*$ exist but the mode at $s_0 = 0$ corresponds to the
global maximum of the effective potential $M(s)$. Hence, the
structure of kink modes corresponds to a {\it subcritical
pitchfork bifurcation} in the parameter $\kappa$, such that three
solutions exist (at $s_0=0$ and $s_0=\pm s_*$) for $0 < \kappa <
\kappa_0$ and only one solution persists for $\kappa > \kappa_0$.
We point out that the effective potential $M(s)$ gives a different
prediction in comparison with the true potential $V_2(x)$ which
possesses a minimum at $x=0$ and two maxima at $x=\pm 2/\kappa$,
{\it for all $\kappa$}.} \label{example-bifurcations}
\end{example}

\section{Stability analysis of black solitons}

We first consider the spectral stability of black solitons in the
NLS equation (\ref{NLS}). We will obtain conditions for spectral
stability and instability of kinks and bubbles and then extend
these conditions to kink modes of the GP equation (\ref{GP}). In
the end of this section, we will apply these conditions to the
kink modes related to the two potentials
(\ref{potential-explicit}).

\begin{definition}
Let $\phi_0(x)$ be the black soliton of Definition
\ref{definition-black-soliton}. The black soliton is said to be
spectrally unstable in the time evolution of the NLS equation
(\ref{NLS}) if there exists an eigenvector $(u,w) \in
L^2(\mathbb{R},\mathbb{C}^2)$ of the spectral problem
\begin{equation}
\label{spectral-problem} L_+ u = - \lambda w, \qquad L_- w = \lambda
u,
\end{equation}
for the eigenvalue $\lambda$ with ${\rm Re}(\lambda) > 0$, where
\begin{equation}
\label{operators-L-pm} L_+ = -\frac{1}{2} \partial_x^2 + f(\phi_0^2)
- f(q_0) + 2 \phi_0^2 f'(\phi_0^2), \qquad L_- = -\frac{1}{2}
\partial_x^2 + f(\phi_0^2) - f(q_0).
\end{equation}
Otherwise, the black soliton is said to be spectrally stable.
\label{definition-stability-dark}
\end{definition}

\begin{remark}
{\rm The spectral problem (\ref{spectral-problem}) arises in the
linearization of the NLS equation (\ref{NLS}) by using the ansatz
$$
u(x,t) = e^{- i f(q_0) t} \left[ \phi_0(x) + e^{\lambda t} \left[
u(x) + i w(x) \right] + e^{\bar{\lambda} t} \left[ \bar{u}(x) + i
\bar{w}(x) \right] \right].
$$
It will be clear from analysis of the system
(\ref{spectral-problem}) that the spectral instability of black
solitons is always associated with a real positive eigenvalue
$\lambda$, while the spectral stability of black solitons (under a
non-degeneracy constraint) corresponds to the case when the black
soliton is a ground state of an equivalent variational principle.
It is relatively straightforward to develop the nonlinear analysis
for these two cases and to show that the spectral instability and
stability of black solitons (under a non-degeneracy constraint)
correspond to their orbital instability and stability. See
\cite{B95,Lin02,MG06} for developments of this nonlinear analysis.
}
\end{remark}

\begin{proposition}
Let $u(x)$ be a suitable function such that $u(x) = \sqrt{q_0}
e^{i \Theta_{\pm}}$ exponentially fast as $x \to \pm \infty$. Let
the renormalized energy $E_r[u]$ and momentum $P_r[u]$ of the NLS
equation (\ref{NLS}) be defined by
\begin{eqnarray}
\label{Hamiltonian} E_r[u] & = & \frac{1}{2} \int_{\mathbb{R}}
\left[ |u_x|^2 + 2 \int_{|u|^2}^{q_0} \left( f(q_0) - f(q) \right) dq \right] dx, \\
\label{momentum} P_r[u] & = & \frac{i}{2} \int_{\mathbb{R}} \left(
\bar{u} u_x - u \bar{u}_x \right) \left( 1 - \frac{q_0}{|u|^2}
\right) dx.
\end{eqnarray}
The family of dark solitons of Definition
\ref{definition-dark-soliton} is a critical point of the Lyapunov
functional $\Lambda[u] = E_r[u] + v P_r[u]$.
\end{proposition}

\begin{proof}
By direct differentiation, if $u = U(x)$ satisfies the
second-order ODE (\ref{second-order-ode-u}) with $\omega =
f(q_0)$, then the variational derivative $E_r'[u] |_{u = U} + v
P_r'[u] |_{u = U}$ is zero. The second term in $P_r[u]$ is a
Casimir functional, which has {\em zero} variational derivative.
Therefore, the Lyapunov functional $\Lambda[u]$ is defined with
accuracy to an arbitrary constant $C$ in
\begin{equation}
\label{lyapunov} \Lambda[u] = E_r[u] + v P_r[u] + C S[u],
\end{equation}
where
\begin{equation}
\label{total-phase} S[u] = \frac{i}{2} \int_{\mathbb{R}} \left(
\frac{ \bar{u}_x}{\bar{u}} - \frac{u_x}{u} \right) dx = \left[
\arg(u) \right] |_{x \to -\infty}^{x \to \infty}
\end{equation}
is the total phase shift of $u(x)$ on $x \in \mathbb{R}$ subject
to the non-zero boundary conditions on $u(x)$ at infinity. In
order to define the constant $C$ uniquely, we add a constraint on
the variational problem by requiring that if $u = \phi e^{i
\theta}$, $\theta' = v ( 1 - q_0/\phi^2)$, and $\phi = \Phi(x)$
satisfies the second-order ODE (\ref{second-order-ode-Phi}), then
the first variation $\tilde{\Lambda}'[\phi] |_{\phi = \Phi}$ is
zero, where $\tilde{\Lambda}[\phi] = \Lambda[u] |_{u = \phi e^{i
\theta}}$, $\theta' = v ( 1 - q_0/\phi^2)$. By direct
differentiation, it follows immediately that $C = 0$ under this
requirement.
\end{proof}

\begin{remark}
{\rm The renormalized momentum (\ref{momentum}) was constructed in
\cite{KY94} as a difference between the standard momentum $P[u]$
associated with a solution $u(x)$ and the value $P[u_0]$ evaluated
at the background solution $u_0 = \sqrt{q_0} e^{i {\rm sign}(x)
S_0/2}$, where the value $S_0 = S[u]$ is related to the total
phase shift of the solution $u(x)$. It was shown in
\cite{B96,Lin02} that the renormalized momentum $P_r[u]$ computed
at the family of dark solitons $U(x)$ of Definition
\ref{definition-dark-soliton} defines the spectral stability and
instability of dark solitons in the sense that the dark soliton is
spectrally stable if $P_r'(v) \geq 0$ and unstable if $P_r'(v) <
0$ where $P_r(v) = P_r[U]$. (The degenerate case $P'_r(v) = 0$
corresponds to the dark solitons which are spectrally stable and
orbitally unstable. Under the non-degeneracy constraint $P_r'(v)
\neq 0$, the spectral stability and instability corresponds to the
orbital stability and instability, see \cite{Lin02}. In what
follows, we will consider dark solitons under the non-degeneracy
constraint $P_r'(v) \neq 0$ $\forall v \in (-c,c)$.) We will show
that the limit $v \downarrow 0$ is well defined and the quantity
$P_r' |_{v \downarrow 0}$ determines spectral stability and
instability of kinks of Definition \ref{definition-black-soliton}.
The latter point is missed in the recent paper \cite{MG06}, where
stability of kinks is considered. }
\end{remark}

\begin{lemma}
Let $U(x)$ be the family of dark solitons of Definition
\ref{definition-dark-soliton} and $f(q)$ be $C^2(\mathbb{R}_+)$.
Then, (i) the function $P_r(v) = P_r[U]$ is $C^1$ on $v \in (-c,0)
\cup (0,c)$ and (ii) the limiting quantity $P_r' |_{v \downarrow 0}$
is well-defined. \label{lemma-momentum-dark-solitons}
\end{lemma}

\begin{proof}
(i) By construction of dark solitons in Theorem
\ref{lemma-existence-dark-soliton}, the function $P_r(v) = P_r[U]$
is represented by
\begin{eqnarray}
\nonumber P_r(v) & = & \frac{i}{2} \int_{\mathbb{R}} \left(
\bar{U} U' - U \bar{U}' \right) \left( 1 - \frac{q_0}{|U|^2}
\right) dx \\
\label{momentum-relation}  & = & - v \int_{\mathbb{R}} \Phi^2(x)
\left( 1 - \frac{q_0}{\Phi^2(x)} \right)^2 dx  = v N(v) + q_0
S(v),
\end{eqnarray}
where $N(v)$ and $S(v)$ is the total power and phase shift of the
dark solitons:
\begin{equation}
N(v) = \int_{\mathbb{R}} \left( q_0 - \Phi^2(x) \right) dx, \qquad
S(v) = \int_{\mathbb{R}} \Theta'(x) dx = \Theta_+ - \Theta_-.
\end{equation}
By the ODE theory for the system
(\ref{theta-ODE})--(\ref{second-order-ode-Phi}) with $\Phi(x) > 0$
on $x \in \mathbb{R}$, the map $v \mapsto (\Phi,\Theta)$ is $C^1$
on $v \in (-c,0) \cup (0,c)$, such that $N(v)$ and $S(v)$ are
smooth functions and so is $P_r(v)$.

(ii) We will show that the functions $N(v)$, $S(v)$ and $P_r(v)$
remains smooth in the limit $v \downarrow 0$. Let $U(x)$ be a dark
soliton for $v \in (0,c)$ according to Definition
\ref{definition-dark-soliton} and $\phi_0(x)$ be a black soliton
according to Definition \ref{definition-black-soliton}. Let us
consider
$$
\tilde{U}(x) = \frac{U(x) - \phi_0(x)}{v} = \tilde{U}_r(x) + i
\tilde{U}_i(x), \qquad v \in (0,c),
$$
where $\tilde{U}_r(x)$ and $\tilde{U}_i(x)$ are real-valued
functions on $x \in \mathbb{R}$. By the construction of $U(x)$ and
$\phi_0(x)$ in Theorems \ref{lemma-existence-dark-soliton} and
\ref{lemma-existence-black-soliton}, it is clear that $\tilde{U}_r,
\tilde{U}_i \in L^{\infty}(\mathbb{R})$ are continuous in $v$ for $v
\in (0,c)$. We need to prove that these functions remain continuous
in $v$ as $v \downarrow 0$. By separating the real and imaginary
parts in the ODEs (\ref{second-order-ode-u}) and
(\ref{second-order-ode-phi-0}), we obtain the equivalent ODE system
for $\tilde{U}_r(x)$ and $\tilde{U}_i(x)$:
\begin{eqnarray}
\label{equivalent-system-1} v \tilde{U}_i' + \frac{1}{v} \phi_0
\left( f(\phi_0^2) - f(|U|^2) \right) + \frac{1}{2} \tilde{U}_r'' +
\tilde{U}_r \left( f(q_0) - f(|U|^2) \right) = 0, \\
\label{equivalent-system-2} - \phi_0' - v \tilde{U}_r' + \frac{1}{2}
\tilde{U}_{i}'' + \tilde{U}_i \left( f(q_0) - f(|U|^2) \right) = 0,
\end{eqnarray}
where $|U|^2 = (\phi_0 + v \tilde{U}_r)^2 + v^2 \tilde{U}_i^2$. Let
us rewrite the ODE (\ref{equivalent-system-2}) as an inhomogeneous
problem:
\begin{equation}
\label{inhomogeneous-problem-2} ( L_- + \tilde{L}_-) \tilde{U}_i =
F_-,
\end{equation}
where operator $L_-$ is defined by (\ref{operators-L-pm}) and
$$
\tilde{L}_- =  f(|U|^2) - f(\phi_0^2), \qquad F_- = - \phi_0' - v
\tilde{U}_r'.
$$
Since $|U|^2(x)$, $\phi_0^2(x)$ approach $q_0$ and
$\tilde{U}_r(x)$, $\tilde{U}_i(x)$, $\phi_0(x)$ approach some
constants exponentially fast as $|x| \to \infty$, it is clear that
$\tilde{L}_-$ is a relatively compact perturbation to $L_-$ and
$\tilde{L}_-, F_- \in L^2(\mathbb{R})$ for $v \in (0,c)$. By
continuity of the solution $\phi_0(x) = \lim\limits_{v \downarrow
0} U(x)$ in Theorem \ref{lemma-existence-black-soliton}, we know
that $\| \tilde{L}_- \|_{L^{\infty}} = {\rm o}(1)$ and $v \|
\tilde{U}_r' \|_{L^{\infty}} = {\rm o}(1)$ as $v \downarrow 0$.

Since $L_- \phi_0 = 0$ and $\phi_0 \in L^{\infty}(\mathbb{R})$,
then $\forall f \in L^2(\mathbb{R})$ there exists $L_-^{-1} f \in
L^{\infty}(\mathbb{R})$ if and only if $(\phi_0,f) = 0$. The
following computation shows that this condition for $f = F_- -
\tilde{L}_- \tilde{U}_i$ is equivalent to the ODE
(\ref{equivalent-system-2}) and is thus satisfied on $v \in
(0,c)$:
\begin{eqnarray*}
&& - (\phi_0,\phi_0') - v (\phi_0,\tilde{U}_r') +
 \left( \phi_0(f(\phi_0^2) - f(q_0), \tilde{U}_i \right) + \left( \phi_0,
\tilde{U}_i
\left( f(q_0) - f(|U|^2) \right) \right) \\
&& = - (\phi_0,\phi_0') - v (\phi_0,\tilde{U}_r') + \frac{1}{2}
(\phi_0'', \tilde{U}_i) +  \left( \phi_0, \tilde{U}_i \left( f(q_0) - f(|U|^2) \right)
\right) \\
&& =  \left( \phi_0, - \phi_0' - v \tilde{U}_r' + \frac{1}{2}
\tilde{U}_{i}'' + \tilde{U}_i \left( f(q_0) - f(|U|^2) \right)
\right) = 0.
\end{eqnarray*}
Therefore, the operator $L_-^{-1} : L^2(\mathbb{R}) \mapsto
L^{\infty}(\mathbb{R})$ is well-defined for the inhomogeneous
problem (\ref{inhomogeneous-problem-2}) on $v \in (0,c)$, which can
be rewritten as follows:
$$
\tilde{U}_i = - \left( L_- + \tilde{L}_- \right)^{-1} \left( \phi_0'
+ v \tilde{U}_r' \right).
$$
Since $\| \tilde{L}_- \|_{L^{\infty}} = {\rm o}(1)$ and $v \|
\tilde{U}_r'\|_{L^{\infty}} = {\rm o}(1)$ as $v \downarrow 0$,
there exists a solution $\tilde{U}_i \in L^{\infty}(\mathbb{R})$
uniformly in $v \in [0,c)$, such that $\tilde{U}_i |_{v \downarrow
0} = -L_-^{-1} \phi_0' \in L^{\infty}(\mathbb{R})$. Therefore, the
function ${\rm Im} U(x)$ is smooth as $v \downarrow 0$ and ${\rm
Im} \partial_v U(x) |_{v \downarrow 0} = \tilde{U}_i |_{v
\downarrow 0} = -L_-^{-1} \phi_0'$.

We can now use the fact that $v \| \tilde{U}_i \|_{L^{\infty}} =
{\rm O}(v)$ as $v \downarrow 0$. Since $f(q)$ is $C^2(\mathbb{R}_+)$
and $v \| \tilde{U}_r \|_{L^{\infty}} = {\rm o}(v)$ as $v \downarrow
0$, there exists a function $g(\phi_0,v\tilde{U}_r)$ for small $v
\tilde{U}_r$ such that
$$
f((\phi_0 + v \tilde{U}_r)^2) - f(\phi_0^2) - 2 v \phi_0
\tilde{U}_r f'(\phi_0^2) = v \tilde{U}_r g(\phi_0,v \tilde{U}_r),
$$
where $\| g(\phi_0,v\tilde{U}_r) \|_{L^{\infty}} = {\rm o}(v)$ as
$v \downarrow 0$. Using these facts, we rewrite the ODE
(\ref{equivalent-system-1}) as an inhomogeneous problem:
\begin{equation}
\label{inhomogeneous-problem-1} ( L_+ + \tilde{L}_+) \tilde{U}_r =
F_+,
\end{equation}
where operator $L_+$ is defined by (\ref{operators-L-pm}) and
$$
\tilde{L}_+ =  f(|U|^2) - f(\phi_0^2) + \phi_0 g(\phi_0,v
\tilde{U}_r), \qquad F_+ = v \tilde{U}_i' + \frac{1}{v} \phi_0
\left( f((\phi_0 + v \tilde{U}_r)^2) - f(|U|^2) \right).
$$
It is clear that $\tilde{L}_+, F_+ \in L^{\infty}(\mathbb{R})$ for
$v \in (0,c)$. Since $L_+ \phi_0' = 0$ and $\phi'_0(x) \in
L^2(\mathbb{R})$, then $\forall f \in L^{\infty}(\mathbb{R})$
there exists $L_+^{-1} f \in L^{\infty}(\mathbb{R})$ if and only
if $(\phi_0',f) = 0$ (by the Fredholm's Alternative). The
following computation shows that this condition for $f = F_+ -
\tilde{L}_+ \tilde{U}_r$ is equivalent to the ODE
(\ref{equivalent-system-1}) and is thus satisfied on $v \in
(0,c)$:
\begin{eqnarray*}
&& v (\phi_0',\tilde{U}_i') + \frac{1}{v} \left(\phi_0' \phi_0,
\left( f(\phi_0^2) - f(|U|^2) \right) \right) + \left( \phi_0'
\left( f(\phi_0^2)
+ 2 \phi_0^2 f'(\phi_0^2) - f(|U|^2) \right),\tilde{U}_r \right)  \\
&& = v (\phi_0',\tilde{U}_i') + \frac{1}{v} \left( \phi_0' \phi_0,
\left( f(\phi_0^2) - f(|U|^2) \right) \right) + \frac{1}{2}
(\phi_0''', \tilde{U}_r) + \left( \phi_0', \tilde{U}_r
\left( f(q_0) - f(|U|^2) \right) \right) \\
&& = \left( \phi_0', v \tilde{U}_i' + \frac{1}{v} \phi_0 \left(
f(\phi_0^2) - f(|U|^2) \right) + \frac{1}{2} \tilde{U}_r'' +
\tilde{U}_r \left( f(q_0) - f(|U|^2) \right) \right) = 0.
\end{eqnarray*}
Therefore, the operator $L_+^{-1} : L^{\infty}(\mathbb{R}) \mapsto
L^{\infty}(\mathbb{R})$ is well-defined for the inhomogeneous
problem (\ref{inhomogeneous-problem-1}) on $v \in (0,c)$, which can
be rewritten as follows:
$$
\tilde{U}_r = - \left( L_+ + \tilde{L}_+ \right)^{-1} \left( v
\tilde{U}_i' + \frac{1}{v} \phi_0 \left( f((\phi_0 + v
\tilde{U}_r)^2) - f(|U|^2) \right) \right).
$$
Since $\| \tilde{L}_+ \|_{L^{\infty}} = {\rm o}(v)$ and $v \|
\tilde{U}_i \|_{L^{\infty}} = {\rm O}(v)$ as $v \downarrow 0$,
there exists a solution $\tilde{U}_r \in L^{\infty}(\mathbb{R})$
uniformly in $v \in [0,c)$, such that $\tilde{U}_r |_{v \downarrow
0} = 0$ (the homogeneous solution $\phi_0'(x)$ is removed from
$\tilde{U}_r(x)$ due to the symmetry in $U(x)$). Therefore, the
function ${\rm Re} U(x)$ is smooth as $v \downarrow 0$ and ${\rm
Re} \partial_v U(x) |_{v \downarrow 0} = \tilde{U}_r |_{v
\downarrow 0} = 0$.

As a result, the map $v \mapsto U$ is $C^1$ on $v \in [0,c)$, such
that $N(v)$ and $S(v)$ are smooth functions as $v \downarrow 0$ and
so is $P_r(v)$.
\end{proof}

\begin{corollary}
The following identities are true for $v \in (-c,0) \cup (0,c)$
\begin{eqnarray}
\label{momentum-identity-1} P_r'(v) = i \int_{\mathbb{R}} \left( U'
\partial_v \bar{U} - \bar{U}' \partial_v U \right) dx = 2
\int_{\mathbb{R}} \left( {\rm Re} U' \; {\rm Im} \partial_v U -
{\rm Im}U' \; {\rm Re} \partial_v U \right) dx
\end{eqnarray}
and
\begin{equation}
\label{momentum-identity-2} P_r' |_{v \downarrow 0} = 2 \left(
\phi_0', {\rm Im} \partial_v U |_{v \downarrow 0} \right) = N |_{v
\downarrow 0} + q_0 S' |_{v \downarrow 0}.
\end{equation}
\end{corollary}

\begin{proof}
By Lemma \ref{lemma-momentum-dark-solitons}, the quantity $P_r'(v)$
is continuous on $v \in [0,c)$. The first identity in
(\ref{momentum-identity-1}) follows by direct differentiation:
\begin{eqnarray*}
P_r'(v) & = & \frac{i}{2} \int_{\mathbb{R}} \left( U'
\partial_v \bar{U} + \bar{U} \partial_v U' - \bar{U}' \partial_v U - U \partial_v \bar{U}' \right) dx
+ \frac{i q_0}{2} \int_{\mathbb{R}} \partial_v \left(
\frac{\bar{U}'}{\bar{U}} - \frac{U'}{U} \right) dx \\ & = & i
\int_{\mathbb{R}} \left( U'
\partial_v \bar{U} - \bar{U}' \partial_v U \right) dx +
\frac{i}{2} \left( \bar{U} \partial_v U - U \partial_v \bar{U}
\right) |_{x \to -\infty}^{x \to \infty} + \frac{i q_0}{2}
\partial_v \log \left( \frac{\bar{U}}{U} \right) \biggr|_{x \to -\infty}^{x \to \infty} \\
& = & i \int_{\mathbb{R}} \left( U'
\partial_v \bar{U} - \bar{U}' \partial_v U \right) dx.
\end{eqnarray*}
Other identities follow by the substitution $U(x) = {\rm Re} U(x)
+ i {\rm Im} U(x)$, by the smoothness of $U(x)$ with respect to $v
\in [0,c)$ and by the relation (\ref{momentum-relation}).
\end{proof}

\begin{example}
{\rm Following Example \ref{example-bifurcations}, we consider the
cubic NLS with $f(s) = s$ and $q_0 = 1$. By using the exact solution
(\ref{dark-soliton-exact}), we find for $v \in [0,1)$
$$
N(v) = 2 \sqrt{1 - v^2}, \quad S(v) = - 2 {\rm arctan}
\frac{\sqrt{1 -v^2}}{v},
$$
such that
$$
P_r'(v) = 4 \sqrt{1 - v^2}, \qquad S'(v) = \frac{2}{\sqrt{1-v^2}},
$$
and $P_r' |_{v \downarrow 0} = 4$, $S' |_{v \downarrow 0} = 2$. }
\label{example-bifurcations-2}
\end{example}

\begin{lemma}
(i) Let $\phi_0(x)$ be the kink of Definition
\ref{definition-black-soliton}. Then, the spectrum of $L_+$ in
$L^2(\mathbb{R})$ consists of the positive continuous spectrum
bounded away from zero by $2c^2$, the kernel with the
eigenfunction $\phi'_0(x)$ and, possibly, a finite number of
positive eigenvalues in $(0,2c^2)$. The spectrum of $L_-$ in
$L^2(\mathbb{R})$ consists of the non-negative continuous spectrum
and a single negative eigenvalue.

(ii) Let $\phi_0(x)$ be the bubble of Definition
\ref{definition-black-soliton}. Then, the spectrum of $L_+$ in
$L^2(\mathbb{R})$ consists of the positive continuous spectrum
bounded away from zero by $2c^2$, the kernel with the
eigenfunction $\phi'_0(x)$, a single negative eigenvalue and,
possibly, a finite number of positive eigenvalues in $(0,2c^2)$.
The spectrum of $L_-$ in $L^2(\mathbb{R})$ consists of the
non-negative continuous spectrum. \label{lemma-spectrum-in-L2}
\end{lemma}

\begin{proof}
Since $\phi^2_0(x) \to q_0$ exponentially fast as $|x| \to
\infty$, $L_{\pm}$ in (\ref{operators-L-pm}) are self-adjoint
Schr\"{o}dinger operators on the domain $H^2(\mathbb{R}) \subset
L^2(\mathbb{R})$, which have absolutely continuous spectrum
$\sigma_c(L_{\pm})$, a finite number of isolated eigenvalues of
finite multiplicities $\sigma_p(L_{\pm})$, and no embedded
eigenvalues or residual spectrum \cite{Sigal}. By the Weyl's
Essential Spectrum Lemma, $\sigma_c(L_+) \geq 2 q_0 f'(q_0) = 2
c^2 > 0$ and $\sigma_c(L_-) \geq 0$, such that the continuous
spectrum of $L_+$ is bounded away from zero and the continuous
spectrum of $L_-$ touches zero. Moreover, $L_+ \phi_0'(x) = 0$ and
$L_- \phi_0(x) = 0$ due to the translational and gauge symmetries
of the NLS equation (\ref{NLS}), such that $L_+$ has a simple
kernel in $L^2(\mathbb{R})$ while $L_-$ has no kernel in
$L^2(\mathbb{R})$.

(i) In the case of kinks, $\phi_0(x)$ has a single zero on $x \in
\mathbb{R}$. By the Sturm Nodal Theorem, $\sigma_p(L_+)$ contains
no negative eigenvalues and $\sigma_p(L_-)$ contains exactly one
negative eigenvalue.

(ii) In the case of bubbles, $\phi_0(x)$ has no zeros on $x \in
\mathbb{R}$. By the Sturm Nodal Theorem, $\sigma_p(L_+)$ contains
exactly one negative eigenvalue and $\sigma_p(L_-)$ contains no
negative eigenvalues.
\end{proof}

\begin{lemma}
Consider the constrained space
\begin{equation}
\label{constrained-space} X_c = \left\{ w \in H^2(\mathbb{R}) :
\quad (w, \phi_0') = 0 \right\},
\end{equation}
where $\phi_0(x)$ is a black soliton of Definition
\ref{definition-black-soliton}. In the case of kinks, the operator
$L_-$ has exactly one negative eigenvalue in $X_c$ if $P_r' |_{v
\downarrow 0} < 0$ and no negative eigenvalues if $P_r' |_{v
\downarrow 0} > 0$, where $P_r' |_{v \downarrow 0}$ is defined by
Lemma \ref{lemma-momentum-dark-solitons}. In the case of bubbles,
the operator $L_-$ is non-negative in $X_c$.
\label{lemma-constrained-space}
\end{lemma}

\begin{proof}
We consider a constrained variational problem:
\begin{equation}
\label{variational-problem} (L_- - \mu) w = - \nu \phi'_0, \qquad
w \in H^2(\mathbb{R}), \;\; \mu \notin \sigma(L_-),
\end{equation}
where $\mu$ is the spectral parameter and $\nu$ is the Lagrange
multiplier. By Lemma \ref{lemma-spectrum-in-L2}, there exists a
unique solution $w \in H^2(\mathbb{R})$ for any $\mu \in
(\mu_0,0)$, where $\mu_0$ is the only negative eigenvalue of $L_-$
in the case of kinks and $\mu_0 = -\infty$ in the case of bubbles.
By the standard variational theory (e.g. see \cite{B96,MG06}), the
smooth function $g(\mu) = -(\phi'_0,(L_- - \mu)^{-1} \phi_0')$ is
decreasing on $\mu \in (\mu_0,0)$ from $\lim\limits_{\mu
\downarrow \mu_0} g(\mu) = +\infty$ in the case of kinks or
$\lim\limits_{\mu \to -\infty} g(\mu) = 0$ in the case of bubbles.
Therefore, operator $L_-$ is non-negative in $X_c$ in the case of
bubbles. In the case of kinks, operator $L_-$  is non-negative in
$X_c$ if $\lim\limits_{\mu \uparrow 0} g(\mu) > 0$ and it has a
single zero on $\mu \in (\mu_0,0)$ if $\lim\limits_{\mu \uparrow
0} g(\mu) < 0$. We need to show that $\lim\limits_{\mu \uparrow 0}
g(\mu) = P_r' |_{v \downarrow 0}$.

It is clear from the solution of the variational problem
(\ref{variational-problem}) that $w \in H^2(\mathbb{R}) \subset
L^{\infty}(\mathbb{R})$ uniformly in $\mu \in (\mu_0,0]$. Since
$(\phi_0,\phi_0') = 0$, there exists a solution of the
inhomogeneous problem $L_- w_0 = - \phi_0'$ in $w_0 \in
L^{\infty}(\mathbb{R})$ and, moreover, it follows from smoothness
by Lemma \ref{lemma-momentum-dark-solitons} that
$$
w_0 = \tilde{U}_i |_{ v \downarrow 0} = {\rm Im} \partial_v U(x)
|_{v \downarrow 0} + c \phi_0,
$$
where $c$ is arbitrary. By the relation
(\ref{momentum-identity-2}), we obtain $\lim\limits_{\mu \uparrow
0} g(\mu) = (\phi_0',{\rm Im}
\partial_v U |_{v \downarrow 0}) = \frac{1}{2} P_r' |_{v
\downarrow 0}$.
\end{proof}

\begin{remark}
{\rm It is interesting to note that the inhomogeneous equation
$L_- w_0 = - \phi_0'$ admits another solution $w_0 = x \phi_0(x)
\notin L^{\infty}(\mathbb{R})$, which may lead to miscalculation
$$
\lim_{\mu \uparrow 0} g(\mu) \neq (\phi_0',x \phi_0) = \frac{1}{2}
N_r |_{v \downarrow 0} > 0.
$$
The linearly growing solution $x \phi_0(x) \notin
L^{\infty}(\mathbb{R})$ is however irrelevant for the variational
problem (\ref{variational-problem}). }
\end{remark}

\begin{theorem}
(i) Let $\phi_0(x)$ be the kink of Definition
\ref{definition-black-soliton}. Then, it is spectrally stable if
$P_r' |_{v \downarrow 0} > 0$ and unstable if $P_r' |_{v \downarrow
0} < 0$. In the latter case, the spectral problem
(\ref{spectral-problem}) has exactly one real positive eigenvalue
$\lambda$.

(ii) Let $\phi_0(x)$ be the bubble of Definition
\ref{definition-black-soliton}. Then, it is spectrally unstable with
exactly one real positive eigenvalue $\lambda$ in the spectral
problem (\ref{spectral-problem}). \label{theorem-stability-kink}
\end{theorem}

\begin{proof}
Let $\lambda$ be a non-zero eigenvalue of the spectral problem
(\ref{spectral-problem}) with $(u,w) \in
H^2(\mathbb{R},\mathbb{C}^2)$. Then, $L_+$ is invertible in $X_c$
defined by (\ref{constrained-space}) and the component $w(x) \in
X_c$ can be found from the generalized eigenvalue problem
\begin{equation}
\label{generalized-eigenvalue-problem} L_- w = \gamma L_+^{-1} w,
\qquad \gamma = - \lambda^2, \qquad w \in X_c.
\end{equation}
Due to the equivalence above, all {\em non-zero} eigenvalues
$\lambda$ of the spectral problem (\ref{spectral-problem}) can be
recovered from the {\em non-zero} eigenvalues $\gamma$ of the
generalized eigenvalue problem
(\ref{generalized-eigenvalue-problem}). The operators $L_{\pm}$
satisfy properties P1--P2 of the recent paper \cite{Chugunova}. Even
though $L_-$ has no  spectral gap near the origin, one can shift
the generalized eigenvalue problem to the equivalent form,
\begin{equation}
\label{generalized-eigenvalue-problem-modified} (L_- + \delta
L_+^{-1}) w = (\gamma + \delta) L_+^{-1} w, \qquad 0 < \delta <
\delta_0,
\end{equation}
where $\delta_0$ is the distance from $\gamma = 0$ to the first
negative eigenvalue $\gamma$ if it exists or $\delta_0 = \infty$
if not. Now
$$
\sigma_c(L_- + \delta L_+^{-1}) \geq \frac{\delta}{2 c^2} > 0,
$$
and the operator $\tilde{L}_- = L_- + \delta L_+^{-1}$ has the
spectral gap near the origin.

(i) In the case of kinks, the operator $L_+^{-1}$ is positive in
$X_c$. By Theorem 3 of \cite{Chugunova}, the problem
(\ref{generalized-eigenvalue-problem-modified}) has no eigenvalues
$\gamma \in \mathbb{C}$ with ${\rm Im}(\gamma) \neq 0$, has no
eigenvalues $\gamma \in \mathbb{R}$ such that $(w,L_+^{-1} w) \leq
0$ and has exactly $N = {\rm dim}(H^-_{L_- + \delta L_+^{-1}})$
eigenvalues $\gamma < 0$, where $H^-_{L_- + \delta L_+^{-1}}
\subset X_c$ is the invariant negative subspace of $X_c$ with
respect to $L_- + \delta L_+^{-1}$. We will prove that $N = {\rm
dim}(H^-_{L_- + \delta L_+^{-1}}) = {\rm dim}(H^-_{L_-})$ for
sufficiently small $\delta > 0$. Indeed, continuity of isolated
negative eigenvalues of $L_-$ in $\delta$ follows by the
perturbation theory since $\delta L_+^{-1}$ is a relatively
compact perturbation to $L_-$ in $X_c$. Therefore, ${\rm
dim}(H^-_{L_- + \delta L_+^{-1}}) \geq {\rm dim}(H^-_{L_-})$ for
sufficiently small $\delta > 0$.  Consider a splitting $X_c =
H^-_{L_-} \oplus H^+_{L_-}$, where $H^-_{L_-}$ ($H^+_{L_-}$) is
negative (non-positive) invariant subspace of $X_c$ with respect
to $L_-$, such that ${\rm dim}(H^-_{L_-}) < \infty$ and ${\rm
dim}(H^+_{L_-}) = \infty$. Since $L_+^{-1}$ is strictly positive
operator in $X_c$, we have
$$
\forall \delta > 0, \;\; \forall w \in H^+_{L_-} : \;\;\; (w, (L_- +
\delta L_+^{-1}) w) \geq \delta (w, L_+^{-1} w) > 0.
$$
Therefore, the operator $(L_- + \delta L_+^{-1})$ is strictly
positive on $H^+_{L_-}$ and ${\rm dim}(H^-_{L_- + \delta
L_+^{-1}}) \leq {\rm dim}(H^-_{L_-})$. We have thus proved that
${\rm dim}(H^-_{L_- + \delta L_+^{-1}}) = {\rm dim}(H^-_{L_-})$.
By Lemma \ref{lemma-constrained-space}, ${\rm dim}(H^-_{L_-}) = 1$
if $P_r' |_{v \downarrow 0} < 0$ and ${\rm dim}(H^-_{L_-}) = 0$ if
$P_r' |_{v \downarrow 0} > 0$. In the former case, $N = 1$ and the
kink is spectrally unstable. In the latter case, $N  = 0$ and the
kink is spectrally stable.

(ii) In the case of bubbles, the operator $L_-$ has no negative
eigenvalues in $X_c$ by Lemma \ref{lemma-constrained-space}.
Therefore, Theorem 3 of \cite{Chugunova} guarantees that the
generalized eigenvalue problem
(\ref{generalized-eigenvalue-problem-modified}) has no eigenvalues
$\gamma \in \mathbb{C}$ with ${\rm Im}(\gamma) \neq 0$, has no
eigenvalues $\gamma \in \mathbb{R}_+$ such that $(w,L_+^{-1} w)
\leq 0$ and has exactly $N = {\rm dim}(H^-_{L_+^{-1}})$
eigenvalues $\gamma \in \mathbb{R}_-$ with $(w,L_+^{-1} w) \leq
0$, where $H^-_{L_+^{-1}} \subset X_c$ is the invariant negative
subspace of $L_+^{-1}$. Since all eigenvectors of $L_+$ for
non-zero eigenvalues are orthogonal to $\phi_0'$ and belong to
$X_c$, it follows immediately that $N = {\rm dim}(H^-_{L_+^{-1}})
= 1$.
\end{proof}

\begin{remark}
{\rm The statement (i)  of Theorem \ref{theorem-stability-kink}
extends the stability--instability theorem in \cite{Lin02} from
dark solitons with $v \neq 0$ to kinks with $v = 0$. This
connection is missed in the recent paper \cite{MG06} where the
same result was obtained by the Vakhitov--Kolokolov method (used
in \cite{B96}) and the variational principle (used in \cite{B95})
without the proof that $\lim\limits_{\mu \uparrow 0} g(\mu) =
\frac{1}{2} P_r' |_{v \downarrow 0}$.

The statement (ii) of Theorem \ref{theorem-stability-kink} was
proved differently in \cite{B95} by using a variational technique.
We note that $P_r(v) < 0$ for $v \in (0,c)$ and $\lim\limits_{v
\downarrow 0} P_r(v) = 0$ in the case of bubbles of Definition
\ref{definition-black-soliton}. Therefore, if $P_r'|_{v \downarrow
0} \neq 0$, then $P_r' |_{v \downarrow 0} < 0$, such that the
statement (i) for kinks extends formally to the statement (ii) for
bubbles. }
\end{remark}

\begin{definition}
Let $\phi_{\epsilon}(x)$ be the kink mode of Definition
\ref{definition-kink}. The kink mode is said to be spectrally
unstable in the time evolution of the GP equation (\ref{GP}) if
there exists an eigenvector $(u,w) \in
L^2(\mathbb{R},\mathbb{C}^2)$ of the spectral problem
\begin{equation}
\label{spectral-problem-GP} {\cal L}_+ u = - \lambda w, \qquad {\cal
L}_- w = \lambda u,
\end{equation}
for the eigenvalue $\lambda$ with ${\rm Re}(\lambda) > 0$, where
\begin{equation}
\label{operators-L-pm-GP} {\cal L}_+ = -\frac{1}{2} \partial_x^2 +
f(\phi_{\epsilon}^2) - f(q_0) + 2 \phi_{\epsilon}^2
f'(\phi_{\epsilon}^2) + \epsilon V(x), \quad {\cal L}_- =
-\frac{1}{2} \partial_x^2 + f(\phi_{\epsilon}^2) - f(q_0) +
\epsilon V(x).
\end{equation}
Otherwise, the kink mode is said to be spectrally stable.
\label{definition-stability-kink-mode}
\end{definition}

\begin{theorem}
Let $\phi_{\epsilon}(x)$ be the kink mode of Definition
\ref{definition-kink}. Assume that the operators ${\cal L}_{\pm}$
have $n_{\pm}$ negative eigenvalues and empty kernels in
$L^2(\mathbb{R})$. Assume that all embedded (purely imaginary)
eigenvalues of the spectral problem (\ref{spectral-problem-GP})
are algebraically simple. Then, the spectral problem
(\ref{spectral-problem-GP}) for $(u,w) \in
L^2(\mathbb{R},\mathbb{C}^2)$ has exactly $N_c$ complex
eigenvalues $\lambda$ in the first quadrant, $N_i^-$ purely
imaginary eigenvalues $\lambda$ with ${\rm Im}(\lambda) > 0$ and
$(w,{\cal L}_+^{-1} w) \leq 0$, and $N_r = N_r^+ + N_r^-$ real
positive eigenvalues $\lambda$, where $N_r^+$ corresponds to
eigenvalues with $(w,{\cal L}_+^{-1} w) \leq 0$ and $N_r^-$
corresponds to eigenvalues with $(w,{\cal L}_+^{-1} w) \geq 0$,
such that
\begin{eqnarray}
\label{count-eigenvalues} N_r^+ + N_i^- + N_c = n_+, \qquad N_r^- +
N_i^- + N_c = n_-,
\end{eqnarray}
where multiple eigenvalues are accounted up to their algebraic
multiplicities. \label{theorem-stability-kink-mode}
\end{theorem}

\begin{proof}
Since $V(x) \to 0$ and $\phi_{\epsilon}^2(x) \to q_0$ exponentially
fast as $|x| \to \infty$, operators ${\cal L}_{\pm}$ have the
absolutely continuous spectrum such that $\sigma_c({\cal L}_+) \geq
2 c^2 > 0$ and $\sigma_c({\cal L}_-) \geq 0$. Since the kernel of
${\cal L}_+$ is empty in $L^2(\mathbb{R})$ by assumption, the
generalized eigenvalue problem
(\ref{generalized-eigenvalue-problem}) for operators ${\cal
L}_{\pm}$ is rewritten in the unconstrained space:
\begin{equation}
\label{generalized-eigenvalue-problem-Lpm} {\cal L}_- w = \gamma
{\cal L}_+^{-1} w, \qquad \gamma = - \lambda^2, \qquad w \in
H^2(\mathbb{R}).
\end{equation}
The kernel of ${\cal L}_-$ is empty in $L^2(\mathbb{R})$ since
${\cal L}_- \phi_{\epsilon} = 0$ and $\phi_{\epsilon} \in
L^{\infty}(\mathbb{R})$, $\phi_{\epsilon} \notin L^2(\mathbb{R})$.
The generalized eigenvalue problem
(\ref{generalized-eigenvalue-problem-Lpm}) can be rewritten in the
equivalent form,
\begin{equation}
\label{generalized-eigenvalue-problem-modified-Lpm} ({\cal L}_- +
\delta {\cal L}_+^{-1}) w = (\gamma + \delta) {\cal L}_+^{-1} w,
\end{equation}
where $\delta > 0$ is sufficiently small. Properties P1--P2 of
\cite{Chugunova} are satisfied and Theorem 3 of \cite{Chugunova}
gives the relations
\begin{eqnarray*}
N_r^+ + N_i^- + N_c = {\rm dim}(H^-_{{\cal L}_+^{-1}}), \qquad
N_r^- + N_i^- + N_c = {\rm dim}(H^-_{{\cal L}_- + \delta {\cal
L}_+^{-1}})
\end{eqnarray*}
for sufficiently small $\delta > 0$, where $H^-_{{\cal L}_+^{-1}}$
and $H^-_{{\cal L}_- + \delta {\cal L}_+^{-1}}$ are invariant
negative subspaces of $L^2(\mathbb{R})$ with respect to ${\cal
L}_+^{-1}$ and ${\cal L}_- + \delta {\cal L}_+^{-1}$ respectively.
It follows immediately that ${\rm dim}(H^-_{{\cal L}_+^{-1}}) =
n_+$. By continuity of eigenvalues and the relative compactness of
${\cal L}_+^{-1}$ with respect to ${\cal L}_-$, it follows that
${\rm dim}(H^-_{{\cal L}_-}) \leq {\rm dim}(H^-_{{\cal L}_- +
\delta {\cal L}_+^{-1}})$. We shall prove that ${\rm
dim}(H^-_{{\cal L}_- + \delta {\cal L}_+^{-1}}) = {\rm
dim}(H^-_{{\cal L}_-})$. The operator ${\cal L}_- + \delta {\cal
L}_+^{-1}$ may have additional negative eigenvalues compared to
operator ${\cal L}_-$ if and only if some eigenvalues bifurcate as
$\delta \neq 0$ from the end point of the continuous spectrum of
${\cal L}_-$ by means of the edge bifurcation
\cite{CP05,KS04,S00}. In order to analyze the edge bifurcation, we
rewrite the eigenvalue problem $({\cal L}_- + \delta {\cal
L}_+^{-1}) w = \mu w$ in the equivalent form:
$$
\left({\cal L} + \delta {\cal M}\right) w = \mu w, \qquad w \in
L^2(\mathbb{R}),
$$
where
$$
{\cal L} = {\cal L}_- + \delta \left( 2 c^2 - \frac{1}{2}
\partial_x^2 \right)^{-1}, \quad {\cal M} = {\cal L}_+^{-1}
\left( f(\phi_{\epsilon}^2) - f(q_0) + 2 \phi_{\epsilon}^2
f'(\phi_{\epsilon}^2) + \epsilon V(x) \right) \left( 2 c^2 -
\frac{1}{2}  \partial_x^2 \right)^{-1}
$$
where ${\cal M}$ is a relatively compact perturbation to the
unbounded operator ${\cal L}$. The continuous spectrum of ${\cal
L}$ is bounded from below by $\sigma_c({\cal L}) \geq
\frac{\delta}{2c^2}$. By the theory of edge bifurcations (see
review in \cite{KS04}), the new eigenvalue $\mu = \mu_{\delta}$,
if it bifurcates from the end point of $\sigma_c({\cal L})$, has
the expansion $\mu_{\delta} = \frac{\delta}{2c^2} - \alpha
\delta^2 + {\rm O}(\delta^3)$, where $\alpha$ is positive
constant. Therefore, there exists sufficiently small $\delta > 0$,
such that $\mu_{\delta} > 0$. As a result, the edge bifurcation
does not change the number of negative eigenvalues of ${\cal L}_-
+ \delta {\cal L}_+^{-1}$ compared to ${\cal L}_-$ and ${\rm
dim}(H^-_{{\cal L}_- + \delta {\cal L}_+^{-1}}) = {\rm
dim}(H^-_{{\cal L}_-}) = n_-$.
\end{proof}

\begin{theorem}
Let $\phi_0(x)$ be the kink of Definition
\ref{definition-black-soliton} and $M'(s)$ be defined by
(\ref{M-function}). Let $\phi_{\epsilon}(x)$ be a unique
continuation of the kink $\phi_0(x-s_0)$ in Theorem
\ref{theorem-continuation} from the root $s_0$ such that $M'(s_0) =
0$ and $M''(s_0) \neq 0$. Then, the operators ${\cal L}_{\pm}$ have
$n_{\pm}$ negative eigenvalues and empty kernels in
$L^2(\mathbb{R})$ for sufficiently small $\epsilon$ with $n_+ = 1$,
$n_- = 1$ for $M''(s_0) > 0$ and $n_+ = 0$, $n_- = 1$ for $M''(s_0)
< 0$. \label{theorem-persistence}
\end{theorem}

\begin{proof}
It follows by Theorem \ref{theorem-continuation} that
$\phi_{\epsilon} = \phi_0(x - s) + \epsilon \varphi_1(x) +
\tilde{\varphi}(x,\epsilon,s)$ and $s = s_0 + \tilde{s}(\epsilon)$,
where $\|\tilde{\varphi}\|_{L^{\infty}} = {\rm o}(\epsilon)$ and
$|\tilde{s}| = {\rm o}(1)$ as $\epsilon \to 0$. Therefore, operators
${\cal L}_{\pm}$ in (\ref{operators-L-pm-GP}) are represented by
$$
{\cal L}_{\pm} = L_{\pm} + \epsilon M_{\pm} + \tilde{M}_{\pm},
$$
where $L_{\pm}$ are given by (\ref{operators-L-pm}), $M_{\pm}$ are
given by
\begin{eqnarray}
\label{operators-M-pm} M_+ = V(x) + 6 \phi_0 \varphi_1 f(\phi_0^2) +
4 \phi_0^3 \varphi_1 f''(\phi_0^2), \qquad M_- = V(x) + 2 \phi_0
\varphi_1 f(\phi_0^2),
\end{eqnarray}
and $\| \tilde{M}_{\pm} \|_{L^{\infty}} = {\rm o}(\epsilon)$ as
$\epsilon \to 0$. We note that $\epsilon M_+ + \tilde{M}_+ =
\epsilon V(x) + D_{\varphi} N(\varphi,s,\epsilon)$, where
$D_{\varphi} N$ is the Jacobian of the nonlinear function in
(\ref{lyapunov-schmidt}). By using the inhomogeneous equation
(\ref{varphi-1-equation}) for the correction term $\varphi_1(x)$,
we compute
\begin{equation}
\label{splitting-zero} (\phi'_0(x-s_0), M_+ \phi'_0(x-s_0)) = -
(\phi_0',V' \phi_0) - (\phi'_0, L_+ \varphi_1') = -\frac{1}{2}
M''(s_0).
\end{equation}
By the regular perturbation theory, the zero eigenvalue of $L_+$
becomes a non-zero eigenvalue $\lambda_{\epsilon}$ of ${\cal L}_+$
for small $\epsilon$, such that
$$
\lambda_{\epsilon} = \epsilon \lambda_1 + \tilde{\lambda}, \qquad
\lambda_1 = -\frac{M''(s_0)}{2 \| \phi_0'\|_{L^2}^2},
$$
where $\tilde{\lambda} = {\rm o}(\epsilon)$ as $\epsilon \to 0$.
Since the zero eigenvalue of $L_+$ is simple and positive
eigenvalues of $L_+$ are bounded away from zero, the kernel of
${\cal L}_+$ is empty, such that $n_+ = 1$ for $M''(s_0) > 0$ and
$n_+ = 0$ for $M''(s_0) < 0$ for sufficiently small $\epsilon$. By
the Implicit Function Theorem applied to $\phi_{\epsilon}(x)$
(since $\phi_0(x)$ has only one simple zero at $x = 0$, the
function $\phi_{\epsilon}(x)$ has only one node on $x \in
\mathbb{R}$ for sufficiently small $\epsilon$. We recall that
${\cal L}_- \phi_{\epsilon} = 0$ and $\phi_{\epsilon} \in
L^{\infty}(\mathbb{R})$, $\phi_{\epsilon} \notin L^2(\mathbb{R})$.
By the Sturm Nodal Theorem, the kernel of ${\cal L}_-$ is empty
and $n_- = 1$ for sufficiently small $\epsilon$.
\end{proof}

\begin{remark}
\label{remark-smoothness} {\rm Let $f \in C^2(\mathbb{R}_+)$. Then
$\| \tilde{M}_{\pm} \|_{L^{\infty}} = {\rm O}(\epsilon^2)$ and
$\tilde{\lambda} = {\rm O}(\epsilon^2)$ as $\epsilon \to 0$. }
\end{remark}

\begin{corollary}
The kink mode with $M''(s_0) < 0$ is spectrally unstable with
exactly one real positive eigenvalue $\lambda$ in the spectral
problem (\ref{spectral-problem-GP}) for sufficiently small
$\epsilon$. The kink mode with $M''(s_0) > 0$ may have up to two
unstable eigenvalues $\lambda$ in the spectral problem
(\ref{spectral-problem-GP}).   \label{corollary-persistence}
\end{corollary}

\begin{proof}
If $M''(s_0) < 0$, then $n_+ = 0$, $n_- = 1$ and the count of
eigenvalues (\ref{count-eigenvalues}) gives $N_r^+ = N_i^- = N_c =
0$ and $N_r^- = 1$. If $M''(s_0) > 0$, then $n_+ = n_- = 1$ and the
count of eigenvalues may give either $N_i^- + N_c = 1$, $N_r^+ =
N_r^- = 0$ or $N_i^- = N_c = 0$, $N_r^+ = N_r^- = 1$. In the cases
$N_c = 1$ or $N_r^+ = N_r^- = 1$, there are two unstable and no
embedded eigenvalues in the spectral problem
(\ref{spectral-problem-GP}). In the case $N_i^- = 1$, the pair of
embedded eigenvalues is simple, such that the last assumption of
Theorem \ref{theorem-stability-kink-mode} is satisfied.
\end{proof}

\begin{remark}
{\rm Asymptotic approximations of eigenvalues $\lambda$ and
precise statements on unstable eigenvalues in the case $M''(s_0)
> 0$ are obtained in Section 4 under
non-degeneracy assumptions $P_r' |_{v \downarrow 0} \neq 0$ and $S'
|_{v \downarrow 0} \neq 0$. By Corollary \ref{corollary-splitting},
the case $N_i^- = N_c = 0$, $N_r^+ = N_r^- = 1$ occurs for $P_r'
|_{v \downarrow 0} < 0$ and the case $N_r^+ = N_r^- = N_i^- = 0$,
$N_c = 1$ occurs for $P_r' |_{v \downarrow 0} > 0$. }
\label{remark-persistence}
\end{remark}

\begin{example}
{\rm Continuing Examples \ref{example-bifurcations} and
\ref{example-bifurcations-2}, we consider the cubic NLS equation
with $f(s) = s$, $q_0 = 1$ and $P_r' |_{v \downarrow 0} = 4 > 0$.
When the potential $V(x)$ is even with $V(-x) = V(x)$, the function
$M'(s)$ is represented by $M'(s) = L'(s) - L'(-s)$, where $L(s)$ is
given by (\ref{L-function}). One family of kink modes with $s_0 = 0$
always bifurcates with $M''(0) = 2 L''(0)$.

When $V = V_1(x)$, it follows from Fig. \ref{fig0} (top left panel)
that $L''(0) > 0$ for any $\kappa \neq 0$. The kink mode with $s_0 =
0$ corresponds to the minimum of $M(s)$ and it is unstable with two
complex conjugate eigenvalues, according to Remark
\ref{remark-persistence}.

When $V = V_2(x)$, it follows from Fig. \ref{fig0} (top right panel)
that there exists $0 < \kappa_0 < \infty$ such that $L''(0) > 0$ for
$0 < \kappa < \kappa_0$ and $L''(0) < 0$ for $\kappa > \kappa_0$.
For $0 < \kappa < \kappa_0$, a pair of kink modes bifurcates from
$s_0 = \pm s_*$ with $M''(s_0) < 0$. These modes correspond to the
maxima of the effective potential $M(s)$ and they are unstable with
one real eigenvalue, according to Corollary
\ref{corollary-persistence}. The kink mode with $s_0 = 0$
corresponds to the minimum of the effective potential $M(s)$ and it
is unstable with two complex eigenvalues for $0 < \kappa <
\kappa_0$. On the other hand, it corresponds to the maximum of
$M(s)$ and it is unstable with a simple real positive eigenvalue for
$\kappa > \kappa_0$. This scenario indicates the subcritical
pitchfork bifurcation at $\kappa=\kappa_0$. We will illustrate this
bifurcation in Section \ref{Numerics1}.}
\label{example-bifurcations2}
\end{example}

\section{Eigenfunctions and eigenvalues of kinks}

We report here asymptotic analysis of the spectral problem
(\ref{spectral-problem-GP}) in the limit of small $\lambda$ and
$\epsilon$. This asymptotic analysis is needed to complete the
stability analysis of kink modes with $M''(s_0) > 0$ which is not
conclusive in Corollary \ref{corollary-persistence}. We will show
that if the kink is stable in the linear problem
(\ref{spectral-problem}) for $\epsilon = 0$, then the pair of zero
eigenvalues of the spectral problem (\ref{spectral-problem-GP}) at
$\epsilon = 0$ splits into a pair of purely imaginary eigenvalues
at ${\rm O}(\epsilon)$ and bifurcates into a quartet of four
complex eigenvalues (two of which are unstable) at  ${\rm
O}(\epsilon^{3/2})$. These eigenvalues $\lambda$ for small
$\epsilon$ correspond to the eigenvectors $(u,w) \in
L^2(\mathbb{R},\mathbb{C}^2)$, persistence of which in $\epsilon$
follows by Theorem \ref{theorem-stability-kink-mode}.

From a technical point of view, our analysis is complicated by the
fact that the eigenvalues $\lambda \in i \mathbb{R}$ are embedded
into the continuous spectrum of the non-self-adjoint problem
(\ref{spectral-problem-GP}). Since we are interested in specific
information about bifurcations of eigenvalues near the point
$\lambda = 0$, we have to abandon the generalized eigenvalue
problem (\ref{generalized-eigenvalue-problem-Lpm}) and to work
directly with the non-self-adjoint eigenvalue problem
(\ref{spectral-problem-GP}).

One way to deal with this problem is to introduce exponential
weights which move branches of the continuous spectrum from the
imaginary axis (see \cite{PY} and references therein). However,
there are two branches of the continuous spectrum, and,
independently of the weight parameter, one branch moves to the left
and the other branch moves to the right of the imaginary axis. Any
eigenvalues that bifurcate off the imaginary axis may become
resonant poles when the weight parameter is sent to zero unless
specific information about the decay rate of eigenfunctions is
available. However, if this information were available, one could
avoid the technique of exponential weights and perform a direct
analysis of the eigenfunctions and eigenvalues of the problem
(\ref{spectral-problem-GP}).

Another way to deal with this problem is to consider the Evans
function with careful analysis of fast and slow decaying solutions
(see \cite{KR00} and reference therein). By using the Gap Lemma,
the Evans function can be appropriately extended across the
continuous spectrum with a full account of the branch points on
the imaginary axis. Information about small eigenvalues is drawn
from the derivatives of the Evans function with respect to
$\lambda$ and $\epsilon$ near $\lambda = 0$ and $\epsilon = 0$.
However, computational formulas become more and more involved when
higher-order derivatives of the Evans function are needed and this
has caused some miscalculations in the past (see Appendix A in
\cite{SS05}).

Our treatment of the problem brings together the analysis of fast
and slow decaying solutions in the two approaches above. To avoid
complications, it is based on direct analysis of eigenfunctions
and eigenvalues of the spectral problem
(\ref{spectral-problem-GP}) expanded in powers of
$\epsilon^{1/2}$. We will obtain a characteristic equation for
small eigenvalue $\lambda$ versus small parameter $\epsilon$.

\begin{lemma}
Let $\phi_0(x)$ be the kink of Definition
\ref{definition-black-soliton} and $U(x)$ be the dark soliton of
Definition \ref{definition-dark-soliton} for $v > 0$. Let
operators $L_{\pm}$ be defined by (\ref{operators-L-pm}). The
uncoupled homogeneous problems
$$
L_+ u_0 = 0, \qquad L_- w_0 = 0
$$
admit four linearly independent solutions:
\begin{itemize}
\item[(i)] exponentially decaying eigenfunction $u_0 = \phi_0'(x)$ in
$L^2(\mathbb{R})$

\item[(ii)] bounded eigenfunction $w_0 = \phi_0(x)$ in
$L^{\infty}(\mathbb{R})$

\item[(iii)] unbounded linearly growing solution $w_0 = x \phi_0(x) -
{\rm Im} \partial_v U(x) |_{v \downarrow 0}$

\item[(iv)] and an unbounded exponentially growing solution $u_0$
\end{itemize}
The uncoupled inhomogeneous problems
$$
L_+ u_1 = -w_0, \qquad L_- w_1 = u_0
$$
admit solutions in the same order:
\begin{itemize}
\item[(i)] bounded eigenfunction $w_1 = - {\rm Im} \partial_v U(x) |_{v \downarrow 0}$ in
$L^{\infty}(\mathbb{R})$

\item[(ii)] a bounded eigenfunction $u_1$ in $L^{\infty}(\mathbb{R})$

\item[(iii)] an unbounded exponentially growing solution $u_1$ if
$S' |_{v \downarrow 0} \neq 0$

\item[(iv)] and an unbounded exponentially growing solution $w_1$
\end{itemize}
The uncoupled inhomogeneous problems
$$
L_+ u_2 = -w_1, \qquad L_- w_2 = u_1
$$
admit no solutions in $L^{\infty}(\mathbb{R})$ if $P_r' |_{v
\downarrow 0} \neq 0$. \label{lemma-kernel}
\end{lemma}

\begin{proof}
It follows from the proofs of Lemmas \ref{lemma-spectrum-in-L2}
and \ref{lemma-constrained-space} that
\begin{equation}
L_+ \phi'_0 = 0, \quad L_- \phi_0 = 0, \quad L_- x \phi_0 = -
\phi_0', \quad L_- {\rm Im} \partial_v U(x) |_{v \downarrow 0} =
-\phi_0',
\end{equation}
which proves (i)--(iii) for $u_0$ and $w_0$ and (i) for $w_1$.
Existence of an exponentially growing solution $u_0$ in (iv)
follows from the fact that the Wronskian determinant of two
linearly independent solutions of $L_+ u_0 = 0$ is constant in
$x$. Existence of bounded solution $u_1$ in (ii) follows from the
fact that $(\phi_0',\phi_0) = 0$, such that $L_+^{-1} \phi_0 \in
L^{\infty}(\mathbb{R})$. The solution $u_1$ in (iii) grows
exponentially since the Fredholm Alternative is not satisfied for
$w_0$ in (iii):
\begin{equation}
\label{zero-projection} \left( \phi_0', x \phi_0(x) - {\rm Im}
\partial_v U(x) |_{v \downarrow 0} \right) = \frac{1}{2} N |_{v
\downarrow 0} - \frac{1}{2} P_r' |_{v \downarrow 0} = -
\frac{q_0}{2} S' |_{v \downarrow 0} \neq 0,
\end{equation}
where the relation (\ref{momentum-identity-2}) has been used.
Existence of an exponentially growing solution $w_1$ in (iv)
follows from the fact that $L_-^{-1} u_0$ has the same exponential
growth in $x$ as $u_0$ in (iv). The solution $u_2$ of $L_+ u_2 =
{\rm Im} \partial_v U(x) |_{v \downarrow 0}$ grows exponentially
since the Fredholm Alternative is not satisfied:
$$
\left( \phi_0', {\rm Im} \partial_v U(x) |_{v \downarrow 0}
\right) = \frac{1}{2} P_r' |_{v \downarrow 0} \neq 0
$$
The solution $w_2$ of $L_- w_2 = u_1$ in (ii) grows linearly due
to the same reason since
$$
(\phi_0, u_1) = - (L_+ u_1, u_1) \neq 0.
$$
The last inequality is due to the non-negativity of $L_+$ for
kinks and the orthogonality of the odd function $u_1$ to the even
function $\phi_0'$ of the kernel of $L_+$.
\end{proof}

\begin{definition}
Let $\lambda$ be fixed in the strip $\{ \lambda \in \mathbb{C} : 0
< {\rm Re}\lambda < c^2 \}$ and define $\kappa_{\pm}(\lambda)$
from the roots of the characteristic equations
\begin{equation}
{\rm Re} \kappa_{\pm} > 0 : \quad  \kappa_{\pm}^2 = 2 c^2 \left( 1
\pm \sqrt{1 - \frac{\lambda^2}{c^4}} \right),
\label{equation-kappa}
\end{equation}
such that $\kappa_+ \kappa_- = 2 \lambda$ and $\kappa_+ = \sqrt{4
c^2 - \kappa_-^2}$.
\end{definition}

\begin{remark}
{\rm The roots $\kappa_{\pm}$ can be expanded in the Taylor series
near $\lambda = 0$, such that
\begin{equation}
\label{expansion-kappa-pm} \kappa_+(\lambda) = 2c \left( 1 -
\frac{\lambda^2}{8 c^4} + {\rm O}(\lambda^4)\right), \qquad
\kappa_-(\lambda) = \frac{\lambda}{c} \left( 1 +
\frac{\lambda^2}{8 c^4} + {\rm O}(\lambda^4) \right).
\end{equation}
} \label{remark-expansion-kappa}
\end{remark}

\begin{lemma}
Let $\phi_{\epsilon}(x)$ be the kink mode of Definition
\ref{definition-kink} in the domain $-\epsilon_0 < \epsilon <
\epsilon_0$ for some $\epsilon_0 > 0$. There exist four
fundamental solutions $(u,w)$ of the spectral problem
(\ref{spectral-problem-GP}) for any $\lambda$ in the strip $\{
\lambda \in \mathbb{C} : 0 < {\rm Re}\lambda < c^2 \}$, such that
\begin{equation}
\label{boundary-conditions-1} \left( \begin{array}{cc} u \\ w
\end{array} \right)_{\pm} \to \left(
\begin{array}{cc} \kappa_{\pm} \\ -\kappa_{\mp}
\end{array} \right) e^{\kappa_{\pm} x} \qquad \mbox{as} \qquad x \to - \infty
\end{equation}
and
\begin{equation}
\label{boundary-conditions-2} \left( \begin{array}{cc} \tilde{u} \\
\tilde{w} \end{array} \right)_{\pm} \to \left(
\begin{array}{cc} \kappa_{\pm} \\ -\kappa_{\mp}
\end{array} \right) e^{-\kappa_{\pm} x} \qquad \mbox{as} \qquad x \to +\infty
\end{equation}
\label{lemma-fundamental-solutions}
\end{lemma}

\begin{proof}
For any $\lambda$ in the strip $\{ \lambda \in \mathbb{C} : 0 <
{\rm Re}\lambda < c^2 \}$, the two roots $\kappa_+(\lambda)$ and
$\kappa_-(\lambda)$ of the characteristic equations
(\ref{equation-kappa}) are distinct and ${\rm Re} \kappa_{\pm} >
0$. Existence of four linearly independent solutions with the
exponential tails in (\ref{boundary-conditions-1}) and
(\ref{boundary-conditions-2}) follows by the
Coddington--Levinson's Theorem for ODEs \cite{Levenson} under the
condition that $V(x) \to 0$ and $\phi^2_{\epsilon}(x) \to q_0$
exponentially fast as $|x| \to \infty$.
\end{proof}

\begin{definition}
The determinant of the four fundamental solutions in Lemma
\ref{lemma-fundamental-solutions} at any $x \in \mathbb{R}$ is
called the Evans function $E(\lambda,\epsilon)$ of the spectral
problem (\ref{spectral-problem-GP}), namely
\begin{equation}
\label{Evans-determinant} E(\lambda,\epsilon) = {\rm det} \left|
\begin{array}{cccc} u_+ & \tilde{u}_+ & u_- & \tilde{u}_- \\
u_+' & \tilde{u}_+' & u_-' & \tilde{u}_-' \\  w_+ & \tilde{w}_+ & w_- & \tilde{w}_- \\
w_+' & \tilde{w}_+' & w_-' & \tilde{w}_-' \end{array} \right|.
\end{equation}
\label{definition-Evans-function}
\end{definition}

\begin{remark}
{\rm Because the Wronskian determinant of any four particular
solutions of the ODE (\ref{spectral-problem-GP}) is independent of
$x$, the values of $E(\lambda,\epsilon)$ are independent of $x$. }
\end{remark}

\begin{lemma}
\label{lemma-Evans-analyticity} Let $\phi_{\epsilon}(x)$ be the
kink mode of Definition \ref{definition-kink}, $\lambda =
\frac{1}{2} \kappa_- \kappa_+$ and $\kappa_+ = \sqrt{4 c^2 -
\kappa_-^2}$. The four fundamental solutions and the Evans
function $E(\lambda,\epsilon)$ of the spectral problem
(\ref{spectral-problem-GP}) are analytically continued in variable
$\kappa_-$ near $\kappa_- = 0$ for sufficiently small $\epsilon$.
\end{lemma}

\begin{proof}
Let us unfold the branch point $\lambda = 0$ with the
transformation
\begin{equation}
\label{unfolding-transformation} \lambda = \frac{\kappa_+
\kappa_-}{2}, \qquad c^2 = \frac{\kappa_+^2 + \kappa_-^2}{4}.
\end{equation}
The spectral problem (\ref{spectral-problem-GP}) is rewritten
explicitly as follows:
\begin{equation}
\label{coupled-system-help} \left[ - \partial_x^2 + \kappa_+^2 +
\kappa_-^2 + 2 V_+(x) \right]
 u = - \kappa_+ \kappa_- w, \qquad
\left[ - \partial_x^2 + 2 V_-(x) \right] w = \kappa_+ \kappa_- u,
\end{equation}
where
$$
V_+(x) = f(\phi_{\epsilon}^2) - f(q_0) + 2 \phi_{\epsilon}^2
f'(\phi_{\epsilon}^2) - 2 q_0 f'(q_0) + \epsilon V(x), \qquad
V_-(x) = f(\phi_{\epsilon}^2) - f(q_0) + \epsilon V(x).
$$
Since the ODE system (\ref{coupled-system-help}) depends
analytically on $(\kappa_+,\kappa_-) \in \mathbb{C}^2$ and the
boundary conditions
(\ref{boundary-conditions-1})--(\ref{boundary-conditions-2}) are
also analytic in variables $(\kappa_+,\kappa_-)$, the four
fundamental solutions are analytic on $(\kappa_+,\kappa_-) \in
\mathbb{C}^2$ and so is the Evans function $E(\lambda,\epsilon)$
as a determinant of analytic functions at any fixed $x \in
\mathbb{R}$. The unfolding transformation
(\ref{unfolding-transformation}) implies that the parameter $c \in
\mathbb{C}$ is arbitrary. Since $c \in \mathbb{R}_+$ is fixed,
this leads to the constraint $\kappa_+ = \sqrt{4 c^2 -
\kappa_-^2}$, which is locally analytic near $\kappa_- = 0$. The
analytic functions on $(\kappa_+,\kappa_-) \in \mathbb{C}^2$ with
the locally analytic constraint $\kappa_+ = \sqrt{4 c^2 -
\kappa_-^2}$ are locally analytic functions in a neighborhood of
$\kappa_- = 0$ for any sufficiently small $\epsilon$.
\end{proof}

\begin{remark}
{\rm The Evans function was constructed in \cite{KR00} for the
cubic NLS equation with a perturbation. It was discussed in
\cite{KR00} under a general set of assumptions that the function
$E(\lambda)$ is analytic in a small domain near $\lambda = 0$ with
${\rm Re} \lambda > 0$, its zeros coincide with eigenvalues
$\lambda$ with the account of their algebraic multiplicities, and
it is analytically continued in the variable $\kappa_-$ near the
point $\lambda = 0$ ($\kappa_- = 0$). Our analysis of Lemma
\ref{lemma-Evans-analyticity} is different. It is based on the
unfolding transformation (\ref{unfolding-transformation})
similarly to the recent work \cite{DPC06}. }
\end{remark}

\begin{example}
{\rm Continuing Example \ref{example-bifurcations}, we compute the
Evans function $E(\lambda)$ explicitly for the cubic NLS with
$f(s) = s$ and $q_0 = 1$. There exist explicit solutions of the
spectral problem (\ref{spectral-problem}) for the cubic NLS (see,
e.g. \cite{KT88}). By using these solutions, we obtain the
explicit representation of the eigenvectors $(u,w)$ satisfying the
boundary conditions (\ref{boundary-conditions-1}) and
(\ref{boundary-conditions-2}):
\begin{equation}
\label{explicit-formulas-1} u_{\pm} = -\frac{2}{2 + \kappa_{\pm}}
e^{\kappa_{\pm} x} \left( {\rm sech}^2 x + \kappa_{\pm} \tanh x -
\frac{1}{2} \kappa_{\pm}^2 \right), \qquad w_{\pm} =
-\frac{\kappa_{\mp}}{2 + \kappa_{\pm}} e^{\kappa_{\pm} x} \left(
\kappa_{\pm} - 2 \tanh x \right)
\end{equation}
and
\begin{equation}
\label{explicit-formulas-2} \tilde{u}_{\pm} = -\frac{2}{2 +
\kappa_{\pm}} e^{- \kappa_{\pm} x} \left( {\rm sech}^2 x -
\kappa_{\pm} \tanh x - \frac{1}{2} \kappa_{\pm}^2 \right), \qquad
\tilde{w}_{\pm} = - \frac{\kappa_{\mp}}{2 + \kappa_{\pm}}
e^{-\kappa_{\pm} x} \left( \kappa_{\pm} + 2 \tanh x \right).
\end{equation}
The Evans function $E(\lambda)$ is computed explicitly as the
determinant of the four fundamental solutions in the form
\begin{equation}
\label{explicit-Evans} E(\lambda) = \frac{4 \kappa_+^3 \kappa_-^3
( \kappa_+^2 - \kappa_-^2)^2}{(\kappa_+ + 2)^2 (\kappa_- + 2)^2},
\end{equation}
such that $E(\lambda) = 8 \lambda^3 \left( 1 - \lambda + {\rm
O}(\lambda^2) \right)$ as $\lambda \to 0$ with ${\rm Re} \lambda >
0$. The validity of all explicit formulas has been confirmed by
means of the Wolfram's Mathematica. } \label{example-Evans-cubic}
\end{example}

\begin{theorem}
\label{theorem-analyticity} Let $f(q)$ be
$C^{\infty}(\mathbb{R}_+)$ and $V(x)$ be $C^2(\mathbb{R})$
satisfying (\ref{potential}). Let $P_r' |_{v \downarrow 0} \neq 0$
and $M''(s_0) = 0$ in Theorems \ref{theorem-continuation} and
\ref{theorem-stability-kink}. Let $\gamma = \Gamma(\epsilon)$ be
an eigenvalue of the generalized eigenvalue problem
(\ref{generalized-eigenvalue-problem-Lpm}) for sufficiently small
$\epsilon$, such that $w \in L^2(\mathbb{R})$, ${\rm Im} \Gamma
\leq 0$, and $\lim\limits_{\epsilon \to 0} \Gamma(\epsilon) = 0$.
Then, the spectral problem (\ref{spectral-problem-GP}) for
sufficiently small $\epsilon$ admits two eigenvalues $\lambda =
\pm \Lambda(\epsilon) = \pm \sqrt{-\Gamma(\epsilon)}$ with $(u,w)
\in L^2(\mathbb{R},\mathbb{C}^2)$, ${\rm Re} \Lambda {\rm Im}
\Lambda \geq 0$, and $\lim\limits_{\epsilon \to 0}
\Lambda(\epsilon) = 0$, such that
\begin{itemize}
\item[(i)] $\Lambda(\epsilon)$ is infinitely smooth with respect to
$\epsilon^{1/2}$.

\item[(ii)] $(u,w)$ is infinitely smooth with respect to $\epsilon^{1/2}$ and
\begin{equation}
\label{limiting-relation} \lim_{\epsilon \to 0} u(x) = \phi_0'(x),
\qquad \lim_{\epsilon \to 0} w(x) = 0,
\end{equation}
up to an arbitrary multiplicative factor.

\item[(iii)] $(u,w)$ admits an asymptotic expansion as $\lambda \to 0$,
${\rm Re} \lambda
> 0$ for large $\pm x \gg 1$:
\begin{equation}
\label{decay-slow} \left( \begin{array}{cc} u \\ w \end{array}
\right) \to a_{\pm} \left( 1 + {\rm O}(\lambda) \right) \left(
\begin{array}{cc}  \lambda/c + {\rm O}(\lambda^3) \\ -2c + {\rm O}(\lambda^2)
\end{array} \right) \left(1 \mp \frac{\lambda x}{c} + {\rm O}(\lambda x)^2 \right),
\end{equation}
where $a_{\pm}$ are some constants and $\lambda =
\Lambda(\epsilon)$.
\end{itemize}
\end{theorem}

\begin{proof}
Under the conditions of the theorem, there must exist a zero of the
Evans function $E(\lambda,\epsilon)$ at $\lambda =
\Lambda(\epsilon)$ for sufficiently small $\epsilon$, such that
${\rm Re} \Lambda {\rm Im} \Lambda \geq 0$ and
$\lim\limits_{\epsilon \to 0} \Lambda(\epsilon) = 0$.

(i) By Lemma \ref{lemma-Evans-analyticity}, the Evans function
$E(\lambda,\epsilon)$ is analytically continued in $\kappa_- =
\lambda/c + {\rm O}(\lambda^3)$ near $\lambda = 0$. It is also
infinitely smooth in $\epsilon$ near $\epsilon = 0$. (Indeed, the
potential terms $V_{\pm}(x)$ in the representation
(\ref{coupled-system-help}) are infinitely smooth in $\epsilon$ and
exponentially decaying as $|x| \to \infty$.) In addition, the Evans
function $E(\lambda,\epsilon)$ has the following properties:
$$
E(\lambda,0) = \alpha \lambda^3 + {\rm O}(\lambda^4), \qquad
E(0,\epsilon) = 0,
$$
where $\alpha$ is a numerical constant. According to Lemma
\ref{lemma-kernel}, the triple root of $E(\lambda,0)$ corresponds
to the generalized kernel (i) due to translational invariance and
the bounded solutions (ii) due to the gauge invariance. The former
subspace results in the double root $\lambda = 0$ of
$E(\lambda,0)$ (as a proper eigenvalue), while the latter subspace
results in a single root $\kappa_- = 0$ ($\lambda = 0$) of
$E(\lambda,0)$ (as a proper resonance) \cite{KS04}. By Lemma
\ref{lemma-kernel}, $\alpha \neq 0$ is equivalent to the condition
$P_r' |_{v \downarrow 0} \neq 0$. The constraint $E(0,\epsilon) =
0$ follows from existence of $\phi_{\epsilon} \in
L^{\infty}(\mathbb{R})$ such that ${\cal L}_- \phi_{\epsilon} =
0$. As a result, the Evans function is expanded near $\lambda = 0$
and $\epsilon = 0$ as follows:
\begin{equation}
\label{power-expansion} E(\lambda,\epsilon) = \lambda \left(
\alpha \lambda^2 + \beta \epsilon + {\rm O}(\lambda^3, \lambda
\epsilon, \epsilon^2) \right),
\end{equation}
where $\beta$ is another numerical constant. Since the gauge
invariance is preserved while the translational invariance of dark
solitons is destroyed by Theorem \ref{theorem-persistence}, $\beta
\neq 0$ is equivalent to the condition that $M''(s_0) \neq 0$. It
follows from expansion (\ref{power-expansion}) and the smoothness
of $E(\lambda,\epsilon)$ that the root $\Lambda(\epsilon)$ is
infinitely smooth in $\epsilon^{1/2}$.

(ii) By the construction of the Evans function
$E(\lambda,\epsilon)$, the eigenvector $(u,w)$ is spanned by the
four fundamental solutions in Lemma
\ref{lemma-fundamental-solutions}. By Lemma
\ref{lemma-Evans-analyticity}, these solutions are analytic in
$\kappa_-$ near $\kappa_- = 0$. By Remark
\ref{remark-expansion-kappa}, $\kappa_- = \lambda/c + {\rm
O}(\lambda^3)$ near $\lambda = 0$ and by (i) of Theorem
\ref{theorem-analyticity}, the root $\lambda = \Lambda(\epsilon)$
of $E(\lambda,\epsilon) = 0$ is infinitely smooth in
$\epsilon^{1/2}$. Therefore, the eigenvector $(u,w)$ is also
infinitely smooth in $\epsilon^{1/2}$. By Lemma
\ref{lemma-kernel}, the kernel of operators ${\cal L}_+$ and
${\cal L}_-$ is one-dimensional in $L^2(\mathbb{R})$, such that
the limiting relation (\ref{limiting-relation}) follows by
smoothness (eigenvectors are defined up to an arbitrary
multiplicative factor).

(iii) It follows from the decay (\ref{boundary-conditions-1}) and
(\ref{boundary-conditions-2}) for the four fundamental solutions
that there exist constants $A_{\pm}(\lambda)$ and
$B_{\pm}(\lambda)$, such that
\begin{equation}
\label{decay-slow-2} \left( \begin{array}{cc} u \\ w \end{array}
\right) \to A_{\pm} \left( \begin{array}{cc} \kappa_- \\
- \kappa_+ \end{array} \right) e^{\mp \kappa_- x} + B_{\pm} \left(
\begin{array}{cc} \kappa_+ \\ - \kappa_-
\end{array} \right) e^{\mp \kappa_+ x} \quad \mbox{as} \quad x \to \pm
\infty,
\end{equation}
where $\kappa_{\pm}$ are defined by the characteristic equation
(\ref{equation-kappa}) with $\lambda = \Lambda(\epsilon)$. By
expansion (\ref{expansion-kappa-pm}), the expansion of slowly
decaying solutions $e^{\mp \kappa_- x}$ gives (\ref{decay-slow}),
while expansion of fast decaying solutions $e^{\mp \kappa_+ x}$ is
not written.
\end{proof}

\begin{theorem}
Let $f(q)$ be $C^2(\mathbb{R}_+)$ and $V(x)$ satisfy
(\ref{potential}). Let $P_r' |_{v \downarrow 0} \neq 0$ and $S'
|_{v \downarrow 0} \neq 0$ for the black soliton $\phi_0(x)$ of
Definition \ref{definition-black-soliton}. Let $\lambda =
\Lambda(\epsilon)$ be a small eigenvalues of the spectral problem
(\ref{spectral-problem-GP}) for sufficiently small $\epsilon$ with
${\rm Re} \Lambda > 0$. Then, the value of $\Lambda(\epsilon)$ is
given by the root of the characteristic equation:
\begin{equation}
\label{characterstic-equation-s} {\rm Re} \Lambda > 0 : \quad
\left( P_r' |_{v \downarrow 0}\right) \Lambda^2 - \epsilon
\frac{q_0 (S' |_{v \downarrow 0})^2 M''(s_0)}{2 c (P_r' |_{v
\downarrow 0})} \Lambda + \epsilon M''(s_0) = {\rm O}(\epsilon^2).
\end{equation}
\label{theorem-splitting}
\end{theorem}

\begin{proof}
By Remark \ref{remark-smoothness}, the spectral problem
(\ref{spectral-problem-GP}) for sufficiently small $\epsilon$ can be
written in the form:
$$
\left[ L_+ + \epsilon M_+ + {\rm O}(\epsilon^2) \right] u = -
\lambda w, \qquad \left[ L_- + \epsilon M_- + {\rm O}(\epsilon^2)
\right] w = \lambda u.
$$
By Theorem \ref{theorem-analyticity}, the eigenvector $(u,w)$ and
the eigenvalue $\lambda = \Lambda(\epsilon)$ can be expanded in
powers of $\epsilon^{1/2}$:
$$
u = \phi_0'(x) + \epsilon^{1/2} u_1 + \epsilon u_2 +
\epsilon^{3/2} u_3 + {\rm O}(\epsilon^2), \quad w = \epsilon^{1/2}
w_1 + \epsilon w_2 + \epsilon^{3/2} w_3 + {\rm O}(\epsilon^2),
$$
and $\lambda = \epsilon^{1/2} \lambda_1 + \epsilon \lambda_2 +
\epsilon^{3/2} \lambda_3 + {\rm O}(\epsilon^2)$. The first-order
corrections terms satisfy the system
\begin{equation}
\label{help-equation-1} L_+ u_1 = 0, \qquad L_- w_1 = \lambda_1
\phi'_0.
\end{equation}
By the expansion (\ref{decay-slow}), we are looking for a solution
with $u_1  \in L^2(\mathbb{R})$ and $w_1 \in
L^{\infty}(\mathbb{R})$. By Lemma \ref{lemma-kernel}, the explicit
solution is
$$
u_1 = c_1 \phi'_0(x), \qquad w_1 = a_1 \phi_0(x) - \lambda_1 {\rm
Im} \partial_v U(x) |_{v \downarrow 0},
$$
where $(a_1,c_1)$ are parameters. Without loss of generality, we
can set $c_1 = 0$. The second-order corrections terms satisfy the
system
\begin{equation}
\label{help-equation-2} L_+ u_2 = - \lambda_1 w_1 - M_+ \phi_0',
\qquad L_- w_2 = \lambda_2 \phi'_0
\end{equation}
By the Fredholm alternative for $L_+$, we obtain the constraint
$$
-\lambda_1 (\phi_0', w_1) - (\phi_0',M_+ \phi_0') = 0.
$$
By using (\ref{momentum-identity-2}) and (\ref{splitting-zero}), the
constraint is equivalent to the characteristic equation
\begin{equation}
\label{charactersitic-help-1} \frac{1}{2} (P_r' |_{v \downarrow
0}) \lambda_1^2 + \frac{1}{2} M''(s_0) = 0.
\end{equation}
Since $w_1 \in L^{\infty}(\mathbb{R})$ and the Fredholm constraint
is satisfied, there exists a solution $u_2 \in
L^{\infty}(\mathbb{R})$ of the first inhomogeneous equation
(\ref{help-equation-2}). There exists also a solution $w_2 \in
L^{\infty}(\mathbb{R})$ of the second inhomogeneous equation
(\ref{help-equation-2}) similarly to the solution $w_1 \in
L^{\infty}(\mathbb{R})$. By the expansion (\ref{decay-slow}), we
shall add a homogeneous linearly growing solution (iii) of Lemma
\ref{lemma-kernel}  which is not in $L^{\infty}(\mathbb{R})$, such
that $w_2 \notin L^{\infty}(\mathbb{R})$. As a result, the
second-order corrections terms are written in the form
\begin{eqnarray*}
u_2 & = & c_2 \phi'_0(x) + \tilde{u}_2(x), \\ w_2 & = & a_2
\phi_0(x) - \lambda_2 {\rm Im} \partial_v U(x) |_{v \downarrow 0}
+ b_2 ( x \phi_0(x) - {\rm Im} \partial_v U(x) |_{v \downarrow
0}),
\end{eqnarray*}
where $(a_2,b_2,c_2)$ are parameters ($c_2 = 0$ without loss of
generality) and $\tilde{u}_2 \in L^{\infty}(\mathbb{R})$ is a
solution of  the inhomogeneous equation
$$
L_+ \tilde{u}_2 = - \lambda_1 a_1 \phi_0 - \frac{M''(s_0)}{P_r'
|_{v \downarrow 0}} {\rm Im} \partial_v U(x) |_{v \downarrow 0} -
M_+ \phi_0'.
$$
The third-order corrections terms satisfy the system
\begin{equation}
\label{help-equation-3} L_+ u_3 = - \lambda_2 w_1 - \lambda_1 w_2,
\qquad L_- w_3 = \lambda_3 \phi'_0 + \lambda_1 u_2 - M_- w_1.
\end{equation}
By the Fredholm alternative for $L_+$, we obtain the constraint
$$
-\lambda_2 (\phi_0', w_1) - \lambda_1 (\phi_0',w_2) = 0
$$
which is equivalent by virtue of (\ref{zero-projection}) to the
characteristic equation
\begin{equation}
\label{charactersitic-help-2} (P_r' |_{v \downarrow 0}) \lambda_1
\lambda_2 + \frac{q_0 (S' |_{v \downarrow 0})}{2}  \lambda_1 b_2 =
0.
\end{equation}
Combining two corrections $\lambda_1$ and $\lambda_2$ in
$\Lambda(\epsilon) = \epsilon^{1/2} \lambda_1 + \epsilon \lambda_2
+ {\rm O}(\epsilon^{3/2})$, we rewrite the characteristic
equations (\ref{charactersitic-help-1}) and
(\ref{charactersitic-help-2}) in the form
\begin{equation}
\label{characteristic-equation} (P_r' |_{v \downarrow 0}) \Lambda^2
+ \epsilon q_0 (S' |_{v \downarrow 0}) b_2 \Lambda + \epsilon
M''(s_0) = {\rm O}(\epsilon^2).
\end{equation}
In order to find $b_2$ for ${\rm Re} \Lambda > 0$, we need to
consider $w(x) = \epsilon^{1/2} w_1 + \epsilon w_2 + {\rm
O}(\epsilon^{3/2})$ for large $\pm x \gg 1$:
\begin{equation}
\label{decay-slow-3} w(x) \to \pm \sqrt{q_0} \left( \epsilon^{1/2}
\left( a_1 - \lambda_1 \partial_v \Theta^{\pm} |_{v \downarrow 0}
\right) + \epsilon \left( a_2 - \lambda_2 \partial_v \Theta^{\pm}
|_{v \downarrow 0}  + b_2 ( x - \partial_v \Theta^{\pm} |_{v
\downarrow 0} ) \right) + {\rm O}(\epsilon^{3/2}) \right)
\end{equation}
where $\Theta^{\pm} = \lim\limits_{x \to \pm \infty} \Theta(x)$
and $\lim\limits_{x \to \pm \infty} \phi_0(x) = \pm \sqrt{q_0}$
have been used. Matching the asymptotic expansions
(\ref{decay-slow}) and (\ref{decay-slow-3}), we find a linear
system on parameters $(a_1,b_2)$:
$$
c b_2 = \mp \lambda_1 \left( a_1 - \lambda_1  \partial_v
\Theta^{\pm} |_{v \downarrow 0} \right).
$$
The linear system has the explicit solution
$$
2 c b_2 = \lambda_1^2 \partial_v \left( \Theta^+ - \Theta^-
\right) |_{v \downarrow 0}, \qquad 2 \lambda_1 a_1 = \lambda_1^2
\partial_v \left( \Theta^+ + \Theta^- \right) |_{v \downarrow 0},
$$
such that
$$
b_2 = \frac{(S' |_{v \downarrow 0})}{2c}  \lambda_1^2 = -
\frac{(S' |_{v \downarrow 0}) M''(s_0)}{2 c (P_r' |_{v \downarrow
0})},
$$
where $S = \Theta^+ - \Theta^-$. As a result, the characteristic
equation (\ref{characteristic-equation}) reduces to the form
(\ref{characterstic-equation-s}).
\end{proof}

\begin{corollary}
\label{corollary-splitting} Let $P_r' |_{v \downarrow 0} \neq 0$ and
$S' |_{v \downarrow 0} \neq 0$ for the kink mode of Theorems
\ref{theorem-continuation}, \ref{theorem-stability-kink-mode},
\ref{theorem-persistence} and \ref{theorem-splitting} for
sufficiently small $\epsilon$. Then,
\begin{itemize}
\item If $P_r' |_{v \downarrow 0} > 0$, the kink mode with $M''(s_0) > 0$ has
precisely one quartet of small complex eigenvalues ($N_r = N_i^- =
0$, $N_c = 1$), while the kink mode with $M''(s_0) < 0$ has
precisely one pair of small real eigenvalues ($N_r = 1$, $N_i^- =
N_c = 0$) in the spectral problem (\ref{spectral-problem-GP}).

\item If $P_r' |_{v \downarrow 0} < 0$, the kink mode with $M''(s_0) > 0$ has
precisely one pair of small real eigenvalues and one pair of finite
real eigenvalues ($N_r = 2$, $N_i^- = N_c = 0$), while the kink mode
with $M''(s_0) < 0$ has precisely one pair of finite real
eigenvalues and no small eigenvalues ($N_r = 1$, $N_i^- = N_c = 0$)
in the spectral problem (\ref{spectral-problem-GP}).
\end{itemize}
\end{corollary}

\begin{remark}
{\rm Comparison of the characteristic equation
(\ref{characterstic-equation-s}) and the linearized version of the
Newton's particle law (\ref{Newton-law}) shows that
$$
\mu_0 = P_r' |_{v \downarrow 0}, \qquad \lambda_0 = \frac{q_0 (S'
|_{v \downarrow 0})^2}{2 c (P_r' |_{v \downarrow 0})}.
$$
Both constants are positive if $P_r' |_{v \downarrow 0} > 0$, i.e.
if the kink is stable in the spectral problem
(\ref{spectral-problem}).} \label{remark-constants}
\end{remark}

\begin{remark}
\label{remark-contradiction} {\rm Characteristic equation
(\ref{characterstic-equation-s}) can be derived from the power
expansion of the Evans function $E(\lambda,\epsilon)$ of
Definition \ref{definition-Evans-function}:
\begin{equation}
\label{power-expansion-2} E(\lambda,\epsilon) = \lambda \left(
\alpha \lambda^2 + \tilde{\alpha} \lambda^3 + \beta \epsilon +
\tilde{\beta} \lambda \epsilon + {\rm O}(\lambda^4, \lambda^2
\epsilon, \epsilon^2) \right),
\end{equation}
where $(\alpha,\beta)$ are constants from the expansion
(\ref{power-expansion}) and $(\tilde{\alpha},\tilde{\beta})$ are new
constants. Computations of these constants from derivatives of
$E(\lambda,\epsilon)$ are technically involved (see
\cite{KR00,SS05}). These computations are replaced in Theorem
\ref{theorem-splitting} with direct expansions of eigenvectors and
eigenvalues of the spectral problem (\ref{spectral-problem-GP}) in
powers of $\epsilon^{1/2}$. }
\end{remark}

\begin{example}
{\rm Continuing Example \ref{example-bifurcations-2} we consider
the cubic NLS with $f(s) = s$ and $q_0 = 1$, where $P_r' |_{v
\downarrow 0} = 4$ and $S' |_{v \downarrow 0} = 2$. As a result,
the characteristic equation (\ref{characterstic-equation-s}) is
written explicitly by
\begin{equation}
\label{approximated-eigenvalues} {\rm Re} \Lambda > 0 : \quad
\Lambda^2 + \frac{\epsilon}{4} M''(s_0) \left( 1 -
\frac{\Lambda}{2} \right) = {\rm O}(\epsilon^2).
\end{equation}
This equation has only one real-valued root $\Lambda(\epsilon) > 0$
for $M''(s_0) < 0$ and two complex-conjugate roots with ${\rm
Re}\Lambda(\epsilon) > 0$ for $M''(s_0) > 0$ provided that $\epsilon
> 0$ is sufficiently small. The validity of the expansion
(\ref{approximated-eigenvalues}) will be tested in Section
\ref{Numerics1}.}

\end{example}

\begin{remark}
{\rm If the characteristic equation
(\ref{approximated-eigenvalues}) is formally applied to the cubic
GP equation (\ref{GP}) with $f(s) = s$, $q_0 = 1$ and $V(x) =
x^2$, we obtain $M''(s) = 2 \int_{\mathbb{R}} {\rm sech}^2(x) dx =
4$, such that the characteristic equation
(\ref{approximated-eigenvalues}) is
$$
\Lambda^2 - \frac{\epsilon}{2} \Lambda + \epsilon = {\rm
O}(\epsilon^2).
$$
This characteristic equation was derived in \cite{PFK05} with a
formal method for slow dynamics of dark solitons in parabolic
potentials subject to radiative boundary conditions. The validity
of radiative boundary conditions for parabolic potentials $V(x)$
can not be verified by the present analysis. }
\end{remark}

\begin{remark}
{\rm The characteristic equation (\ref{characterstic-equation-s})
can be rewritten in the form
$$
\left( P_r' |_{v \downarrow 0}\right) \Lambda^2 + \frac{q_0 (S'
|_{v \downarrow 0})^2}{2 c} \Lambda^3 = - \epsilon M''(s_0) + {\rm
O}(\epsilon^2).
$$
The left-hand-side of this equation was derived in Eqs. (19)-(20) of
\cite{PKA96} and Eqs. (2.37)--(2.38) of \cite{PG97} in a more
general context of dark solitons with $v \in (-c,c)$. The method of
\cite{PKA96} was based on asymptotic theory for slow dynamics of
dark solitons, while the method of \cite{PG97} was based on slow
decay conditions for eigenfunctions of the linearized problem. Here
we have replaced these formal methods with rigorous proof in the
framework of Theorems \ref{theorem-analyticity} and
\ref{theorem-splitting}. }
\end{remark}

\section{Numerical approximations of eigenvalues}
\label{Numerics1}

We test here the predictions of the characteristic equation
(\ref{approximated-eigenvalues}) for the cubic GP equation
(\ref{GP}) with $f(s) = s$ and the two potentials $V_1(x)$ and
$V_2(x)$ in (\ref{potential-explicit}). In particular, we focus on
examining the dependencies of small unstable eigenvalues of the
spectral problem (\ref{spectral-problem-GP}) versus parameters
$\epsilon$ and $\kappa$. The numerical approximations of the kink
mode $\phi_{\epsilon}(x)$ are obtained by means of fixed point
iterations of the ODE (\ref{second-order-ode-phi-epsilon}). The
iterations are applied to a finite-difference discretization of the
computational domain $x \in [-L,L]$ (typically $L = 10$) on a grid
of $N$ nodes (typically $N = 1600$) with a spacing $\Delta x$
(typically $\Delta x = 0.2$). Subsequently, the spectral problem
(\ref{spectral-problem-GP}) is discretized in a matrix eigenvalue
problem that, in turn, is solved through standard numerical linear
algebra routines.

In the case of the potential $V_1(x)$, as is considered in Examples
\ref{example-bifurcations} and \ref{example-bifurcations2}, the
positive-definite sign of $L''(0)$ (and hence $M''(0)$) leads to a
sole kink mode bifurcating from $s_0=0$ (and staying at $s = 0$ by
Remark \ref{remark-symmetry-kink-mode}). The kink mode is located at
the minimum of the effective potential $M(s)$ and is unstable due to
a complex quartet of eigenvalues according to the characteristic
equation (\ref{approximated-eigenvalues}).

\begin{figure}
\begin{center}
\includegraphics[height=6cm]{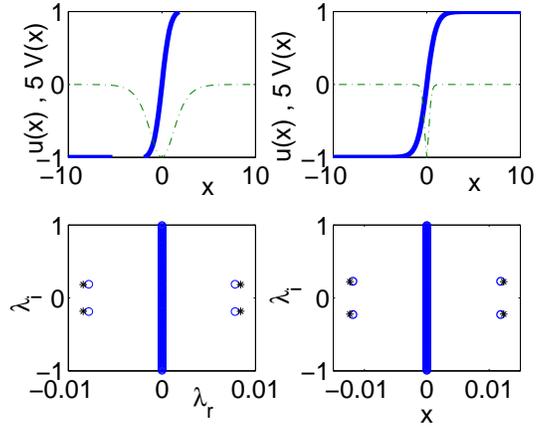}
\end{center}
\caption{Bifurcation results for the potential $V_1(x)$. The top
panels show the kink mode (solid line) and the potential $V_1(x)$
amplified by a factor of 5 (dash-dotted line) for
$(\epsilon,\kappa)=(0.2,1.1)$ (left) and
$(\epsilon,\kappa)=(0.2,6.4)$ (right). The bottom panels show the
corresponding spectrum of the linearized problem (the numerical
result is shown by circles, while the theoretical prediction is
shown by stars).} \label{dep_fig2}
\end{figure}

\begin{figure}
\begin{center}
\includegraphics[height=6cm]{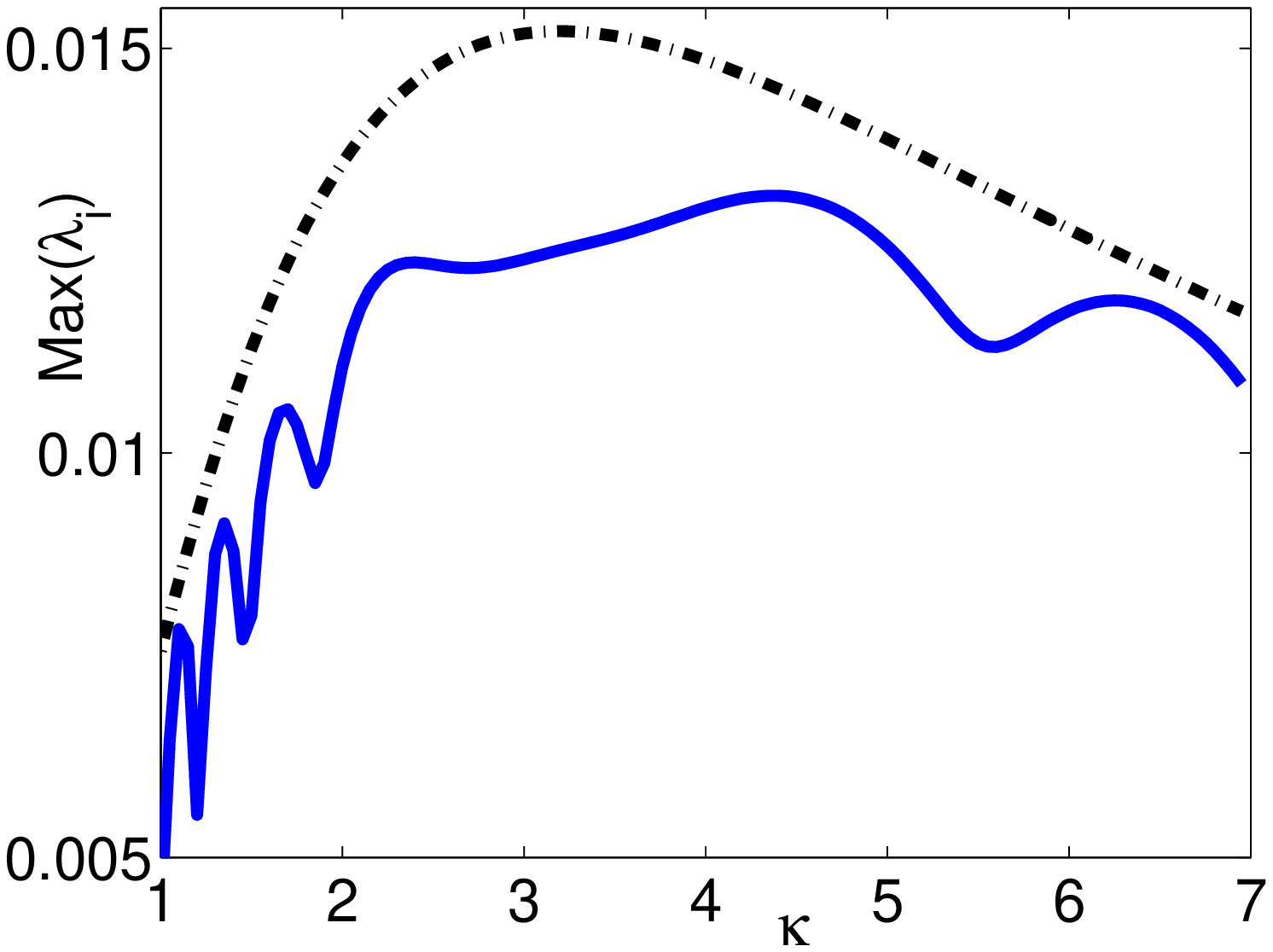}
\includegraphics[height=6cm]{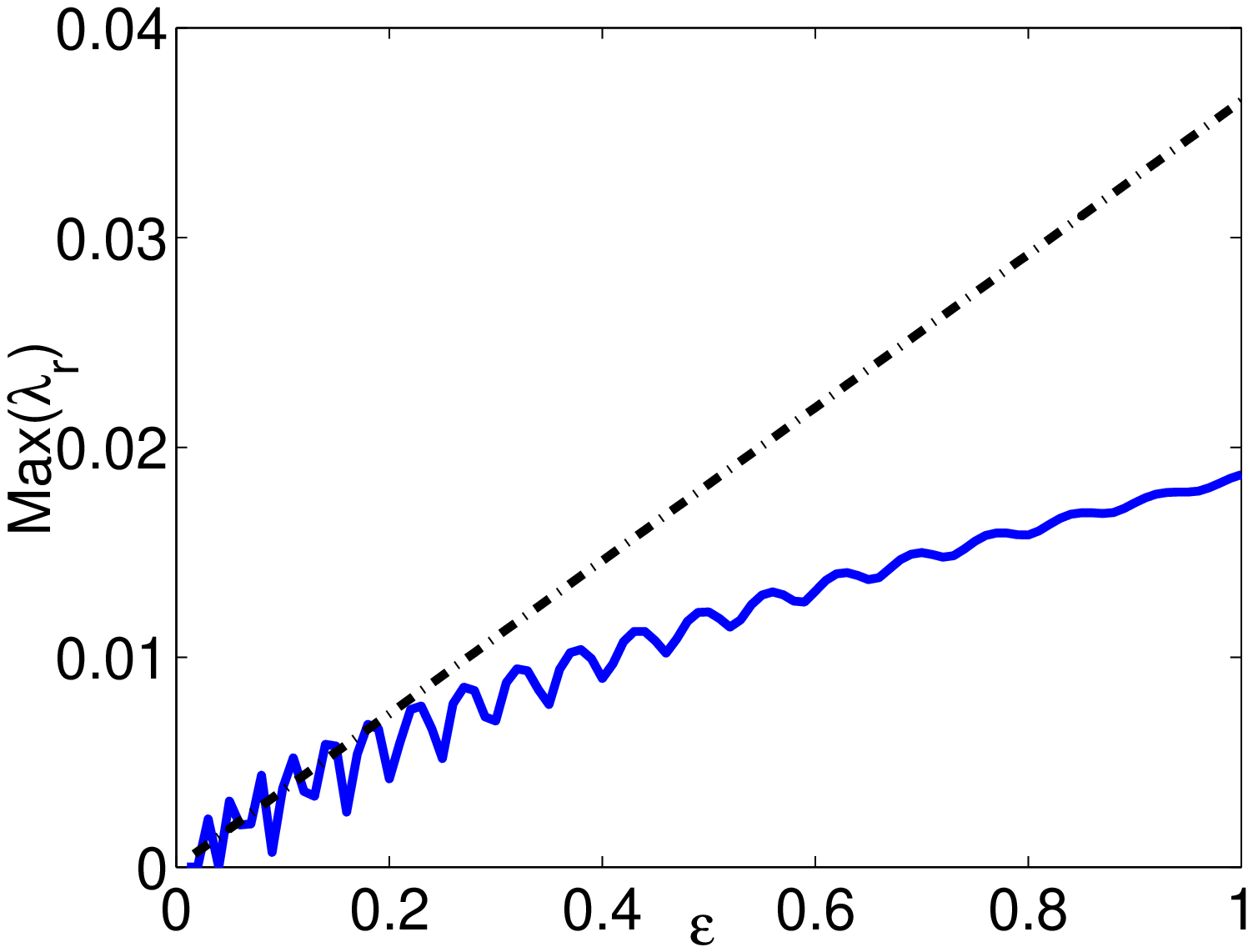}
\end{center}
\caption{The left panel shows the theoretical (dash-dotted line) and
numerical (solid line) dependence on $\kappa$ of the real part of
the unstable complex eigenvalue for fixed $\epsilon=0.2$. The right
panel shows the dependence of the same quantity on $\epsilon$ for
fixed $\kappa=1$.} \label{dep_fig2b}
\end{figure}

The numerical results on Figures \ref{dep_fig2} and
\ref{dep_fig2b} fully confirm the above picture. Fig.
\ref{dep_fig2} shows only one solution $\phi_{\epsilon}(x)$ of the
ODE (\ref{second-order-ode-phi-epsilon}) with the potential
$V_1(x)$ and a unique quartet of complex eigenvalues $\lambda =
\lambda_r + i \lambda_i$ in the spectral problem
(\ref{spectral-problem-GP}). The left panel of Fig.
\ref{dep_fig2b} shows the real part of this quartet as a function
of $\kappa$ for a given $\epsilon =0.2$, while the right panel
shows the relevant real part as a function of $\epsilon$ for a
given $\kappa = 1$. The predictions of the characteristic equation
(\ref{approximated-eigenvalues}) are shown by dashed-dotted lines,
while the numerically obtained eigenvalues are shown by thick
lines.

The non-monotonic behavior of the real part of complex eigenvalues
is produced by the truncation of the computational domain $x \in
[-L,L]$ and subsequent discretization on a finite grid. This
numerical phenomenon is explained in \cite{JK99} (see their Figure
2) as follows. The continuous spectrum of the spectral problem
(\ref{spectral-problem-GP}) becomes a finite spectral band along
the imaginary axis near $\lambda = 0$ due to the truncation and
the band is represented by isolated eigenvalues due to the
discretization. The quartet of eigenvalues bifurcates from the
point $\lambda = 0$ in the direction of the imaginary axis with
small real part for small $\epsilon$ and interferes with
eigenvalues from the discretized continuous spectrum. This
interference leads to the non-monotonic behavior of the real part
of the relevant eigenvalues on Fig. \ref{dep_fig2b}. We note that
this effect is {\it not} present for real eigenvalues bifurcating
from the point $\lambda = 0$.

In the case of the potential $V_2(x)$, we have a more interesting
phenomenology. While the potential always has two maxima at $x =
\pm 2/\kappa$, the {\it effective potential} $M(s)$ possesses two
maxima at $s_0= \pm s_*$, $s_* \neq 0$ and a minimum at $s_0 = 0$
for $\kappa < \kappa_0 \approx 3.21$ and only one maximum at $s_0
= 0$ for $\kappa > \kappa_0$. Branches of solutions with $M''(s_0)
< 0$ are unstable due to a pair of real eigenvalues, while the
branch of solutions with $M''(s_0) > 0$ is unstable due to a
quartet of complex eigenvalues, according to the characteristic
equation (\ref{approximated-eigenvalues}). The transition of the
kink mode $s_0 = 0$ from $\kappa < \kappa_0$ (when $M''(0) > 0$)
to $\kappa > \kappa_0$ (when $M''(0) < 0$) resembles the
subcritical pitchfork bifurcation at $\kappa = \kappa_0$.

\begin{figure}
\begin{center}
\includegraphics[height=6cm]{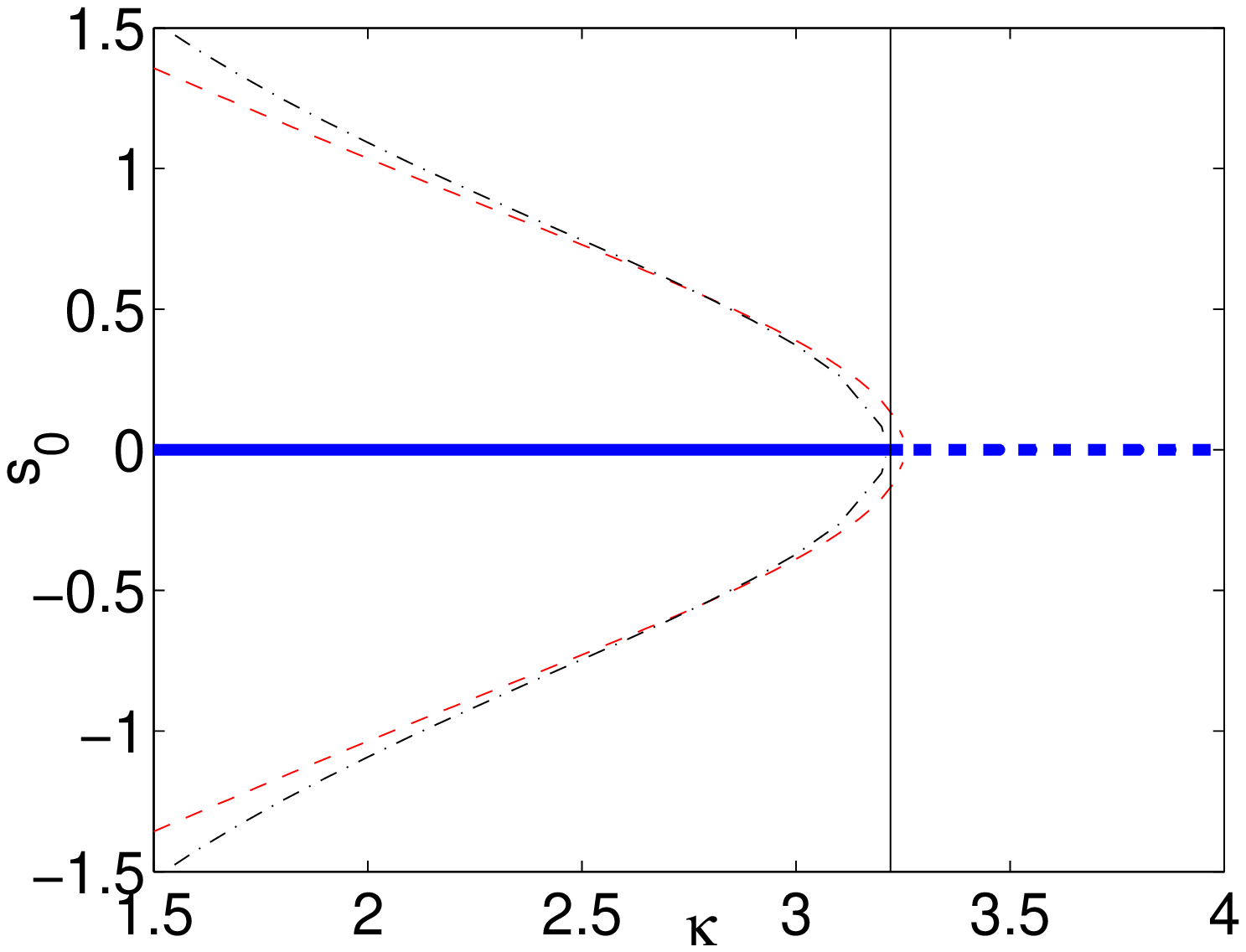}
\includegraphics[height=6cm]{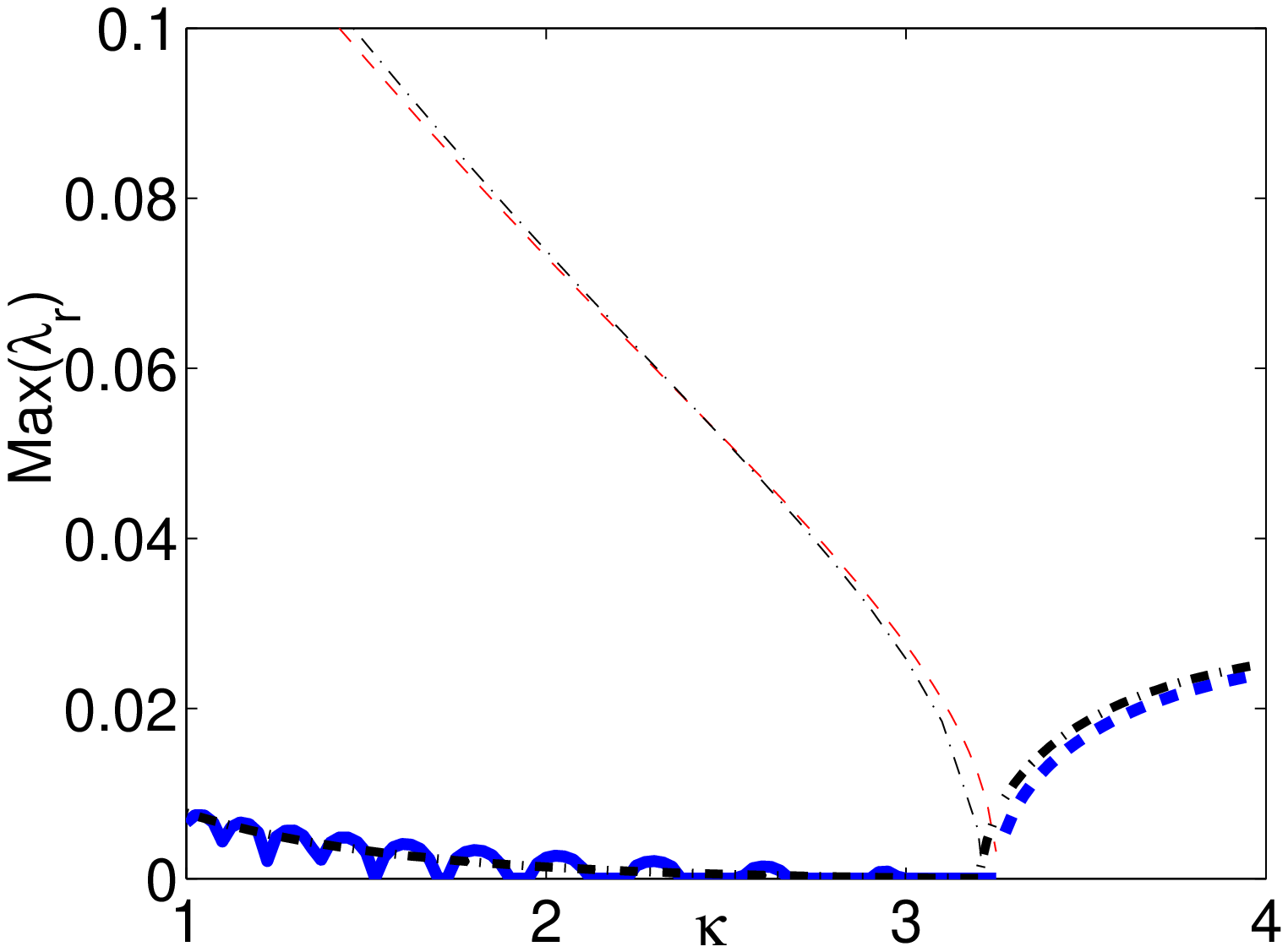}
\end{center}
\caption{The subcritical pitchfork bifurcation in parameter
$\kappa$ for the potential $V_2(x)$ for fixed $\epsilon = 0.2$.
The left panel shows the center of mass $s_0$ of the kink modes
($s_0 \neq 0$ by dashed line, $s_0 = 0$ by thick solid and dashed
lines). The theoretical predictions of $s_0$ are shown by
dash-dotted line. The vertical line gives the theoretical
prediction for the bifurcation point $\kappa = \kappa_0$. The
right panel shows the real part of the unstable eigenvalues for
the relevant kink modes, using the same symbolism as the left
panel.  The theoretical predictions of eigenvalues are shown by
thick and thin dash-dotted lines, respectively for the branches
with $s_0=0$ and $s_0 \neq 0$.} \label{dep_fig3}
\end{figure}

\begin{figure}
\begin{center}
\includegraphics[height=6cm]{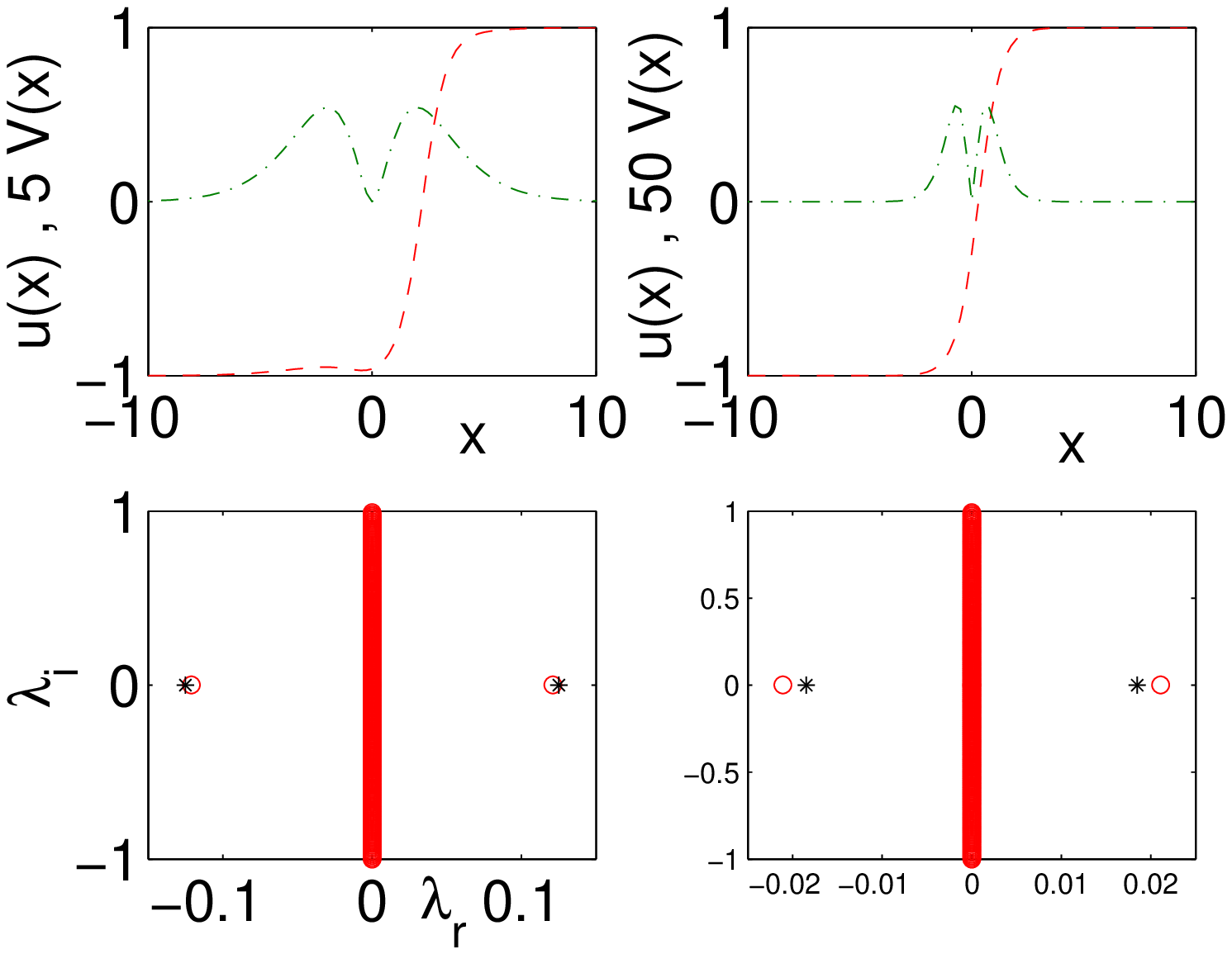}
\includegraphics[height=6cm]{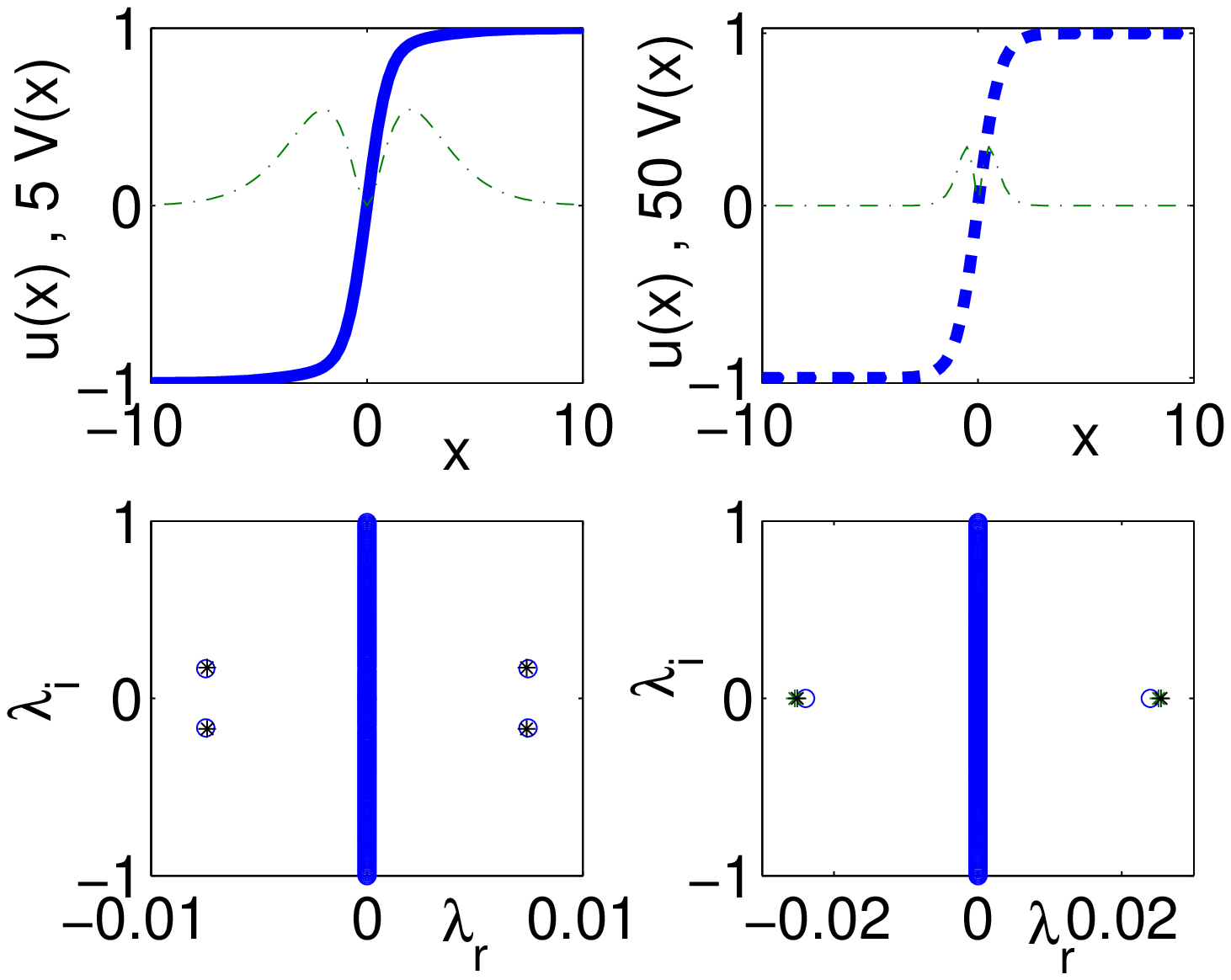}
\end{center}
\caption{The left quartet of panels shows the solutions with $s_0
\neq 0$ (dashed lines) and the potential $V_2(x)$ (dash-dotted line)
for $(\epsilon,\kappa)=(0.2,1.0)$ and $(\epsilon,\kappa)=(0.2,3.1)$.
The corresponding spectrum features a pair of real eigenvalues
(numerical results are shown by circles and the theoretical
predictions are shown by stars). The right quartet of panels shows
the similar picture for the kink mode with $s_0=0$ for
$(\epsilon,\kappa)=(0.2,1.025)$ and $(\epsilon,\kappa)=(0.2,3.975)$.
The corresponding spectrum features either a quartet of complex
eigenvalues or a pair of real eigenvalues.} \label{dep_fig4}
\end{figure}

\begin{figure}
\begin{center}
\includegraphics[height=6cm]{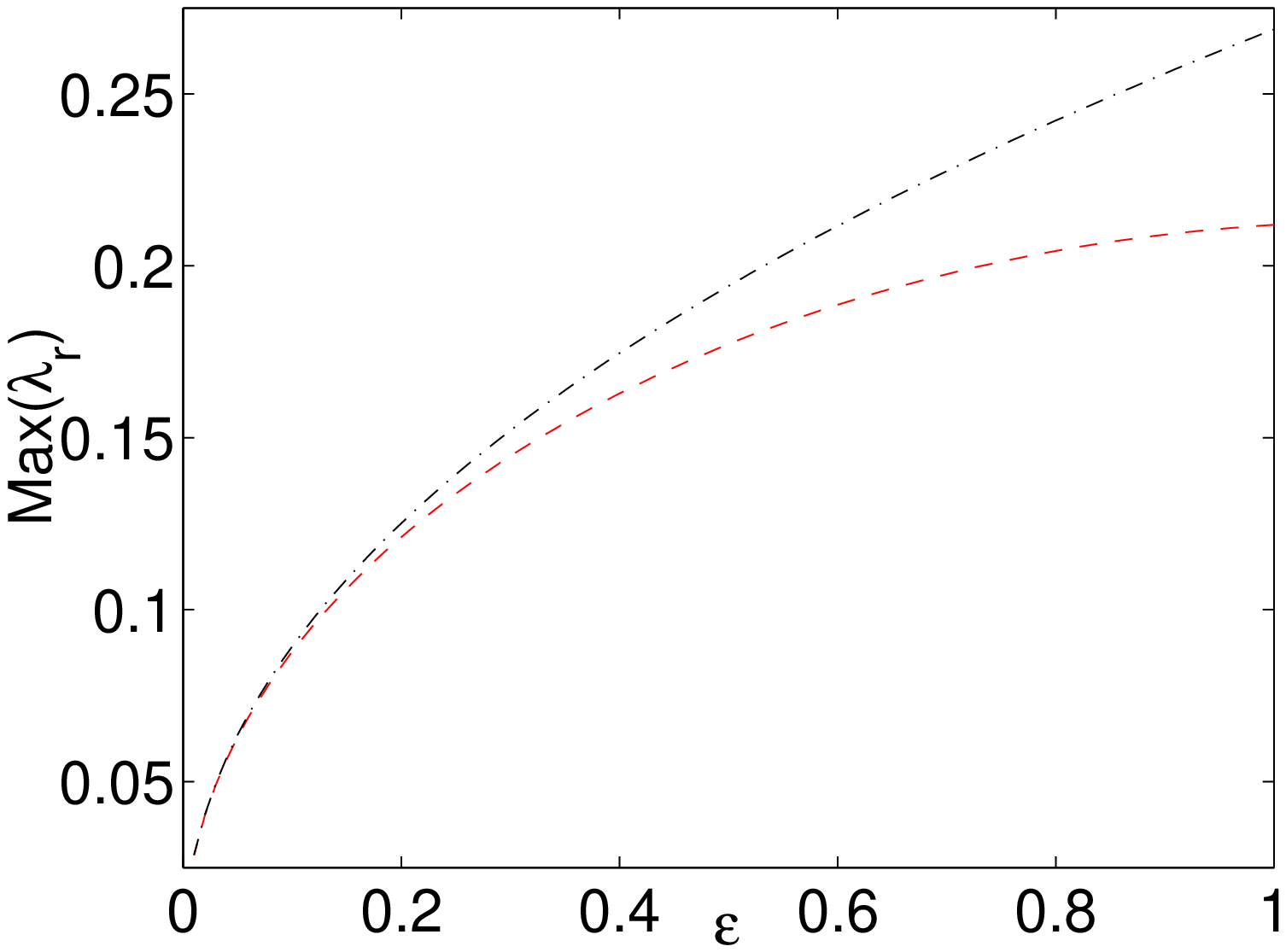}
\includegraphics[height=6cm]{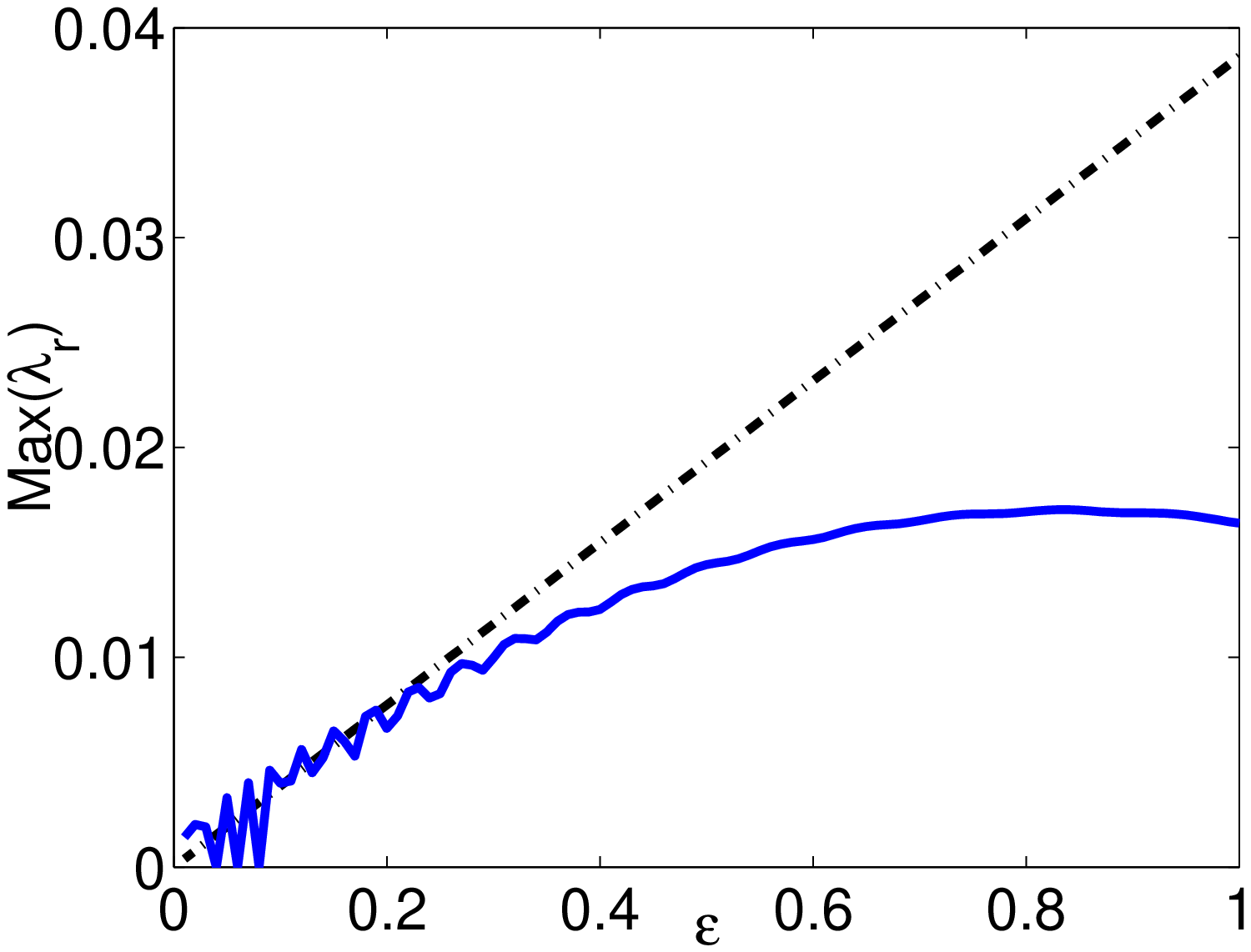}
\end{center}
\caption{The real part of the unstable eigenvalues versus $\epsilon$
for $\kappa = 1$ for the solution branches with $s_0 \neq 0$ (left
panel) and with $s_0=0$ (right panel). The numerical results are
shown by dashed and thick solid lines, while the theoretical
predictions are shown by dash-dotted lines.} \label{dep_fig5}
\end{figure}

The numerical results for the potential $V_2(x)$ are shown on
Figures \ref{dep_fig3}, \ref{dep_fig4} and \ref{dep_fig5}. Branches
of the solutions with $s_0 = \pm s_*$ are denoted by a dashed line,
the branch of the solution $s_0 = 0$ with $M''(s_0) > 0$ is denoted
by thick solid line and the same branch with $M''(s_0) < 0$ is
denoted by thick dashed line. The corresponding predictions from the
extremal points of the effective potential $M(s)$ and from the
characteristic equation (\ref{approximated-eigenvalues}) are shown
by dash-dotted lines (thick for $s_0 = 0$ and thin for $s_0 = \pm
s_*$).

The bifurcation point is found numerically to be $\kappa_0 \approx
3.26$ for $\epsilon = 0.2$ in a very good agreement with the value
$\kappa_0 = 3.21$ obtained from $M(s)$ with $M''(0) = 0$. The
computational error is approximately $1.5 \%$. The values of $s_0$
for the kink mode $\phi_{\epsilon}(x)$ are obtained numerically
from its "center of mass" defined by
\begin{equation}
\label{center-mass} s_0 = \frac{\int_{-\infty}^{\infty} x
(1-|\varphi_{\epsilon}|^2) dx}{\int_{-\infty}^{\infty}
(1-|\varphi_{\epsilon}|^2) dx}.
\end{equation}
The values of $s_0$ are plotted on the left panel of Fig.
\ref{dep_fig3} for $\epsilon = 0.2$ in a good agreement with the
value $s_*$ obtained from $M(s)$ with $M'(s_*) = 0$.

The solution profile $\phi_{\epsilon}(x)$ and the corresponding
linearization spectra for the different branches and for
particular choices of $(\epsilon,\kappa)$ are shown on Fig.
\ref{dep_fig4}. In agreement with the characteristic equation
(\ref{approximated-eigenvalues}), the kink modes with $s_0 = \pm
s_*$ for $\kappa < \kappa_0$ and with $s_0 = 0$ for $\kappa >
\kappa_0$ has a pair of small real eigenvalues, while the kink
mode with $s_0 = 0$ for $\kappa < \kappa_0$ has a quartet of small
complex eigenvalues. The quartet of complex eigenvalues for the
kink mode with $s_0=0$ and the pair of real eigenvalues for the
kink mode with $s_0 \neq 0$ exists for $\kappa < \kappa_0$, merge
at the origin at $\kappa = \kappa_0$ and lead to a pair of real
eigenvalues for the kink mode with $s_0=0$ and $\kappa>\kappa_0$.
The right panel of Fig. \ref{dep_fig3} shows the real parts of
unstable eigenvalues for each kink mode.

Fig. \ref{dep_fig5} shows the dependence of the relevant
eigenvalues versus $\epsilon$ for a fixed $\kappa = 1 < \kappa_0$.
In this case, three branches of kink modes exist and the branch
with $s_0 \neq 0$ has a pair of real eigenvalues, while the branch
with $s_0=0$ has a quartet of complex eigenvalues. We can see that
the non-monotonic behavior of the real part of unstable
eigenvalues is only observed for the quartet of complex
eigenvalues. We can also see from all figures of this section that
the agreement between numerical and theoretical results is
excellent for $\epsilon < 0.3$ and deteriorate for $\epsilon >
0.3$ due to the truncation error of the order of ${\rm
O}(\epsilon^2)$ in the characteristic equation
(\ref{approximated-eigenvalues}).

\section{Numerical simulations of the GP equation}
\label{Numerics2}

We examine here the dynamics of the unstable kink modes in the
full GP equation (\ref{GP}) by using direct numerical simulations.
The time-evolution problem is approximated by the fourth-order
Runge-Kutta method applied to the spatial discretization of the GP
equation. The output of the fixed point iteration was used as
input in the time evolution integrator with the time step $\Delta
t$ (typically $\Delta t = 0.001$). The  results of the time
evolution are compared against the effective Newton's particle
equation (\ref{Newton-law}) for the position $s(t)$ of the center
of dark soliton $\phi_{\epsilon}(x-s(t))$.

In order to test the theoretical result, we have to use the
following numerical technique. The initial condition $u(x,0)$ of
the GP equation (\ref{GP}) is specified in the form of the kink
mode $\phi_{\epsilon}(x)$ plus a small (typically $10^{-4}$)
perturbation multiple of its most unstable eigenmode. The
time-evolution problem is integrated for an initial period $0 < t
< t_0$, during which the dark soliton acquires a small speed due
to instabilities, which quickly grows for $t > t_0$. At the time
instance $t = t_0$, we approximate the values of $s(t_0)$ and
$\dot{s}(t_0)$ by using the center of mass (\ref{center-mass}) at
$t = t_0$ and at earlier time instances. The Newton's particle
equation (\ref{Newton-law}) is initialized at $t = t_0$ with given
$s(t_0)$ and $\dot{s}(t_0)$ and then integrated with the
fourth-order Runge--Kutta method.

\begin{figure}
\begin{center}
\includegraphics[height=6cm]{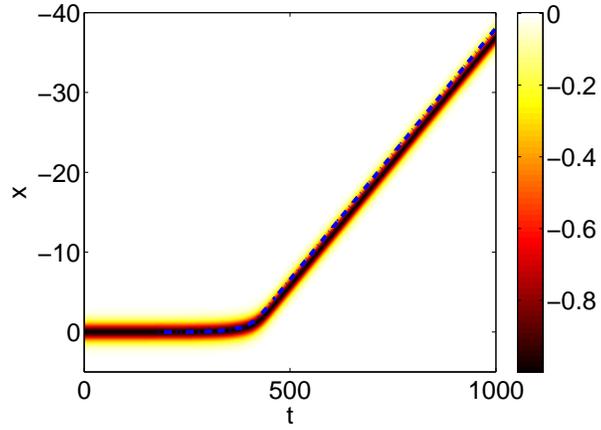}
\end{center}
\caption{The unstable evolution of the kink mode with $s_0=0$ for
the potential $V_2(x)$ with $\kappa=4$ and $\epsilon=0.2$. The
dashed-dotted line shows the result of the Newton's particle law
initialized around $t=180$.} \label{dep_fig6}
\end{figure}

\begin{figure}
\begin{center}
\includegraphics[height=6cm]{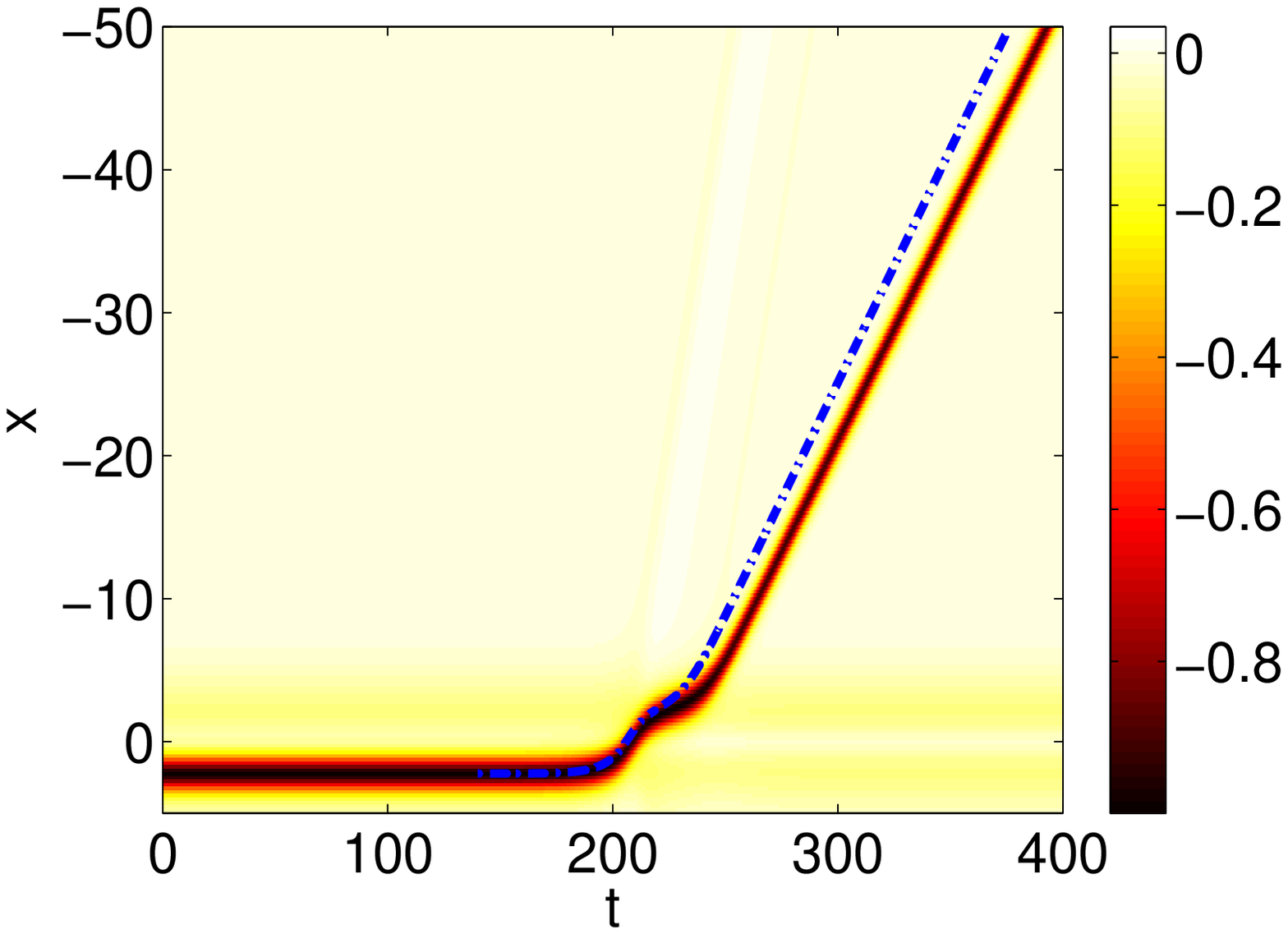}
\end{center}
\caption{The unstable evolution of the kink mode with $s_0 \approx
2.23$ for the potential $V_2(x)$ with $\kappa=1$ and $\epsilon=0.2$.
The dashed-dotted line shows the result of the Newton's particle law
initialized around $t=140$.} \label{dep_fig7}
\end{figure}

Figures \ref{dep_fig6} and \ref{dep_fig7} illustrate the time
evolution of an unstable dark soliton, which possesses a pair of
real eigenvalues in the linearization spectrum. Fig.
\ref{dep_fig6} corresponds to the potential $V_2(x)$ with $\kappa
= 4 > \kappa_0$, when the kink mode with $s_0=0$ has a real
eigenvalue $\lambda = 0.0241$ (the theoretical prediction of the
linearized Newton's particle equation is $\lambda \approx
0.0253$). We observe from the figure that the unstable kink mode
undertakes a monotonic transition to a stable dark soliton, which
escapes the double-humped potential $V_2(x)$ and travels with an
asymptotically constant speed. The predictions of the Newton's
particle equation shown by thick dash-dotted line captures the
entire process accurately but slightly precedes the full
time-evolution of the GP equation. The latter discrepancy can be
attributed to the larger values of $\lambda$ for the unstable
eigenvalues.

Fig. \ref{dep_fig7} shows a similar monotonic transition for the
potential $V_2(x)$ with $\kappa = 1$, when the kink mode with $s_0 =
s_* \approx 2.23$ has a pair of real unstable eigenvalues. In this
case, the theoretical prediction $\lambda \approx 0.1251$ exceeds
the numerically obtained value $\lambda = 0.1211$ too and the
prediction of the Newton's particle equation precedes its
counterpart from the GP equation. It is worth to note the
qualitative agreement between the two time evolutions, including the
small ``leg'' formed in the trajectory as the dark soliton passes
the unstable kink mode with $s_0=- s_* \approx -2.23$.

\begin{figure}
\begin{center}
\includegraphics[height=6cm]{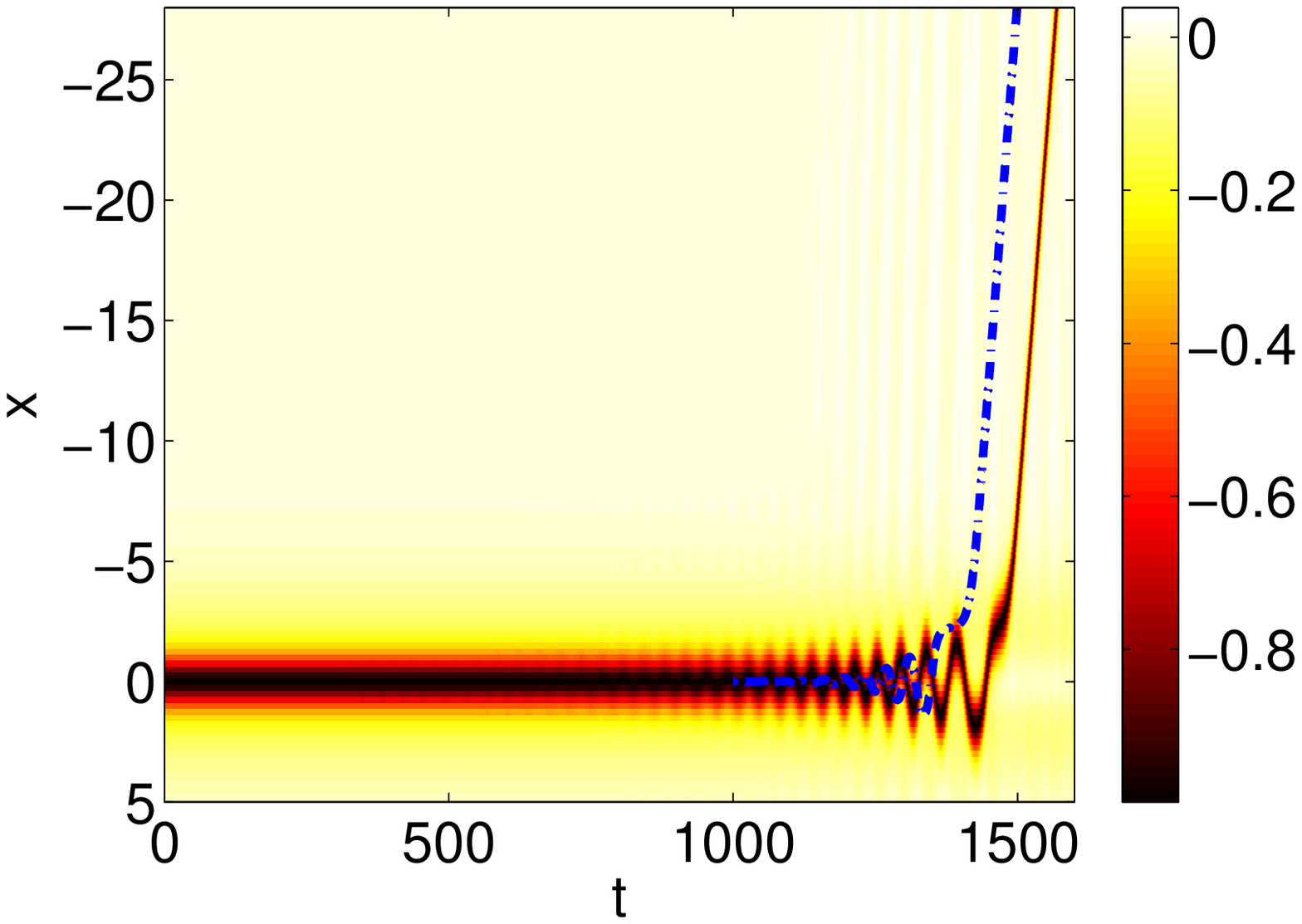}
\end{center}
\caption{The unstable evolution of the kink mode with $s_0 = 0$ for
the potential $V_2(x)$ with $\kappa=1$ and $\epsilon=0.2$. The
dashed-dotted line shows the result of the Newton's particle law
initialized at $t \approx 1000$.} \label{dep_fig8}
\end{figure}

\begin{figure}
\begin{center}
\includegraphics[height=6cm]{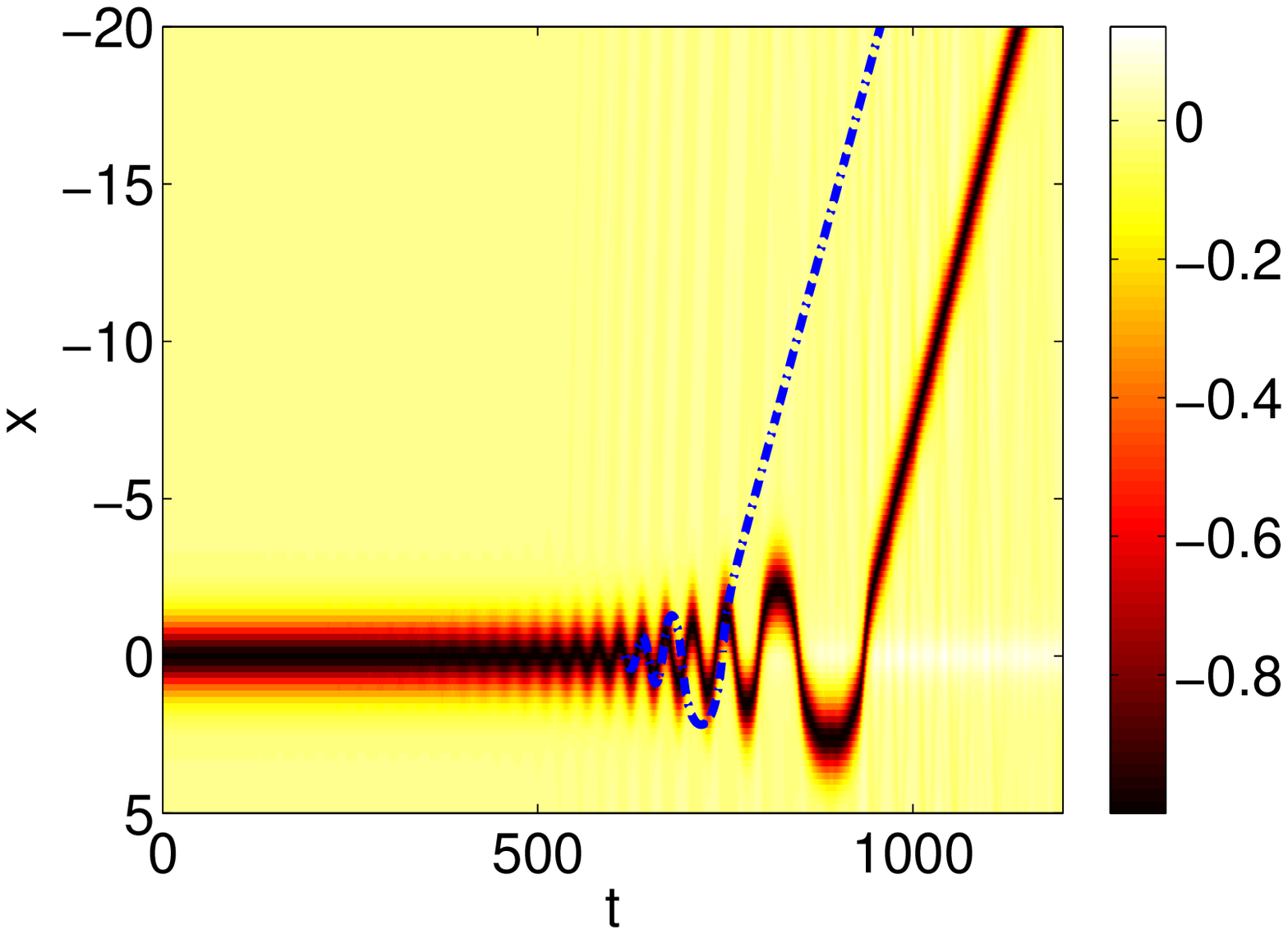}
\end{center}
\caption{The unstable evolution of the kink mode with $s_0 = 0$
for the potential $V_1(x)$ with $\kappa = 6.4$ and $\epsilon=0.2$.
The dash-dotted line shows the result of the Newton's particle law
initialized around $t \approx 600$.} \label{dep_fig9}
\end{figure}

Figures \ref{dep_fig8} and \ref{dep_fig9} illustrate the time
evolution of an unstable dark soliton, which possesses a quartet of
complex eigenvalues in the linearization spectrum. Fig.
\ref{dep_fig8} corresponds to the potential $V_2(x)$ with $\kappa =
1$, when the kink mode with $s_0=0$ has a quartet of complex
eigenvalues with the real part ${\rm Re}\lambda = 0.06606$ (the
theoretical prediction of the linearized Newton's particle equation
is ${\rm Re} \lambda \approx 0.07724$). We observe from the figure
that the unstable kink mode oscillates in a local potential well of
the double-humped potential $V_2(x)$ with an increasing amplitude
due to unstable complex eigenvalues. When the oscillations reach a
large amplitude, the dark soliton escapes the maximum of the
effective potential and transforms to a steadily moving soliton. The
predictions of the Newton's particle equation represent this
dynamics correctly with a larger deviation from the full GP equation
in comparison with the case of monotonic transitions.

Fig. \ref{dep_fig9} shows a similar oscillatory behavior of an
unstable kink mode with $s_0 = 0$ for the potential $V_1(x)$ with
$\kappa = 6.4$. In this case, the theoretical prediction ${\rm Re}
\lambda \approx 0.0123$ exceeds again the numerically obtained value
${\rm Re} \lambda = 0.0118$ and the prediction of the Newton's
particle equation precedes its counterpart from the GP equation.

In the end, we mention that the rigorous derivation of the
Newton's particle equation for slow dynamics of a bright soliton
in an external potential has been reported recently in
\cite{Br,Gus1,Gus2}. Derivation of its counterpart
(\ref{Newton-law}) for slow dynamics of a dark soliton is an open
problem of analysis. Our numerical results suggest that this
Newton's particle equation is highly appropriate for understanding
the nonlinear time-evolution of dark solitons in the GP equation
with small external potentials.

\section{Conclusion}

We have systematically analyzed the persistence and stability of
dark solitons in the presence of small decaying potentials. We
have shown how the effective potential can be used to predict
bifurcations of kink modes in a small potential and to approximate
small unstable eigenvalues of the linearization spectrum. These
theoretical results have been tested against the numerical
bifurcation results indicating excellent qualitative and good
quantitative agreement. We have also conjectured a dynamical
evolution equation (the Newton's particle law) that can be used,
quite successfully, to describe the motion of the kink modes and
the manifestation of their instabilities.

One of the directions of interest for future studies is to expand
the present results to other types of potentials which include
periodic and confining potentials. While, as argued in the text,
we expect many of the qualitative features to persist, periodic or
growing potentials may possess additional interesting properties
due to the presence of spectral bands or purely discrete spectrum
in the linearization problem.

Another open direction would involve extending the present analysis
to the two-dimensional setting and, in particular, to the case of
vortices in the presence of external potentials. While some of the
techniques applied herein (in particular, ones involving
perturbative expansions) would apply to the latter case as well,
others are more geared towards the one-dimensional setting (e.g. the
Evans function technique). It would be especially interesting to
generalize our current results to the two-dimensional GP equation.

{\bf Acknowledgement.}  This work was initiated during the workshop
``Gross-Pitaevskii equation and related equations with non-zero
boundary conditions at infinity'' held at Wolfgang Pauli Institute in
Vienna, Austria in June 2006. D.P. is supported by the NSERC
Discovery and PREA grants. P.G.K. is supported by NSF through the
grants DMS-0204585, DMS-CAREER and DMS-0505663.

\end{document}